\RequirePackage{fix-cm}
\documentclass[natbib,smallextended]{svjour3}       
\smartqed  
\usepackage{graphicx,color}
\usepackage{amssymb}
\newcommand{\alfven}{Alfv{\'e}n }
\newcommand{\RE}{ \rm{Re} }
\newcommand{\RM}{ \rm{Rm} }
\newcommand{\PM}{ \rm{Pm} }
\newcommand{\pd}[2]{ \frac{\partial #1}{\partial #2} }

\newcommand{\vct}[1]{ \mathbf #1 }

\newcommand{\be}{\begin{equation}}
\newcommand{\ee}{\end{equation}}

\begin{document}

\title{Magnetic Field Generation in Stars}

\author{Lilia Ferrario \and Andrew Melatos \and Jonathan Zrake}
\authorrunning{Ferrario, Melatos, Zrake} 

\institute{L. Ferrario \at
              Mathematical Sciences Institute, The Australian National University, ACT2601, Australia  \\     
              \email{Lilia.Ferrario@anu.edu.au} \\
           \and
                 Andrew Melatos \at
                School of Physics, University of Melbourne, Australia \\
  \email{amelatos@unimelb.edu.au} \\
         \and
                Jonathan Zrake \at Kavli Institute of Particle Astrophysics and Cosmology,
               Stanford University, USA\\
\email{zrake@stanford.edu} \\
}

\date{Received: date / Accepted: date}

\maketitle

\begin{abstract}
  Enormous progress has been made on observing stellar
    magnetism in stars from the main sequence (particularly thanks to
    the MiMeS, MAGORI and BOB surveys) through to compact
    objects. Recent data have thrown into sharper relief the vexed
    question of the origin of stellar magnetic fields, which remains
    one of the main unanswered questions in astrophysics. In this
    chapter we review recent work in this area of research. In
    particular, we look at the fossil field hypothesis which links
    magnetism in compact stars to magnetism in main sequence and
    pre-main sequence stars and we consider why its feasibility has
    now been questioned particularly in the context of highly magnetic
    white dwarfs. We also review the fossil versus dynamo debate in
    the context of neutron stars and the roles played by key physical
    processes such as buoyancy, helicity, and superfluid turbulence,
    in the generation and stability of neutron star fields.

 Independent information on the internal magnetic
    field of neutron stars will come from future gravitational wave
    detections. Coherent searches for the Crab pulsar with the Laser
    Interferometer Gravitational Wave Observatory (LIGO) have already
    constrained its gravitational wave luminosity to be $\lesssim 2$\%
    of the observed spin-down luminosity, thus placing a limit of
    $\lesssim 10^{16}$\,G on the internal field.  Indirect
    spin-down limits inferred from recycled pulsars also yield
    interesting gravitational-wave-related constraints.  Thus we may
    be at the dawn of a new era of exciting discoveries in compact
    star magnetism driven by the opening of a new, non-electromagnetic
    observational window.

We also review recent advances in the theory and
    computation of magnetohydrodynamic turbulence as it applies to
    stellar magnetism and dynamo theory. These advances offer insight
    into the action of stellar dynamos as well as processes which
    control the diffusive magnetic flux transport in stars.


\end{abstract}

\section{Introduction}\label{intro_origin}

It was \citet{Hale1908}, one of the greatest astrophysicist of the twentieth century, who built the first spectroheliograph and used it to establish that sunspots are magnetic and grouped in pairs of opposite polarities.  Further pioneering work on solar magnetism conducted by Hale and collaborators in \citeyear{Hale1919} revealed an East-West direction of polarity in the sunspots' magnetic fields exhibiting a mirror symmetry with respect to the solar equator. Such polarity was observed to undergo inversion according to the 11 years solar cycle. This is commonly known as the ``Hale-Nicholson law''.

The first detection of a magnetic field in a star other than our own Sun, 78 Vir, was obtained in \citeyear{Babcock1947} by Babcock. In \citeyear{Babcock1958}, Babcock published the first catalogue of magnetic stars and in 1960 he discovered what is still today the most magnetic main sequence star known, HD\,215441 ($\sim 3.4\times 10^4$\,G). This became known as ``Babcock's star''. Surface fields discovered in more recent years that rival in strength that of Babcock's star are those of HD\,154708 \citep[$\sim 2.45\times 10^4$\,G;][]{Hubrig2005}, HD\,137509 \citep[$\sim 2.9\times 10^4$\,G; ][]{Mathys1995, Kochukhov2006} and HD\,75049 \citep[$\sim 3\times 10^4$\,G;][]{Freyhammer2008, Elkin2010}.

Following the detections of strong fields in main sequence stars, \citet{Blackett1947} suggested that if magnetic flux is conserved, some white dwarfs should exhibit magnetic fields of up to $10^7 -10^8$\,G. However, spectroscopic surveys aimed at detecting magnetic fields in white dwarfs yielded negative results \citep{Preston1970}. \citet{Kemp1970} argued that electrons spiralling in a magnetic field would emit linearly and circularly polarised radiation that should be detectable in the continuum of strongly magnetic white dwarfs. A spectropolarimetric survey of white dwarfs led to the discovery of strong circular polarisation in the continuum of the white dwarf Grw$+70^{\circ} 8247$ \citep{Kempetal1970}.

\citet{BaadeZwicky1934} first suggested that some stars could be made up of neutrons and that a supernova could be the result of a rapid transition of a normal star into a neutron star. \citet{Baade1942} and \citet{Minkowski1942} found unusual emission arising from the central parts of the Crab Nebula. Later on radio pulsations in this nebula were discovered by \citet{Staelin1968}.  Such radio emission had been predicted by \citet{Shklovsky1953} as caused by relativistic electrons spiralling around magnetic field lines.  

\citet{w64} first proposed that under magnetic flux conservation, if a star contracts to the density expected in a neutron star, then the surface field strength could be amplified to values of up to $10^{14}-10^{16}$\,G. The first pulsar, a highly magnetised rapidly spinning neutron star \citep{Pacini1968}, was discovered in 1967 by Jocelyn Bell \citep{Hewish1968}.

In this chapter we review progress made on the origin of magnetic fields in stars. The origin of magnetic fields is still a major unresolved problem in astrophysics. The ``fossil field'' hypothesis is often invoked to link magnetism in compact stars to magnetism on the main sequence (see Sect.\,\ref{fossil_MWDs} and Sect.\,\ref{fossil_NS}).  In the fossil scenario, some fraction of the magnetic flux of the progenitor star is conserved during the collapse process, because the stellar plasma is highly conducting and hence, by Faraday's law, the magnetic field is `frozen in' to the fluid \citep{w64,r72,bs04,fw06,fw07,fw08}. Under these circumstances, the field strength $B$ scales with radius $R$ of the star as $B\propto R^{-2}$. A main sequence B star with radius $5R_\odot$ and field $\sim 1,000$\,G compresses to give a neutron star with a field of $\sim 10^{14}$\,G; a main sequence A star with radius $1.6R_\odot$ and field $\sim 1,000$\,G compresses to give a magnetic white dwarf with a field of $\sim 10^{7}$\,G. The fossil field scenario is indeed quite attractive and can explain the existence of magnetic fields even in the strongest magnets in the universe, the so-called ``magnetars''. However, the feasibility of the fossil field hypothesis does have its problems and has been recently questioned, in the context of highly magnetic white dwarfs, on the basis of recent observational results related to their binarity (see Sect.\, \ref{dynamo_mwds}). The fossil versus dynamo-generated fields debate for neutron stars is analysed in Sect.\, \ref{problems_with_fossil}. Alternative explanations for the origin of fields in stars are presented in Sect.\, \ref{Magnondeg}, \ref{dynamo_mwds} and \ref{dynamo_NS}.

In this chapter we also review recent progress in magnetohydrodynamics (MHD) turbulence theory and computation as it applies to stellar magnetism (see Sect.\,\ref{mhd-review}), which underpins the operation of stellar dynamos, and controls diffusive magnetic flux transport in stars.

\section{Magnetism in non-degenerate stars}\label{Magnondeg}

Magnetic fields in main sequence stars have been measured mainly via spectropolarimetry. Different techniques measure different field properties, e.g., line-of-sight component, volume-averaged field strength, or dipole moment. A thorough review on magnetic field measurements of non-degenerate stars across the Hertzprung-Russell diagram can be found in this book in the chapter by Linsky \& Sch\"oller.

Direct measurements of magnetic fields in the chemically peculiar main sequence Ap and Bp stars, which form about 10\% of stars in the $1.6-5$\,M$_\odot$ range, have revealed the existence of strong ($\sim 3\times 10^2 - 3 \times 10^4$\,G) large scale fields \citep[e.g.][]{Auriere2007, DonatiLandstreet2009}. Landstreet et al. \citeyearpar{letal07} found that the majority of magnetic objects have an average longitudinal field (line of sight component of the surface field) $B > 0.25$ kG in every 1\,$M_\odot$ mass bin between $2M_\odot$ and $9M_\odot$ \citep{betal06,letal07}.  The lack of fields below $300$\,G, value that corresponds to the strength at which the magnetic field is in equipartition with the gas pressure in the stellar photosphere, is not a detection threshold effect since \citet{Auriere2010} have set a $3\sigma$ upper limit of longitudinal fields down to $1-10$\,G depending on stellar brightness. The absence of magnetic stars below this cutoff has been referred to as ``the Ap/Bp magnetic desert''. This indicates that Ap and Bp stars are not simply the high field tail of a continuous field distribution. \citet{Landstreet2008} also find that the field strengths seem to show some decline with stellar age but the field incidence does not. Such large scale fields are observed throughout the main sequence phase and, in more recent years, have also been observed in a small number of stars on the red giant branch \citep{Auriere2008}.
  
The recent detections of sub-gauss fields in Vega \citep{Lignieres2009}, and possibly in the Am stars Sirius A \citep{Petit2011} and $\beta$\,UMa and $\theta$\,Leo \citep{Blazere2014} have unveiled a new class of magnetic stars which is at the 1\,gauss end of the magnetic desert, thus potentially suggesting a dichotomy between strong and ultra-weak magnetic fields among intermediate-mass stars \citep{Lignieres2014}. This intriguing observational result has prompted \citet{BraithwaiteCantiello2013} to investigate the origin of these low fields. They argued that such fields could be the remnants of fields already present or formed during or immediately after the star formation stage. Hence, these fields would still be evolving on a timescale that is comparable to the age of the star. According to these studies, all intermediate and high mass stars lacking strong fields should display sub-gauss field strengths that would slowly decline over their main sequence lifetime. There has been some recent effort to test this prediction by \citet{WadeFolsom2014}, \citet{Neiner2014} and \citet{NeinerFolsom2014} but without any clear conclusion.

\citet{Alecian2008} conducted a spectropolarimetric study of Herbig Ae/Be (HAeBe) objects \citep{Herbig1960}. HAeBe stars are pre-main sequence stars of $2-15$\,M$_\odot$ which are still embedded in their protostellar gas-dust envelope and exhibit emission lines in spectra of type A/B. The observations of \citet{Alecian2008}, \citet{Alecian2009}, \citet{Hubrig2009} and more recently \citet{Alecian2013a} and \citet{HubrigIlyin2013} have revealed that about 7\% of HAeBe objects are magnetic. These studies have also found that they display large scale dipolar fields of strength comparable to that of Ap and Bp stars, under the assumption of conservation of magnetic flux.  Interestingly, although some magnetic HAeBe stars are in the totally radiative evolutionary phase while others have already developed convective cores, they seem to share the same magnetic field strength and structure thus indicating that the convective core is not responsible for generation or destruction of fields \citep{Alecian2013b}.

The Magnetism in Massive Stars \citep[MiMeS,][]{Wade2011} project, the MAGORI \citep{Hubrig2011,Schoeller2011} project and the B fields in OB star (BOB) Collaboration \citep{Hubrig2014} have conducted large surveys of bright and massive Galactic stars of spectral types B and O. It is now clear that about 10\% of all stars with radiative envelopes in the mass range $1.5-50$\,M$\odot$ possess large scale mostly dipolar magnetic fields \citep{Grunhut2013}. Even more interestingly, the MiMes project has revealed that all of the basic field characteristics do not vary significantly from the coolest spectral types F0 ($\sim 1.5$\,M$\odot$) to the hottest spectral types O4 ($\sim 50$\,M$\odot$) stars thus supporting a common formative scenario over a very large range of stellar masses \citep{Wade2013}. Furthermore, the upper dipolar field limits placed on the undetected magnetic O stars sample studied by the MiMeS collaboration are 40\,G at 50\% confidence, and 130\,G at 80\% confidence \citep{Wade2014}. This result seems to indicate that the field distribution of massive stars may also have a magnetic desert, similar to that observed in intermediate-mass stars. However, the recent work of \citet{Fossati2014}, who detected weak longitudinal fields of $\lesssim 30$\,G in the early B-type main sequence stars $\beta$\,CMa and $\epsilon$\,CMa, appears to support a more continuous distribution of fields in massive stars. They also claim that weak fields in massive stars could be more widespread than currently observed because of the numerous observational biases associated with the detection of weak fields in massive stars using current instrumentation and techniques.

The vexed question on the origin of fields in main sequence stars is still unanswered. The two main hypothesis are the fossil field, according to which magnetism in stars is a relic of the interstellar field from which the star was born \citep[e.g.][]{Borra1982, Moss2001}, and dynamo action taking place in the rotating cores of main sequence stars. Both theories are unable to explain why only a small subset of main sequence stars is magnetic, although the fossil field theory does assume that differences such as initial density and magnetic field strength in interstellar clouds could be responsible for this. However, it is peculiar that observations have never revealed the presence of magnetism in both components of a binary system. The BinaMIcS (Binarity and Magnetic Interactions in various classes of Stars) project \citep{Alecian2014, Neiner2013} is a collaborative endeavour that has been recently setup to investigate the phenomenon of magnetism in close (orbital periods shorter than 20\,d) binary systems. The study of the magnetic properties of these systems, at a detection limit of 150\,G, will give us crucial information on the origin of fields in magnetic main sequence stars which are well known to be rare in binaries \citep{Carrier2002}. Since the components of binaries share the same history, the study of the two stars will help us unravel the importance of their physical parameters and their studies should enable us to distinguish effects caused by initial conditions at formation from those caused later on by evolutionary processes. Preliminary studies of 314 massive stars in close binaries have revealed the presence of fields in 6 systems yielding an incidence of magnetism of $\lesssim 2$\%  which is up to 5 times lower than what is observed in single stars. This confirms the results of \citet{Carrier2002} which were based on a smaller sample of A-type stars.

In order to explain the rarity of short period binaries containing magnetic main sequence stars and also the fact that magnetism is only present in a small percentage of single main sequence stars, \citet{Ferrario2009} proposed that magnetic fields in main sequence stars could form when two proto-stellar objects merge late on their approach to the main sequence and when at least one of them has already acquired a radiative envelope.  The N-body simulations of pre-main sequence evolution of \citet{Railton2014} have indicated that in high initial cluster densities the number of collisions between stars is twice as high as that of stars already on the main sequence. Furthermore, they find that massive stars generally form through the merging of lower mass stars.  Thus, the expectation is that the incidence of magnetism should  increase with mass, which seems to be demonstrated by the observations of \citet{Power2007} for a volume limited sample of A and B type stars with $M\le 4$\,M$_\odot$.  Observations also seem to be in general agreement with the predictions of \citet{Ferrario2009} that the Ap and Bp stars should have pre-main sequence progenitors with similar magnetic field flux and structure. 

Another merger scenario has been proposed by \citet{Tutukov2010}. They propose that the coalescence of the two components of a close binary system with masses in the range $0.7-1.5$\,$M_\odot$, which are expected to have convective envelopes and strong magnetic fields, would be responsible for the formation of Ap and Bp stars with $M\lesssim 3$\,M$_\odot$. In this picture, the magnetic Ap and Bp stars with radiative envelopes owe their strong magnetic fields to progenitors with convective envelopes. \citet{Tutukov2010} speculate that systems such as the W UMa-type contact binaries could represent the precursory phase of these merging events.

However all of these predictions are so far only of a very qualitative nature and need to be supported by detailed quantitative calculations in the future.

\section{Magnetic white dwarfs}\label{MWDs}

Magnetism in white dwarfs is revealed either through Zeeman splitting and circular polarisation of spectral lines or, at very high field strengths, as continuum polarisation.  The magnetic white dwarfs, which represent about $8-13$\% of the total white dwarf population \citep{Liebert2003,Kawka2007} and exhibit polar field strengths of $\sim 10^3- 10^9$\,G, have been discovered mainly in optical sky surveys. At spectral resolutions of about $10$\,\AA, the Zeeman triplet in the lower members of the Balmer and Lyman series can be easily recognised at fields of $\sim 10^6 - 10^7$\,G when the splitting in intensity spectra is smaller than other broadening factors, such as pressure broadening.  At higher spectral resolutions, it is possible to detect fields down to $\sim 10^5$\,G.  However, circular spectropolarimetry is a much more sensitive observational technique that is used to measure even lower ($<10^5$\,G) fields, since it can independently measure the $\sigma^+$ and $\sigma^-$ oppositely polarised components of the Zeeman split absorption features. On the other hand, due to the faintness of white dwarfs, fields below $\sim 3\times 10^4$\,G have only been recently detected. Spectropolarimetric observations on the ESO VLT by \citet{AznarCuadrado2004,Jordan2007} revealed fields down to $\sim 10^3$\,G in about 10\% of the surveyed white dwarfs.\citet{Landstreet2012} have confirmed this finding by conducting a survey of objects
randomly drawn from a list of nearby cool ($T_{\rm eff}\lesssim 14,000$\,K) WDs.  

Observations seem to indicate that there is a paucity of white dwarfs with fields in the intermediate field range $10^5-10^6$\,G \citep{Kawka2012}, reminiscent of the magnetic desert of Ap and Bp stars.  However, this finding has not been fully corroborated by observations and future surveys may find objects also in this field range. In this context we need to stress that while magnetic fields in bright main sequence and pre main sequence stars have been mainly found via circular polarisation observations, only a small percentage of magnetic white dwarfs has been discovered using this method. This non-systematic approach has created, over the past 30 years, a sample of stars whose properties are difficult to investigate since the observational biases are not fully understood and thus are difficult to remove.

Another interesting characteristic of strongly magnetic white dwarfs is that their mean mass ($0.85\pm 0.04$\,M$_\odot$) is higher than that of non-magnetic or weakly magnetic white dwarfs ($0.593\pm0.002$\,M$_\odot$, see the chapter on magnetic white dwarfs in this book). This indicate that their progenitors are more massive than those of ordinary white dwarfs. Interestingly, the observations of \citet{Vennes1999} of the EUVE/Soft X-ray selected ultra-massive white dwarfs ($M>1.2$\,M$_\odot$) have found that $\sim 20\,$\% of them are strongly magnetic.  Thus it appears that the incidence of magnetism in the high field group increases with white dwarf mass and hence with progenitor mass, unless they result from a merger \citep[e.g. EUVE J0317-853][]{Barstow1995,Ferrario1997}.

\subsection{Origin of fields in highly magnetic white dwarfs: the fossil hypothesis}\label{fossil_MWDs}

According to the ``strong'' fossil field hypothesis, magnetic stars originate in gas clouds whose fields are at the upper end of the field distribution observed in the interstellar medium \citep[$10^{-6}$ and $10^{-4}$\,G, ][]{Heiles1997}. One of the very first articles on star formation and flux freezing is that of \citet{Mestel1966}. Briefly, in the simple model proposed by \citet{twf04}, a star of mass $M$ that collapses from the interstellar medium entraps a magnetic flux $\Phi \propto B_{\rm ISM} M^{2/3}$\,Mx where $B_{\rm ISM}$ is the interstellar (primordial) field. The variations expected in the interstellar magnetic field will determine the distribution of magnetic fluxes in the protostars. The magnetic flux lines would freeze in the radiative upper layers of the emerging star that will then evolve towards the main sequence. Assuming perfect magnetic flux conservation, a contraction by a $10^7$ factor could in principle create fields of the order of $10^8$\,G in a main sequence star. However during the cloud's collapse most of the initial magnetic flux would be lost via ohmic dissipation or ambipolar diffusion, so that much lower fields than this upper limit can be realised. The survival of significant large-scale magnetic flux through the pre-main sequence evolution has been addressed by \citet{Moss2003}.

According to the ``weak'' fossil field hypothesis, stellar fields could just be the remnants of those generated by the dynamos of active pre-main sequence stars or by late mergers of protostars with at least one of the two having a radiative envelope \citep[see Sect.\,\ref{Magnondeg} and][]{Ferrario2009}. The magnetic field flux would then be conserved during the subsequent evolution from main sequence to the white dwarf stage.

Under the fossil field hypothesis, the magnetic Ap and Bp stars in the mass range $1.5-8$\,M$_\odot$ would be the progenitors of the highly magnetic white dwarfs ($B\ge 10^6$\,G).  The paucity of white dwarfs with fields in the range $10^5- 10^6$\,G could be interpreted as being related to the lower field cut off of Ap and Bp stars.

The main issue is whether magnetic fields can survive complex phases of evolution when a star develops convective and radiative zones that contract and expand in size with time. \citet{twf04} point out that a dynamo-generated field in the convective regions of a star is transient and has no large-scale component. A star can conserve its primordial fossil field as long as this star possesses a radiative zone throughout its life, because the field would be pumped out of any developing convective region into an adjacent radiative region.  The conclusions of \citet{twf04} is that a poloidal fossil field that is present in a main sequence star can appear in the compact star phase with a similar field structure and with its magnetic flux nearly conserved.  

There have been many semi-analytical and numerical calculations that have addressed the issues concerning the evolution of a magnetic field during the main sequence phase. \citet{BraithwaiteSpruit2004} and \citet{BraithwaiteNordlund2006} have shown that stable magnetic fields with roughly equal poloidal and toroidal field strengths can exist in the radiative interior of a star and their exterior appearance would be that of a dipole with minor contributions from higher multipoles \citep[see also][]{MestelMoss2010}. Interestingly, they find that Ohmic dissipation diffuses the field outward over time so that the field strength at the surface of the star increases while the field structure in the stellar interior would switch from being mainly toroidal to poloidal. Hence they predict an increase in the surface field strength with the age of the star. In order to test these theoretical findings, it is necessary to study a sample of Ap and Bp stars with known age and fraction of the main sequence evolution completed.  However, it is quite difficult to estimate age-related quantities for magnetic stars \citep[see][]{Bagnulo2006}.  The evolution of magnetic fields from the zero-age main sequence (ZAMS) to the terminal-age main sequence (TAMS) has been investigated observationally by \citet{Bagnulo2006} and \citet{Landstreet2007} through the study of Ap stars in clusters, since the age of a cluster can generally be established to better than about $\pm0.2$\,dex.  They find that magnetic fields exist at the ZAMS phase for stellar masses $2-5$\,M$_\odot$ when stars have fractional ages (that is, fraction of the main sequence evolution completed) below about 0.05 and for fractional ages of less than about $\lesssim 0.10$ for masses up to 9\,M$_\odot$. This is consistent with the existence of fields in the late pre-main sequence stars Herbig AeBe (see section \ref{Magnondeg}). However, the study of \citet{Landstreet2007} also reveals that the evolutionary time of magnetic fields is less straightforward to interpret. For stars with $M \gtrsim 3$\,M$_\odot$ they find that the field strength decreases on a timescale of about $2−3\times 10^7$\,yr, in agreement with \citet{Kochukhov2006}. However it is not clear whether the magnetic flux really decreases or whether it is the conservation of magnetic flux that causes the surface field to decrease as the star expands. Interestingly, for stars with $M\lesssim, 3$\,M$_\odot$ they find no conclusive evidence of field strength or magnetic flux reduction on a time scale of a few $10^8$\,yr.

Support for the fossil field hypothesis also comes from spectropolarimetric studies of fifty red giants by \citet{Auriere2013}. This work has revealed the existence of four magnetic giants that have been identified as probable descendants of Ap/Bp stars. 

\begin{figure}
\centering
\includegraphics[width=7.5cm]{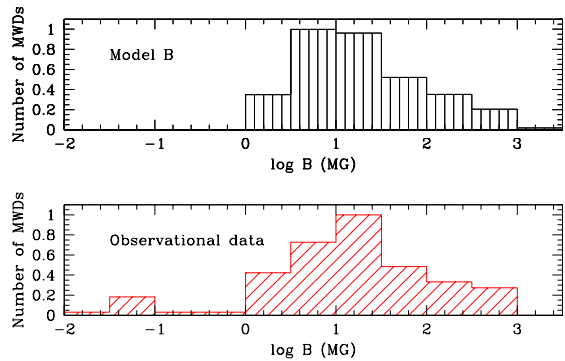}
\caption{Population synthesis calculations of \citet{Wick2005} (their scenario B) for the field distributions of magnetic white dwarfs based on the fossil field hypothesis compared with observations}
\label{scenarioB_field}
\end{figure}

Population synthesis calculations carried out by \citet{Wick2005} have shown that starting with a distribution of magnetic fields on the main sequence, as observed for the Ap and Bp stars and under the assumption of magnetic flux conservation, the predicted magnetic field and mass distribution in white dwarfs are in general agreement with observations, provided that a modified initial to final mass relation is employed for magnetic white dwarfs (their ``scenario A''). That is, in order to fit the mass distribution, one needs to assume that mass loss mechanisms are partly inhibited by the presence of strong fields. However, such models can at most yield an incidence of magnetism of $\sim 5$\% as compared to the observed $\sim 8 - 12$\% \citep[see also][]{Kawka2007}. \citet{Wick2005} have shown that it is possible to achieve a better agreement with observations if one assumes that in addition to the actually observed fields in Ap and Bp stars, about $40$\% of stars more massive than $4.5$\,M$_\odot$ (early B-type stars) have fields of $10-100$\,G and evolve into high fields magnetic white dwarfs (see Figure \ref{scenarioB_field}).  This is their ``scenario B'' which would only be viable if magnetic B-type stars do not have a ``magnetic desert'', which seems to be, albeit not conclusively, supported by the work of \citet{Fossati2014}.

Larger and more sensitive spectropolarimetric surveys of white dwarfs may confirm the classification of white dwarfs within either the very low ($\lesssim 10^4$\,G) or the very high field group ($\gtrsim 10^6$\,G), hence forming a bimodal field distribution. This would point to two different channels for field formation. It is interesting to note that the weak magnetic fields which have recently been detected in Vega and Sirius \citep{Lignieres2009,Petit2011}, would scale up to a surface field of a few $\sim 10^4$\,G at the white dwarf stage under magnetic flux conservation. Such weakly magnetised white dwarfs would belong to the low field component of the bimodal white dwarf field distribution.

\subsection{Origin of fields in highly magnetic white dwarfs: the stellar merger hypothesis} \label{dynamo_mwds}

\begin{figure}
\centering
\includegraphics[width=7.5cm]{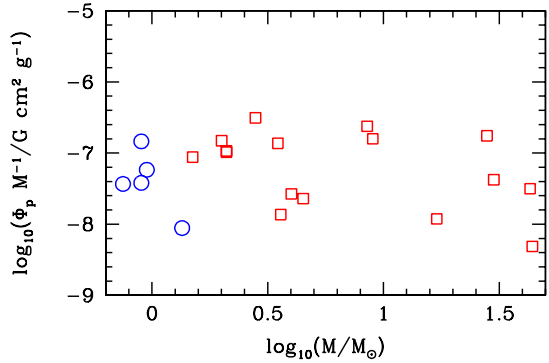}
\caption{The ratio of magnetic flux to mass $\phi_p/M$ for the most magnetic
main sequence stars (squares) and white dwarfs (circles) \citep{Wick2014}}
\label{mostmagnetic_fluxmass}       
\end{figure}

We have plotted in Fig. \ref{mostmagnetic_fluxmass} the ratio of poloidal magnetic flux to mass of the highest field magnetic white dwarfs and of the most magnetic main sequence stars. This diagram shows that the two groups of stars share similar characteristics indicating that their magnetic fields may have a common origin. However, the possible evolutionary link between the two populations, and thus the viability of the fossil field hypothesis, has been questioned by \citet{Tout2008} and \citet{Wick2014}. The main argument hinges on the fact, first highlighted by \citet{Liebert2005}, that there is not a single example of a high field magnetic white dwarf ($B\gtrsim$ a few $10^6$\,G) with a non-degenerate companion star \citep[generally an M dwarf, but see][]{Ferrario2012} in a non-interacting binary. Further searches on a much larger sample of objects conducted by \citet{Liebert2015} have confirmed the hypothesis that magnetic field in white dwarfs and binarity (with M or K dwarfs) are independent at a $9\,\sigma$ level.  However, we cannot invoke some as yet unknown physical mechanism that could inhibit the formation of a strong magnetic field in a white dwarf when a companion star is present, because such a mechanism would also prevent the formation of Magnetic Cataclysmic Variables (MCVs) which are interacting binaries consisting of a magnetic white dwarfs with a low-mass red dwarf companion.  This curious lack of detached white dwarf - non-degenerate star systems indicates that there are no known progenitors of MCVs (for more details see also the chapter on magnetic white dwarfs in this book).

An merger scenario for the generation of fields has also been proposed by \citet{Nordhaus2011}.  Here a low-mass star would be tidally disrupted by its proto-white dwarf companion during common envelope evolution to form an accretion disc. In this disc seed fields would be amplified through turbulence and shear and then advected on to the object that will become an isolated highly magnetic white dwarf.  \citet{garcia2012} conducted three-dimensional hydrodynamic simulations of merging double degenerates and argued that a hot and differentially rotating convective corona would form around the more massive star.  Their population synthesis calculations of double degenerate mergers are in general agreement with observations. Similar population synthesis calculations conducted for a wide range of merging astrophysical objects have also been carried out by \citet{Bogomazov2009}.

Following the work of \citet{Tout2008}, \citet{Wick2014} proposed that the fields are generated by an $\Omega$ dynamo within the common envelope of a binary system where a weak seed poloidal field would wind up by differential rotation. The closer the cores of the two stars are drawn before the envelope is ejected, the stronger the final field of the star emerging from common envelope will be. Therefore the strongest fields occur when the two stars merge, forming an isolated highly magnetic white dwarf. If the two stars do not coalesce but emerge from the common envelope when they are about to transfer mass, they become the MCVs \citep{Tout2008}.

\begin{figure}
\centering
\includegraphics[width=0.6\textwidth]{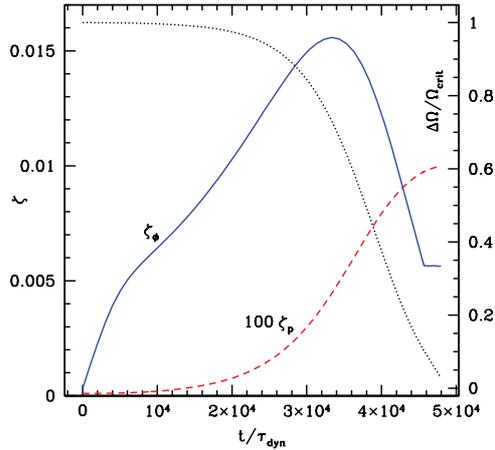} 
\caption{The evolution of the magnetic field components $B_{\rm p}$ and $B_\phi$  ($\zeta = \sqrt{\eta}\approx B$) following a stellar merger which produces a differential rotation equal to the break up spin of the merged object.  The decay of differential rotation $\Delta\Omega$ follows the right hand axis while the fields follow the left hand axis.  Toroidal field decays unless $\eta_{\rm p} > a\eta^2$ and this determines the final ratio of toroidal to poloidal field  \citep{Wick2014}}  
\label{Bp_Bt} 
\end{figure}

Observations indicate that highly magnetic stars are typified by (see Fig.\, \ref{mostmagnetic_fluxmass})
\begin{equation}
10^{-8.5} < \frac{\Phi_{\rm p}/M}{\rm G\,cm^2\,g^{-1}} < 10^{-6.5},
\label{data}
\end{equation}
where $\Phi_{\rm p}=R^2 B_{\rm p}$ is poloidal magnetic flux and $M$ and $R$ are the total mass and radius of the star respectively.

If the dynamo mechanism generates a magnetic field from differential rotation $\Delta\Omega$, then we have
\begin{equation}
0\le \Delta\Omega\le \Omega_{\rm crit}=\frac{1}{\tau_{\rm dyn}}=\sqrt{\frac{GM}{R^3}}.
\end{equation}

Toroidal and poloidal fields are unstable on their own \citep{b09}. The process that gives rise to the decay of toroidal field leads to the generation of poloidal field with one component stabilising the other and thus limiting field growth until a stable configuration is reached. The results of  Fig \ref{mostmagnetic_fluxmass} have been interpreted by \citet{Wick2014} as follows.  If $E_{\rm p}$, $E_\phi$ and $E=E_{\rm p}+E_\phi$  are the poloidal, toroidal and total magnetic energies respectively and $\eta=E/U$, $\eta_{\rm p}=E_{\rm p}/U$ and $\eta_\phi=E_\phi/U$ are, respectively, the ratios of poloidal and toroidal magnetic to thermal energy $U$, then by scaling to the observed maximum poloidal flux, \citet{Wick2014}  find that
\begin{equation}
\eta_{\rm p} = \frac{10^{-8}}{\lambda}\left(\frac{\hat{\Phi}_{\rm p}}
{10^{-6.5}{\rm G\,cm^2\,g^{-1}}}\right)^2,
\label{poloidalmax}
\end{equation}
where $\hat{\Phi}_{\rm p}$ is the ratio of magnetic flux to stellar mass. The observational data in  Fig. \ref{mostmagnetic_fluxmass} show that the maximum $\eta_{\rm p}$ is independent of the mass and type of star. In non-rotating stars, a stable poloidal-toroidal configuration must satisfy \citep{b09}
\begin{equation}
a\eta^2 < \eta_{\rm p} < 0.8\eta,
\end{equation}
where $a\approx 1$ is a buoyancy factor for main sequence stars.  The left hand side inequality is caused by the stabilising effect of a poloidal field on the Taylor instability in purely toroidal fields.  A lower limit to the poloidal field is determined by the relative importance of magnetic to gravitational--thermal energy through buoyancy effects.  The upper limit is due to the stability of poloidal fields which requires they be not significantly larger than the toroidal field.  \citet{b09} argued that the same inequalities are also likely to hold for stable fields in rotating stars. 

Fig.\,\ref{Bp_Bt}, where $\zeta = \sqrt{\eta}\propto B$, shows how $B_{\rm p}$ and $B_\phi$ evolve as $\Delta\Omega$ decreases from its maximum at $\Delta\Omega=\Omega_{\rm crit}$.  Their single model parameter is chosen to give the observed maximum  $\eta_{\rm p}\approx 10^{-8}$, so that $\zeta_{\rm p} \approx 10^{-4}$. Toroidal field is initially generated by the winding up of the seed poloidal field through differential rotation.  As soon as the toroidal field is large enough, the poloidal field starts to grow.  As long as the differential rotation is sufficiently large,  the toroidal field continues to grow up to a maximum value. Then it decreases until it reaches equilibrium with the poloidal field and $a\eta^2 = \eta_{\rm p}$.  The magnetic torque extinguishes $\Delta\Omega$ to yield a final object that rotates as a solid body.  \citet{Wick2014} find that the dynamical timescale for a $2\,M_\odot$ star is of the order of 40\,min so this evolution is completed in about 3.7\,yr which is a lot smaller than the corresponding Kelvin-Helmholtz timescale ($2.3\times 10^6\,$yr for a $2\,M_\odot$ star).  The conclusion of \citet{Wick2014} is that the final poloidal field is stable and proportional to the initial quantity of differential rotation, but is independent of the size of small initial seed field and. 

More sophisticated $\alpha-\Omega$ dynamo models are described in \citet{Nordhaus2007}, although these calculations are aimed at understanding the different problem related to the shaping mechanisms in planetary nebulae.

In any case, the work of \citet{Nordhaus2011} highlights that for the dynamo mechanism of \citet{Wick2014} to be viable, the transport of a strong field generated in the envelope to the proto-white dwarf's surface would require a diffusion coefficient $\ge 10^{21}-10^{22}$\,g\,cm$^{-3}$ which is unphysical.  Hence \citet{Nordhaus2011} propose an alternative scenario whereby a companion star, of sufficiently low mass to avoid a premature ejection of the envelope, would be tidally destroyed by the gravitational field of the primary star and would form a disc extending all the way to the surface of the proto-white dwarf. Accretion would then transport the strong fields that are formed in the disc towards the proto-white dwarf where they would become anchored to its surface.

\begin{figure}[t]
\centering
\includegraphics[width=0.70\textwidth]{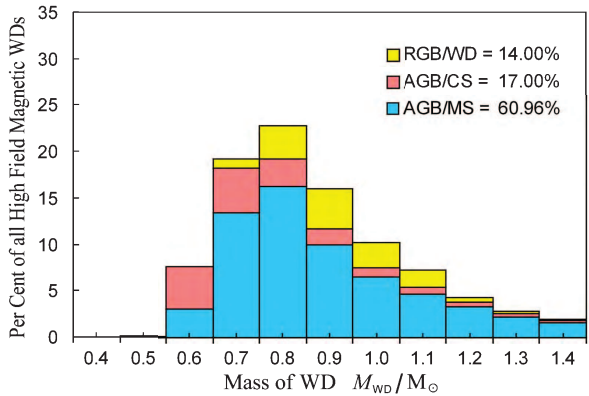}
\caption{Theoretical mass distribution of magnetic white dwarfs for $\alpha=0.1$. The contributions are separated according to their pre-CE progenitors.  AGB= Asymptotic Giant Branch; MS= Main Sequence; RGB= Red Giant Branch; CS= Convective Star and WD=White Dwarf.  Other paths also contribute but are less important than those shown. The Galactic disc age is chosen to be 9.5\,Gyr}
\label{Path}
\end{figure}

The population synthesis calculations carried out by \citet{Briggs2015} have given further support to the stellar merging hypothesis for the origin of fields in the highly magnetic white dwarfs. \citet{Briggs2015} find that the mass distribution and the fraction of stars that merge during a common envelope phase are in good agreement with the observations of magnetic white dwarfs for a wide range of common envelope efficiency parameter $\alpha$. Fig.\,\ref{Path} shows the contributions from the various merging pre-common envelope progenitor pairs for $\alpha$ = 0.1.  They find a theoretical incidence of magnetism among white dwarfs, for $\alpha=0.1-0.3$, of about 13-19\,\%, which is consistent with observational results  \citep[e.g.][]{Kawka2007}.

For the sake of completeness, we remark that similar population synthesis calculations, but related to the viability of the core-degenerate scenario for Type Ia supernovae, have been conducted by \citet{Ilkov2013} (and references therein).

Finally, we would like to bring the attention of the reader to the calculations of \citet{Duez2010a,Duez2010b,Duez2010c} (and references therein) on the stability of non-force-free magnetic equilibria in radiative stellar regions (cores of Sun-like stars, envelopes of intermediate and high mass stars, and compact stars). Through the use of semi-analytic techniques they constructed and tested an axisymmetric non-force-free magnetostatic equilibrium. Their numerical calculations recovered the instabilities which are characteristic to purely poloidal and toroidal magnetic fields and proved computationally, and for the first time, that only a mixed configuration is stable under all types of perturbations. Their important results can be applied to describe the magnetic equilibrium topology in stellar radiative regions (instead of choosing the initial field configuration arbitrarily) and to provide initial conditions for magneto-rotational transport in state-of-the-art stellar evolution codes \citep[e.g., see the models of][for giants and supergiant stars]{Maeder2014}. More general applications of these results to multi-dimensional magnetohydrodynamic computations are also envisaged.

\section{Neutron stars}\label{ns-magnetism}

Magnetic fields of neutron stars at birth are estimated to be in the range $\sim 10^{11} - 10^{15}$\,G. However, measurements of the birth magnetic fields of neutron stars are always indirect. Sometimes the field or its dipole moment is measured today, e.g., from neutron star spin-down or Zeeman spectropolarimetry of suitable progenitors, and then extrapolated backwards or forwards in time respectively in the context of an evolutionary model.  Otherwise, the birth field is inferred indirectly from some relic of the birth event, e.g., calorimetry of the supernova remnant. If the internal magnetic field evolves slowly, so that its strength today approximately equals its strength at birth, then gravitational wave upper limits also provide constraints. We discuss each of these approaches briefly below. More information can be found in other chapters in this volume.

Population synthesis models have a long history of being used to infer the birth fields of neutron stars \citep{go70,hetal97,fk06,kbm08}. Radio timing measurements of the spins, fields, and radio luminosities of the current neutron star population are combined with prescriptions for binary evolution, source kinematics in the Galactic gravitational potential, radio emission properties (e.g., beaming), supernova kicks, and observational selection effects to constrain the spin and field distributions at birth. Population synthesis models have benefited recently from the discovery of many new radio pulsars, both isolated and recycled, in large-sale radio multi-beam surveys \citep{metal01,metal02,ketal03,fetal04,hetal04}, an updated Galactic electron density map \citep{cl02}, and proper motions from very-long-baseline interferometry. The results are that the birth spins are inferred to be normally distributed, with average birth period $\langle P_0\rangle = 0.30$\,s and standard deviation $\sigma_{P_0}=0.15$\,s, while the birth fields $B_0$ are log-normally distributed, with $\langle\log B_0\rangle=12.65$ and $\sigma_{\log B_0}=0.55$ ($B_0$ in gauss) \citep{fk06,kbm08}. The latter studies find no evidence for magnetic field decay over $\lesssim 0.1$\,Gyr, assuming that the radio luminosity scales roughly as the square root of the spin-down luminosity. The no-decay conclusion sits in partial tension with the inference of field decay over $\sim 10^4$\,yr in a different sub-population of neutron stars, e.g., the magnetars and the thermal X-ray sources \citep{pvg12}. Faucher-Gigu{\`e}re and Kaspi \citeyearpar{fk06} also find no evidence for bimodality in the distributions of $P_0$ and space velocity.

The energetics of the birth event, which are related to $B_0$ and $P_0$, leave their imprint on the supernova remnant. From X-ray measurements of the remnant's radius, temperature, and emission measure, one can infer the total blast energy assuming Sedov expansion \citep{rc81,r08}. The blast wave is powered by the core-collapse event, which depends weakly on $B_0$ and $P_0$, and the relativistic wind emitted by the newly born pulsar, whose luminosity scales $\propto B_0^2 P_0^{-4}$. Drawing on X-ray Multi-Mirror Mission Newton (XMM-Newton) data, Vink and Kuiper \citeyearpar{vk06} showed that the pulsar wind played an insignificant role powering the blast wave in two anomalous X-ray pulsars (AXPs) and one soft-gamma-ray repeater (SGR), implying $P_0 \gtrsim 5$\,ms and hence no significant protoneutron star dynamo in these three objects at least. Steep density gradients in the interstellar medium can modify this conclusion, especially if the gradient correlates with the interstellar magnetic field \citep{vetal07}.

Gravitational wave upper limits and future detections furnish independent information on the \emph{internal} magnetic field of a neutron star. At the time of writing, coherent searches for the Crab pulsar with the Laser Interferometer Gravitational Wave Observatory (LIGO) constrain its gravitational wave luminosity to be $\lesssim 2$\% of the observed spin-down luminosity, thereby limiting the internal field to $B_0\lesssim 10^{16}$\,G \citep{aetal10,aetal14}.  An analogous result has been obtained for the Vela pulsar \citep{aetal11,aetal14}.

The gravitational wave strain scales as the square of the spin frequency, so the best constraints come from rapid rotators. Two classes of object are especially interesting in this regard. First, the large deformation of a newly born, fast-spinning ($P_0\sim 1$\,ms) magnetar caused by a super-strong internal magnetic field ($> 10^{16}$\,G) radiates powerful gravitational waves, which should be detectable with Advanced-LIGO-class detectors up to the distance of the Virgo cluster \citep{setal05,ds07,dss09,metal11,melpri14}. A future gravitational-wave detection of a millisecond magnetar would provide a direct measurement of $B_0$, if the distance is known (e.g., from a counterpart whose redshift is measured electromagnetically). Second, recycled millisecond pulsars have low external magnetic dipole moments, but their internal fields may be much stronger, if the external dipole is reduced by accretion-driven diamagnetic screening \citep{pm04}, which leaves the birth field in the interior untouched. Combining this scenario with rapid rotation and tight electromagnetic spin-down limits (ellipticity $\epsilon\lesssim 10^{-8}$ in some cases), recycled millisecond pulsars already yield some of the best constraints on the internal $B_0$ by any method, ruling out $B_0\gtrsim 10^{13}$\,G in some objects \citep{mm12}, depending on the poloidal-toroidal flux ratio and whether the core is superconducting or not. 

The above conclusions change, if the internal field evolves significantly in ordinary pulsars (as opposed to magnetars), cf. Faucher-Gigu{\`e}re and Kaspi \citeyearpar{fk06}. Radio pulse evolution over long time-scales may be a signature of magnetic polar wandering or plate tectonics, where the field wanders inside the star \citep{m74,r91,m12}. The recent discovery of changes in the flux ratio ($0.1$ per century) and component separation (0.6 degrees per century) of the Crab's radio pulses \citep{letal13} is noteworthy in this regard, although other explanations like radiative precession are also possible \citep{m00,bgt13,bgt14}. Stairs et al. \citeyearpar{sls00} discovered switching between two discrete magnetospheric states (and hence two pulse shapes and spin-down rates) in PSR B1828-11.

The central, enduring debate regarding the origin of neutron star magnetic fields revolves, as it does in magnetic white dwarfs (see Sect.\,\ref{MWDs}), around whether the field is a fossilised relic of the progenitor's field or is generated afresh by dynamo action in the protoneutron star in the first $\sim 10$\,s after core collapse. Below, we summarise briefly the pros and cons of each scenario, drawing heavily on an excellent review by Spruit \citeyearpar{s09}. The reader is directed to the latter reference for more discussion. 
\begin{figure}
\centering
\includegraphics[width=7.5truecm]{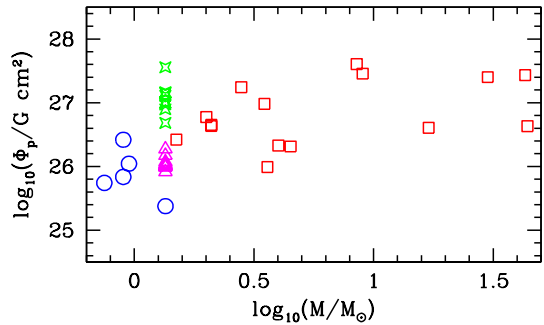}
\caption{The magnetic flux $\phi_p$ for the most magnetic
main sequence stars (squares), white dwarfs (circles), high field radio pulsars (triangles) and magnetars (stars)}
\label{mostmagnetic_flux}
\end{figure}

\subsection{Origin of fields in neutron stars: the fossil hypothesis}\label{fossil_NS}

Although the surface magnetic fields in pre-main sequence stars, Ap/Bp, OB stars, magnetic white dwarfs and neutron stars cover a wide range of magnetic field strength values (from a few $10^2$ to $10^{15}$ Gauss), the magnetic fluxes near the upper limit of the observed field ranges are all a few $10^{27}$\,G\,cm$^2$. We show in Fig.\,\ref{mostmagnetic_flux} the magnetic fluxes of the most magnetic compact stars and non-degenerate stars with radiative envelopes. This finding has been used in support of the fossil field hypothesis for the origin of fields.

The fossil field hypothesis for the origin of fields in neutron stars was first proposed by \citet{w64}.  \citet{r72} also remarked that the similarity of magnetic fluxes in magnetic white dwarfs and neutron stars could be explained through flux conservation during the evolution from main sequence to compact star stage.

\begin{figure}
\centering
\includegraphics[width=7cm]{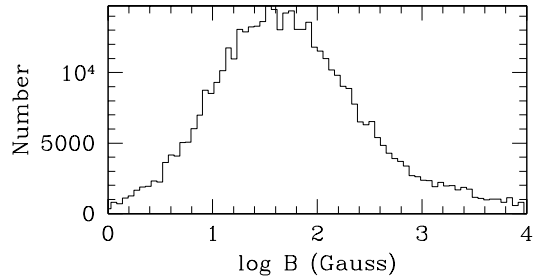}
\caption{Predicted magnetic field distribution of massive stars ($8-45$\,M$_\odot$), according to the population synthesis modelling of \citet{Ferrario2006} under the assumption of fossil fields and magnetic flux conservation}
\label{OBPrediction}
\end{figure}

\citet{Ferrario2006} have investigated the effects of the fossil field hypothesis for the origin of magnetic fields in neutron stars by carrying out population synthesis calculations for different assumptions on the distribution of the magnetic flux of massive ($\ge 8$\,M$_\odot$) main sequence stars and on the dependence of the initial birth period of neutron stars. They used the observed properties of the population of isolated radio pulsars in the $1374$ MHz Parkes Multi-Beam Survey \citep{metal01} to constrain model parameters. These were then used to deduce the required magnetic properties of their progenitor stars. Their conclusion is that the fossil field hypothesis, which does not allow for magnetic flux loss in the post-main sequence evolution, requires a very specific distribution of magnetic fields for massive main sequence stars as shown in Fig.\,\ref{OBPrediction} for an assumed dipolar field structure. In this picture the field distribution in massive stars would be continuous and all massive stars would be magnetic. This distribution is predicted to have a peak near 46\,G with low- and high-field wings covering a field range from $1$ to $10^4$\,G. About 8 per cent of main sequence stars would require to have fields in excess of $10^3$\,G and these would the progenitors of the highest-field neutron stars. This main sequence distribution of B0-O type stars is an observable quantity and thus a prediction of the fossil field model.

\begin{figure}
\centering
\includegraphics[width=7cm]{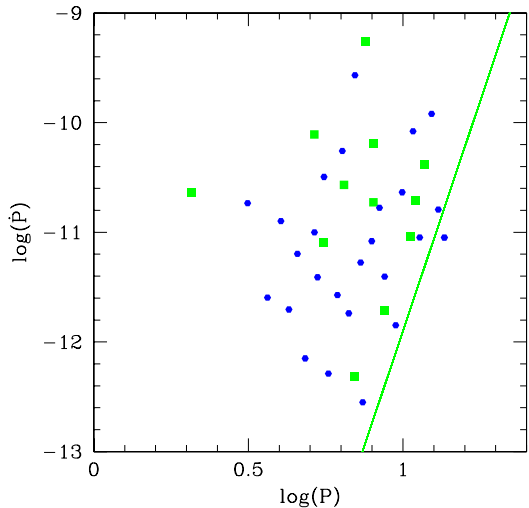}
\caption{Filled squares: observed magnetars; filled circles: magnetars as derived from the 
modelling of \citet{Ferrario2008}. The solid line is an empirically determined boundary (a
``magnetar death line'') given by $\log(\dot P)=8.4\log(P)-20$ \citep{Zhang2003}}
\label{ppdot}
\end{figure}

Further population synthesis calculations of the observed properties of magnetars were carried out by \citet{Ferrario2008} under the assumption that magnetars originate from main sequence stars that are much more massive than those giving rise to normal radio pulsars.  These studies were prompted by (i) the observation that most magnetars have been linked to progenitors in the mass range $\sim 20 -45$\,M$_\odot$ \citep{Gaensler2005,Muno2006} and (ii) the discovery of strong fields in massive O-type stars (see Sect.\,\ref{Magnondeg}).  Thus fossil magnetic fluxes similar to those observed in the magnetars may already be present in stellar cores prior to their collapse to neutron star and would explain the presence of fields of up to $\sim 10^{15}$\,G in magnetars.  According to \citet{Ferrario2008}, the anomalous X-ray emission of magnetars would be caused by the decay of the toroidal field that does not contribute to the spin down of the neutron star.  \citet{Ferrario2008} predict a number of active magnetars (see Fig.\,\ref{ppdot}) that is consistent with the number of sources detected by ROSAT and a Galactic birth rate compatible to that inferred by \citet{ketal98}.

Since stars of spectral types B0 to O are the progenitors of neutron stars, it remains to be seen whether there is a sufficiently large range of magnetic fluxes in these stars to explain the entire range of magnetic fluxes in neutron stars.  So far, observations of magnetism in massive stars by the MiMeS, MAGORI and BOB collaborations have led to the discovery of a large number of new objects. The paper by \citet{Petit2013} lists the physical, rotational and magnetic properties of about 60 highly magnetic massive OB stars. Most of the detected fields are in the $>10^3$\,G regime with masses $>8$\,M$_\odot$ and radii $>5$\,R$_\odot$.  Under magnetic flux conservation, they would be the progenitors of the highest magnetic field neutron stars ($>10^{13}$\,G).

\subsubsection{Problems with the fossil hypothesis}\label{problems_with_fossil}

Taken at face value, the fossil field arguments seem to be reasonable. Certainly, there is no trouble accounting for the magnetar end of the distribution, since typical surface fields measured in Ap and Bp stars compress to give fields $\gtrsim 10^{13}$\,G in neutron stars. If anything, there is a problem at the lower end, where it is hard to produce neutron stars with $B\sim 10^{10}-10^{11}$\,G without some other accretion-related physics. In fact, a main sequence star with a surface field $B \sim 300$\,G and a radius $R\sim 2R_{\odot}$ compresses to give $B\sim 6 \times 10^{12}$\, G. There are three problems with the fossil picture. First, the innermost $\sim 1.4$ $M_\odot$ of the progenitor, which collapses to form a neutron star, occupies $\sim 2\%$ of the progenitor's cross-subsectional area, reducing the maximum compressed field to $B\sim 10^{14}$ G. One can raise the limit somewhat, if the progenitor's field is centrally concentrated, but the three-dimensional MHD simulations of \citet{bn06} (which assume polytropic, spherically symmetric density and temperature profiles and neglect core convection) do not predict much central concentration. Second, the magnetar birth rate is comparable to the ordinary neutron star birth rate \citep{kk08,w08}, yet, as already mentioned in Sect.\,\ref{fossil_NS}, there may be too few known progenitors with $B_0\gtrsim 10$\,kG to create magnetars in the numbers observed. Furthermore, the relatively low incidence of intermediate-strength fields ($10$--$100\,{\rm G}$) \citep{Wade2014} may also represent a problem for the fossil field hypothesis, unless future, more sensitive spectropolarimetric surveys of massive stars discover that intermediate fields \citep[e.g., see][]{Fossati2014} are indeed much more common than current observations show. Third, the core of the progenitor couples magnetically to the giant envelope. The coupling is `frictional' --- the field lines are disrupted into a tangle by instabilities as they shear, behaving like a turbulent magnetic viscosity rather than `bungee cords' --- but it is strong nonetheless at high fields, scaling as $B_0^2$. Consequently, the core decelerates too much to explain the distribution of neutron star spin periods observed today. However, note that this is a problem for any neutron star with a magnetised progenitor, even if the neutron star's field is generated ultimately in a dynamo; it is not an argument against fossil fields specifically. Indeed, a protoneutron star dynamo may never take root, if the progenitor core rotates too slowly. Core-envelope coupling has stimulated the suggestion that neutron star spins are imparted by the birth kick during supernova core collapse \citep{sp98}. Alignment is therefore expected between the spin axis and proper motion, for which some evidence exists in X-ray images of supernova remnants and pulsar radio polarimetry \citep{wlh06}.

\subsection{Origin of fields in neutron stars: dynamo generated fields}\label{dynamo_NS}

In this model, the seed field is amplified by a a dynamo in the proto-neutron star \citep{rs73,dt92,td93,bub05}. There are two main kinds of protoneutron star dynamos: those driven by convection and those driven by differential rotation. Convection-driven dynamos were first analysed in the neutron star context by Thompson and Duncan \citeyearpar{td93}. A thermal luminosity $L$ is transported outwards through the star by turbulent convection with characteristic eddy velocity $V\approx (L/4\pi R^2\rho)^{1/3}$, where $\rho$ is the mean stellar density, and this generates a magnetic field of strength $B_{\mathrm{PNS}}\approx (4\pi\rho V^2)^{1/2}$, if the mechanical and magnetic stresses are in equipartition. For a typical photon luminosity, we obtain $B_{\mathrm{PNS}}\approx 10^9\textrm{ G }(L/10^{38}\textrm{ erg s}^{-1})^{1/3} (R/10^8\textrm{ cm})^{7/6}$. For a typical neutrino luminosity, $B_{\mathrm{PNS}}$ is $\sim 10^2$ times higher ($L\sim 10^{44}$ erg s$^{-1}$). When the protoneutron star collapses from $R=10^8$ cm to $10^6$ cm, the field is compresses further by flux conservation to give $B\approx 10^4 B_{\mathrm{PNS}}$.

A convective dynamo faces several obstacles to explain all neutron star magnetic fields, although it may certainly explain a subset. The most serious obstacle is the assumption of equipartition. Numerical simulations reveal time and again that self-sustaining dynamos do not reach equipartition \citep[e.g.,][]{css03} because then the magnetic stresses would paradoxically self-quench the driving shear. In a variety of contexts, and for various numerical set-ups, the magnetic stresses saturate at $\lesssim 5\%$ of the mechanical stresses \citep{css03,b06}. These theoretical results are consistent with observations of the Sun, whose dipole amounts to $\approx 0.3\%$ of equipartition \citep{Charbonneau2014}. A second obstacle arises when stratification quenches convection $\sim 10$ s after the dynamo begins. Does the field retreat or leave behind a permanent dipole moment? The danger here is that pockets (`domains') of tilted polarity are left behind, if the convective eddies turn over faster than they are quenched, producing a dipole moment lower than one inferred in magnetars \citep[e.g., from spin down ][]{ketal98}, even if the higher-order multipoles are substantial. The consequences for magnetic structure of forming convective pockets (which `insulate') and radiative pockets (which `conduct') has been explored by Tout et al. \citeyearpar{twf04}.

A second type of dynamo is driven by differential rotation. It is sometimes presumed that a sheared, magnetised fluid develops a toroidal field component $B_\phi$, which reacts back to shut off the shear completely. In practice, the back reaction can only ever be partial in a continuously driven system; if the fluid velocity gradient were to vanish momentarily, $B_\phi$ generation would cease, and the gradient would eventually reassert itself in response to the driver \citep{s09}. Microphysically, quenching is avoided because toroidal wind-up is limited to $B_\phi \lesssim 10 B_p$ by instabilities, where $B_p$ is the poloidal field component \citep{s09}. Hence it is necessary to grow $B_\phi$ and $B_p$ together in a differentially driven dynamo. The dynamo loop can be closed by Tayler \citep{b06} or magneto-rotational \citep{mbka06} instabilities, for example. The growth window is open for $\sim 10$ s, before the star stratifies stably, and lasts for a time $\approx R/v_A$ or $(\Delta\Omega)^{-1}$ in the latter scenarios respectively, where $v_A$ is the Alfv\'{e}n speed and $\Delta\Omega$ is the angular velocity shear. Differential rotation can also be driven by stellar mergers; the reader is referred to section \ref{dynamo_mwds} and Wickramasinghe et al. \citeyearpar{Wick2014} for details.

\citet{b06} showed that a shear-driven dynamo can even persist in a stably stratified star.  Fluid is displaced parallel to equipotential surfaces in the Tayler instability. Hence, the condition for dynamo operation reduces to requiring that the maximum hydromagnetic mode wavelength allowed by buoyancy exceeds the minimum wavelength set by ohmic dissipation --- a condition which is easily satisfied in a highly conducting neutron star. Braithwaite \citeyearpar{b06} showed explicitly through simulations that a Tayler dynamo is indeed self-sustaining in the presence of strong stratification and that it saturates with $B_\phi \sim 5 B_P$ and $B_\phi^2/8\pi\approx 10^{-2} \rho V^2$, i.e., at $1\%$ of equipartition. However, \citet{Zahn2007} have strongly contested this result, finding that the Tayler instability cannot sustain a dynamo even at high magnetic Reynolds number.

A mean-field variant of the shear-driven dynamo was analysed in the neutron star context by Mastrano and Melatos \citeyearpar{mm11}. In a mean-field dynamo, the turbulent electromotive force $\langle {\bf{v}}\times{\bf{B}}\rangle$ in Faraday's law is the sum of two terms: the `$\alpha$' term proportional to ${\bf{B}}$, which arises from turbulent vorticity, and the so-called `${\bf{\Omega}}\times{\bf{J}}$' term, which is proportional to $\eta_{ij}(\nabla\times {\bf{B}})_j$ and arises from off-diagonal entries in the resistivity tensor $\eta_{ij}$ \citep{m78,b97}. A natural seat for a mean-field dynamo is a tachocline, where the angular velocity jumps sharply \citep{oetal06,betal07}; in a protoneutron star, this may occur at the boundary between the convection and neutron finger zones \citep{bru03,bub05}. Mastrano and Melatos \citeyearpar{mm12} proved that a tachocline mean-field dynamo operates self-sustainingly in a protoneutron star, if $\eta_{ij}$ is anisotropic and the jumps in `$\alpha$' (vorticity) and angular velocity are coincident, and also (less realistically) if $\eta_{ij}$ is isotropic but the jumps are displaced.

\subsubsection{Generation physics}\label{GenPhysics}

The ultimate fate of a dynamo-generated magnetic field depends on buoyancy --- does the field float out of the star or remain anchored in the core? --- and helicity --- does the field maintain its large-scale integrity against the action of hydromagnetic instabilities, or is it shredded?

Buoyancy exists in a neutron star because, in $\beta$ equilibrium, the ratio of charged to neutral species densities increases with depth \citep{rg92}. Magnetic flux tubes (`light fluid') rise to the surface if the Alfv\'{e}n speed satisfies $v_A \gtrsim HN$, where $H$ is the hydrostatic scale height, and $N$ is the Brunt-V\"{a}is\"{a}l\"{a} frequency \citep{rg92,s09}. The field rises most easily when the protoneutron star is young and neutrinos help to lower $N$ to $\lesssim 10$ rad s$^{-1}$. By contrast, in a normal neutron star, theory predicts $N \gg 10^2$\,rad\,s$^{-1}$ \citep{rg92}. In this way, magnetic loops inside the star penetrate the surface and add up to form a permanent dipole moment, as long as their footpoints remain anchored in the cooling crust or core and do not float away entirely.

Assuming that the field does remain anchored, is it stable? Many analyses have shown that a purely poloidal magnetic field in a highly conducting plasma is always unstable \citep{mt73,w73,fr77}. If we regard the star's eastern and western hemispheres as two bar magnets, it is energetically favourable for the magnets to flip and counteralign, destroying the dipole moment \citep{fr77,mra11}; in practice, the instability proceeds by forming Rayleigh-Taylor `mushrooms' in the toroidal direction. Similarly, a purely toroidal magnetic field is always unstable \citep{t73,aetal13}. However, linked poloidal-toroidal fields appear to be stable under a range of conditions, both in analytic calculations \citep{p56,t73,aetal13} and numerical simulations \citep{bs04,bn06,b09}.

Linked poloidal-toroidal fields are stabilised, because their helicity is conserved approximately in a highly conducting plasma. Helicity is defined as $H=\int \mathrm{d}^3{\bf{x}}\phantom{+} {\bf{A}}\cdot{\bf{B}}$, where ${\bf{A}}$ is the magnetic vector potential; it is a gauge-dependent quantity which describes the number of times one flux tubes winds around another \citep{b97}. In a weakly resistive plasma, $H$ is conserved globally, even when $H$ is scrambled locally on individual flux tubes. Hence, if a dynamo-generated field has $H\neq 0$ initially, the twist can never be undone completely, leading to $B\neq 0$ in the final state. If magnetic energy is minimised during the field's evolution, it tends to a force-free configuration, cf. laboratory spheromaks \citep{bn08,mm08}.

Numerous papers have been written recently reporting stability calculations on poloidal-toroidal fields, e.g., \citep{bs04,bn06,aetal13,aetal13b,cr13,getal13,lj12,metal13,pl13}. A comprehensive review lies outside the scope of this chapter; we simply mention a few highlights. In a barotropic star, it has become apparent that stability depends in subtle ways on the boundary conditions and the entropy distribution within the star \citep{cetal11,lj12}. In a non-barotropic star, magnetic field configurations with an equatorial `torus' are generically stable, when the poloidal-toroidal flux ratio is `modest'. Akg\"{u}n et al. \citeyearpar{aetal13} obtained the stability condition $0.25 < B_\phi^2/B_p^2 < 0.5 (\gamma_{\mathrm{ad}}/\gamma - 1) |E_g|/(B_p^2/8\pi)$, where $\gamma_\mathrm{ad}$ and $\gamma$ are the adiabatic and non-barotropic specific heat ratios respectively, and $|E_g|$ is the total gravitational potential energy. Non-barotropic solutions also match self-consistently to an external dipole (or any arbitrary multipole), extending the power of future observational tests involving gravitational waves \citep{metal11,mm12,mlm13}. Finally, in a superconducting star, the stability problem changes character fundamentally for two reasons: the Lorentz force is much stronger ($|{\bf{F}}_B|\propto H_\mathrm{c1} B\sim 10^3 B^2$, where $H_\mathrm{c1}$ is the critical field in a type II superconductor), and its vectorial structure is more complicated $[{\bf{F}}_B\propto{\bf{B}}\times(\nabla\times {\bf{H}}_\mathrm{c1})+\rho\nabla(|{\bf{B}}|
\partial|{\bf{H}}_\mathrm{c1}|/\partial\rho)]$. The equilibrium magnetic structure and its stability are then controlled by $H_\mathrm{c1}/\langle B\rangle$ at the crust-core boundary \citep{gal12,lag12,l13,l14}. In fact, the internal field configuration now depends very strongly on field strength, suggesting significant differences between the interior fields of pulsars and those of magnetars \citep{l14}.

\subsubsection{Superfluid turbulence}

Does magnetic field evolution in a neutron star conclude $\sim 10$\,s after the protoneutron star is born and stably stratifies? The traditional consensus has been in the affirmative, except for the slow ohmic and Hall evolution described elsewhere in this volume and in \citet{getal12,pvg12,vpm12}. However, this picture may need modification in light of recent work suggesting that the neutron superfluid and charged components to which it is coupled are turbulent. In this scenario, internal magnetic activity may be ongoing, even though the magnetic flux thereby produced struggles to rise buoyantly to the surface. Hints of abrupt magnetospheric changes observed on $\sim 1$ yr time-scales (see Sect.\,\ref{ns-magnetism}) --- much longer than the dynamical time of the magnetosphere, but much shorter than the ohmic and Hall time-scales --- are relevant in this context.

In the standard picture, the vorticity field inside a neutron star is uniform and quasi-static. Macroscopically, this means that there is zero meridional circulation. Microscopically, it means that the superfluid is threaded by a rectilinear array of vortices, each carrying a quantum of circulation $\kappa=h/(2 m_n)$, where $h$ is Planck's constant and $m_n$ is the mass of the neutron. As the star brakes electromagnetically, an angular velocity lag $\Delta\Omega$ builds up between the crust and superfluid, whose value is set by the balance between the Magnus and pinning forces. One finds $\rho\kappa R\Delta\Omega\approx E_\mathrm{pin}(\xi_\mathrm{coh}\xi_\mathrm{pin})^{-1}$ and hence $\Delta\Omega\approx 1$ rad s$^{-1}$ independent of $\Omega$ and $\dot{\Omega}$, where $E_\mathrm{pin}$ is the pinning energy, $\xi_\mathrm{coh}$ is the superfluid coherence length, and $\xi_\mathrm{pin}$ is the pinning site separation \citep{wm13}. Glitches reset $\Delta\Omega$ but only partially, because there is no observed correlation between glitch sizes and waiting times in most pulsars \citep{metal08}.

The lag $\Delta\Omega$ changes the flow structure completely. It is a well-known result of fluid mechanics that a shear flow in a sphere cannot be purely toroidal. The boundary conditions induce Ekman circulation, wherein interior fluid is drawn into a viscous boundary layer, spun down through contact with the crust, then recycled back into the interior \citep{ae96,vem10}. This occurs with or without a rigid core. Furthermore, Ekman circulation is known to be unstable at high Reynolds numbers (see Sect.\,\ref{Dimensionless}) Re $\gtrsim 10^3$, forming unsteady, nonaxisymmetric flow patterns like herringbone waves (Re $\approx 6\times 10^3$) and Taylor-Gortler vortices (Re $\approx 10^4$) \citep{n83}. In a neutron star, where the effective Reynolds number is much greater [Re $\gtrsim 10^7$ for
  $e^-$-$e^-$ shear viscosity modified by Landau damping by transverse plasmons;
  see \citet{sy08}], wave modes resembling fully developed turbulence are
expected \citep{mp07}.

At the microscopic level, the vorticity field is disrupted by vortex-line instabilities. One example is the Donnelly-Glaberson instability, familiar from laboratory experiments with liquid helium and Bose-Einstein condensates \citep{gjo74}. If there is a superfluid counterflow directed along the vortices, Kelvin waves are excited and amplified by mutual friction to produce a reconnecting, self-sustaining vortex tangle \citep{tka03,tkt12}. This process has a low threshold (counterflow $\approx$ mm\,s$^{-1}$, easily exceeded during Ekman pumping) and a fast growth time ($\lesssim$ one spin period) \citep{petal06}. Recently, a related Kelvin-wave instability has been discovered by Link \citeyearpar{l12,l12b}, which does not rely on a counterflow and arises from imperfect pinning, which creates a lag to amplify the Kelvin waves [cf. \citet{gaj09}]. This instability grows over $\sim$ days, which is still fast.

Turbulent, tangled vorticity is an important ingredient in the story of magnetic field generation. The turbulent neutron condensate couples to charged components in the star through mutual friction and entrainment \citep{m98,pca02,hap09,hps12,c13}, as well as through fluxoid-vortex interactions \citep{setal90,jm10}. Consequently, the charged components circulate too \citep{petal05}. Simulations that incorporate the magnetic field dynamics explicitly are required to understand the behaviour of the coupled fluids in detail. There is reason to be hopeful that such simulations will be conducted in the next few years, motivated by the prospect of gravitational wave experiments.

Can superfluid turbulence be quenched by stratification, i.e., by the buoyancy force produced by the charge-neutral ratio increasing with depth (see Sect.\,\ref{GenPhysics})?  Counterintuitively, perhaps, the situation is borderline. Stratified turbulence is characterised by two dimensionless quantities: the Froude number, Fr, which is related to the buoyancy force, and the Reynolds number, Re \citep{betal07b}. Even for Fr$^{-1}\gtrsim 10^2$, the turbulence is not quenched, if Re is sufficiently large. Turbulence persists for Re $\gtrsim$ Fr$^{-2}$ or equivalently $R^2(\Delta\Omega)^3\gtrsim \nu N^2$, where $\nu$ is the kinematic viscosity \citep{itn09,lbm13}. Neutron stars lie near the boundary defined by the above condition in the Fr$^{-1}$--Re plane. In the regime Re $\gtrsim$ Fr$^{-2}$ and Fr$^{-1}$ $\gtrsim 10^2$, the flow is modified strongly away from Kolmogorov isotropy, even though it remains circulatory; the Earth's atmosphere and oceans also occupy this regime \citep{betal07b,cm12}.

Another possibility is that superfluid turbulence is quenched by magnetic stresses. Again, though, the case is unclear. At the microscopic level, the imperfect pinning instability which drives a vortex tangle \citep{l13b} persists even when a magnetic field is present; in fact, Link's \citeyearpar{l13b} analysis assumes magnetic locking of the viscous proton-electron plasma to the crust. Does the $B_\phi$ wound up by flux freezing erase the shear? More simulations are needed, but for now it seems unlikely: the magnetic back reaction saturates at $\lesssim 1\%$ of the mechanical (shear) stress in simulations attempted to date \citep{css03,b06}, and the Tayler instability self-organizes such that ${\bf{B}}\cdot\nabla\Omega=0$, i.e., such that wind-up stops, before $B_\phi$ quenches the process \citep{detal06,retal09}. Magnetised Ekman pumping acts on the rapid time-scale $\approx (\Omega R/v_A)^{2/3}\Omega^{-1}$ (where $V_A$ is the Alfv\'{e}n speed) in simple magnetic geometries, e.g., uniform magnetisation \citep{e79a}. However, it is known to operate much more slowly in complicated geometries and stratified conditions \citep{gp04,m12}. Finally, Easson showed that many --- perhaps most --- realistic magnetic topologies (e.g., those with closed loops) cannot enforce corotation \citep{e79b}; see also \citet{m12}. Physically, this occurs because the toroidal component of the magnetic field perturbation created by spin down satisfies a nonconservative equation of motion in a superconductor with Fermi liquid forces \citep{e79b}. Again, large-scale numerical simulations are needed to determine if the class of magnetic topologies that permit corotation is restricted or generic.

\section{Status of MHD turbulence theory and computation}\label{mhd-review}

The basic physics underlying stellar magnetic fields lies in the theory of
magnetised fluids, or magnetohydrodynamics (MHD). Of particular importance is
the manifestation of magnetohydrodynamic turbulence because its presence
underpins the operation of stellar dynamos, and controls diffusive magnetic flux
transport in stars. This section is intended to give a high-level overview of
recent progress in MHD turbulence theory and computation as it applies to
stellar magnetism.

\subsection{Equations of motion}
The equations of compressible MHD can be written as
\begin{eqnarray}
  \label{eqn:continuity}
  \pd{\rho}{t} =& \nabla \cdot (\rho \vct{u}) \\
  \label{eqn:momentum}
  \pd{\vct{u}}{t} =& -\nabla \cdot (\rho \vct{u} \vct{u}) - \nabla
  p + \rho \nu \nabla^2 \vct{u} \\ \label{eqn:faraday} \pd{\vct{B}}{t} =&
  \nabla \times (\vct{u} \times \vct{B}) + \eta \nabla^2 \vct{B}
\end{eqnarray}
where in Equation \ref{eqn:momentum} $\vct{u}$ is the fluid velocity, $\rho$
is the mass density, and $\nu$ is the kinematic viscosity. Equation
\ref{eqn:faraday} is Faraday's law together with the phenomenological Ohm's law
$\vct{E} = -\vct{u} \times \vct{B} + \eta \vct{J}$. In general an
equation for the transport of energy must also be supplied along with an
equation of state relating the gas pressure to its temperature and density.
However, if incompressibility is assumed, as it is in most analytical
treatments of turbulent dynamo, the energy equation is superfluous.

\subsection{Dimensionless numbers}\label{Dimensionless}
If $L$ and $u$ represent characteristic length and velocity magnitudes of the
system in question, then the ratio $\RE = u L / \nu$ is referred to as the
Reynolds number. It represents the relative importance of inertial and viscous
forces. It can also be seen as the ratio of the viscous time $t_{visc} =
L^2/\nu$ to the advective time $t_{adv} = L/u$. Flows having large $\RE$ are in
general susceptible to turbulence as their dynamics are dominated by the
non-linear coupling in Equation \ref{eqn:momentum}. Due to the enormous length
scales involved in astrophysical flows, it is commonly appropriate to drop the
viscous term altogether and just work with the Euler equation.

Analogously, one can define the \emph{magnetic} Reynolds number $\RM = u L /
\eta$ which measures, correspondingly the relative importance of magnetic
induction and diffusion.It is also the ratio $t_{res} / t_{adv}$ where $t_{res}
= L^2 / \eta$ is the damping time scale for magnetic fluctuations of scale
$L$. The limiting case of infinite $\RM$ corresponds to the case of a perfectly
conducting fluid. This limit has the special property that the magnetic field is
``frozen'' into the fluid. Flux-freezing is a conservation law for the which the
quantity $\int_S \mathbf{B} \cdot d\vct{A}$ is a constant of the motion, when
$S$ represents a surface which is deformed along with the flow. The limit of
vanishing viscosity and resistivity is referred to as ideal MHD.

One can also define the magnetic Prandtl number $\PM = \RM / \RE$ which
characterises the relative importance of viscous and resistive dissipation. In
turbulent flows, it also corresponds to the ratio of length scales where the
velocity and magnetic fields become smooth under the action of their respective
dissipation mechanisms. Thus, for flows having large $\PM$, the velocity field
is smoothed out by viscosity over scales where the magnetic field is still
frozen into the fluid.

As an example, let us estimate the magnetic Prandtl number for various parts of
the sun. The resistively of a fully ionised plasma is given by \citet{Priest2014}
\begin{equation}
  \eta = \frac{m_e}{\mu_0 n_e e^2 \tau_e}
\end{equation}
where $m_e$ is the electron mass, $n_e$ is electron number density, and $\tau_e$
is the electron mean free path. In the solar photosphere this gives $\eta \sim
10^4 \ \rm{m^2 s^{-1}}$. Thus vortices on the solar surface of size $\sim 500
\ \rm{km}$ rotating with typical speed of 1\,km\,s$^{-1}$ \citep{Simon1997} have
magnetic Reynolds number $\RM \sim 5 \times 10^4$. By comparison, the Reynolds
number is of order $10^2$ making the magnetic Prandtl number of order $10^2$. A
similar estimate for the solar convection zone gives $\PM \sim 10^{-2}$. The
magnetic Prandtl number of diffuse astrophysical plasmas is vastly larger, for
example it is $\sim 10^{15}$ \citet{Maron2001a} in the Galaxy and as large as
$10^{22}$ \citep{Schekochihin2002} in the intergalactic medium. These values
reflect their extraordinarily low collisionality.

Other dimensionless numbers of interest include the Lundquist number $S = L v_A
/ \eta$ where $v_A = B / \sqrt{4 \pi \rho}$ is the speed of \alfven wave
propagation. The sonic and \alfven Mach numbers are flow speed measured in
units of their corresponding wave propagation speeds. The sonic Mach number
$\mathcal{M}$ is also roughly the square-root of the ratio of kinetic to
thermal energy densities, and similarly with the \alfven Mach number
$\mathcal{M}_A$ except with the magnetic energy instead of the thermal energy.

\subsection{Conserved quantities}
Conserved quantities in MHD belong to one of three types \citep{Bekenstein1987}.
The first kind, referred to as type A, is where the closed line integral of a
vector field is a constant of the motion, where the path of circulation is
carried along with the local flow. The canonical example is that of Kelvin's
circulation theorem, which states that in a perfect fluid the circulation
$\Gamma = \oint_C \mathbf{u} \cdot \mathbf{ds}$ is a constant of the motion,
where the closed path $C$ is advected and deformed along with the flow. Kelvin's
circulation theorem has been generalised to relativistic hydrodynamics by
\citet{Taub1959}, and to non-relativistic and relativistic MHD by
\citet{Bekenstein2000}. Another example is the \alfven law of magnetic flux
conservation, which states that $\Phi = \oint_C \mathbf{A} \cdot \mathbf{ds}$ is
a constant of the motion. This law can also be stated as the constancy of
magnetic flux through the surface whose boundary is the closed contour $C$. It
formally applies in ideal MHD, where the viscosity and resistivity are
vanishing. This law, commonly referred to as ``flux freezing'', is generally
considered a good approximation in astrophysical plasmas, where the
microphysical diffusion time of magnetic fields is far longer than other
dynamical times. However, for turbulent flows in which the velocity field stays
chaotic down to scales below those of interest (referred to as ``rough''
velocity fields) turbulent diffusivity effectively violates the perfect fluid
assumptions. The violation of Kelvin's circulation theorem in high Reynolds
number hydrodynamic flows has been analysed by \citet{Chen2006}, and similarly
for the \alfven theorem by \citet{Eyink2011}.

The second type of conservation law (type B) is where a local fluid quantity is
a constant of the motion along streamlines. Bernoulli's theorem is the canonical
example of type B conservation laws.

Type C conservation laws apply to the global conservation of scalar or vector
quantities whose transport is strictly local. These are like Gauss's law in that
they relate the conserved quantity's time rate of change in a finite volume to
its flux through the volume's bounding surface. They can thus be stated in
either differential or integral form, the latter of which forms the basis for
modern numerical methods for solving the hydrodynamic and MHD equations. The
prototypical type C conservation law is the continuity equation $\dot{\rho}
+ \nabla \cdot \rho \mathbf{u} = 0$. In the absence of viscous and resistive
effects, momentum and magnetic flux are also conserved in this way, where
$\rho \mathbf{u}$ is replaced with a generic flux function for that conserved
quantity. Similarly, in the absence of heat transport by conduction, the
evolution of total energy can also be written as a continuity equation.

\subsection{Consequences of helicity conservation}
A particularly important conserved quantity is the magnetic helicity $H_M = \int
d^3x \mathbf{A} \cdot \mathbf{B}$. $H_M$ is a global topological
characterisation of the magnetic field, which indicates the degree to which
magnetic field loops are mutually linked. It is analogous to the kinetic
helicity $H_K = \int d^3x \mathbf{u} \cdot \mathbf{\omega}$ (where $\omega =
\nabla \times \mathbf{u}$ is the vorticity) which represents the degree to which
the vortex lines are mutually linked. In a perfect fluid, both helicity measures
are a constant of the motion.

Magnetic helicity is important because it tends to establish magnetic
fields well above the turbulence integral scale. This consequence of
magnetic helicity conservation was first investigated by
\citet{Frisch1975}, where it was predicted that the magnetic energy
would peak around the scale $H_M / E_B$ (where $E_B$ is the total
magnetic energy), even if the injection of magnetic energy took place
at much smaller scales. This process is referred to as inverse
cascading because the magnetic energy cascades from small scales
toward large scales, unlike the direct cascade of energy observed in
hydrodynamic turbulence. It is the underlying mechanism for the
formation of large-scale dynamos which may be responsible for
establishing stellar magnetic fields.

\subsection{Inverse cascade of magnetic energy}
Magnetohydrodynamic turbulence exhibits an unusual
  phenomenon where, at least in its freely decaying state, magnetic
  field correlations arise over distances much longer than the
  original eddy size. This \emph{inverse cascading} effect is
  reminiscent of what is seen in two-dimensional (but not
  three-dimensional) hydrodynamic turbulence. Although this effect may
  be related to the conservation of magnetic helicity
  \citep{Frisch1975}, until recently it was not appreciated that it
  happens even when the field is initially helicity-free. That is,
  when it is topologically equivalent to the uniform (or zero)
  magnetic field.

Inverse cascading of magnetic energy was first seen in numerical simulations by
\citet{Meneguzzi1981}. It has since been confirmed by many others using direct
numerical simulations \citep{Christensson2001, Banerjee2004, Kahniashvili2010,
  Tevzadze2012}, quasi-analytical techniques such as the ``Eddy damped quasi
normal mode'' approximation \citep{Son1999} and also turbulence ``shell models''
\citep{Kalelkar2004}. Such inverse cascading is a process of magnetic field
self-assembly, whereby magnetic energy provided by random motions at very small
scales, may over time establish dynamically important magnetic fields over
scales that are much larger. For example, the sun's convective cells, which set
the turbulence integral scale, are far smaller than the solar diameter. The
ordered magnetic fields that extend over the whole surface of the sun (and many
other stars) may come from such a self-assembly process. Dynamos capable of
establishing magnetic fields on scales far larger than the turbulence integral
scale are generically referred to as large scale dynamos.

The literature to date is still conflicted on whether non-zero magnetic helicity
is a necessary condition for inverse cascading to occur. It was shown by
\citet{Olesen1997} and \citet{Shiromizu1998} that inverse cascading could be
expected even for non-helical magnetic field configurations, as a consequence of
rescaling symmetries native to the Navier-Stokes equations. But no inverse
cascading was seen in numerical studies based on EDQNM theory \citep{Son1999} or
direct numerical simulations with relatively low resolution
\citep{Christensson2001, Banerjee2004}.

Very recently, three numerical studies have appeared which confirm the existence
of inverse cascading in non-helical three-dimensional MHD
turbulence. \citet{Brandenburg2014} carried out high resolution direct numerical
simulations of compressible MHD turbulence with magnetic energy injected at very
small scales. The results confirm the growth of magnetic energy spectrum
$P_M(k,t)$ at low wavenumbers. \citet{Zrake2014} found similar results in high
resolution simulations of freely decaying relativistic MHD turbulence. In the
latter study, the evolution of $P_M(k,t)$ was parametrised around a
self-similar ansatz which may be used to evaluate the time required for magnetic
fields to be established over scales arbitrarily larger than the energy
injection scale. \citet{Berera2014} also found inverse cascades in decaying MHD
turbulence. That study utilised a novel technique, whereby the system was
evolved hydrodynamically with a passive magnetic field. The resulting energy
spectrum also exhibits growth of magnetic field fluctuations at large scales,
suggesting that inverse cascading of passively advected vector fields is a
generic property of hydrodynamic turbulence.

\subsection{Small-scale turbulent dynamo}
In general, the term ``dynamo'' refers to the conversion of kinetic energy into
magnetic energy. In magnetohydrodynamic theory, a number of distinct processes
can facilitate this conversion. The only thing they all have in common is the
presence of turbulence. Many excellent reviews have been written on this the
connection between MHD turbulence and dynamo processes: \citet{Brandenburg2005,
  Lazarian2005, Kulsrud2008, Tobias2011}. Here we will present a very brief
overview of a process known as small-scale turbulent dynamo.

Small-scale turbulent dynamo operates in highly conducting turbulent fluids,
even when the turbulence is isotropic and non-helical. Conceptually it is quite
simple. Magnetic field lines carried along by the turbulent flow become
increasingly chaotic and distorted over time, leading to enhancement of their
energy. When the field is very weak, i.e. $v_A \ll u$, its back-reaction on the
fluid can be ignored, and the field evolves as a passively advected vector field
in otherwise hydrodynamic turbulent flow. This approximation is referred to as
\emph{kinematic} small-scale dynamo, and was first studied by
\citet{Kazantsev1968}, but more modern analytic treatments exist such as that of
\citet{Vincenzi2001}. The original formulation, known as the Kraichnan-Kazantsev
dynamo starts with the assumption of an arbitrary spectrum of velocity
fluctuations that are delta-correlated in time. Despite its simplicity, this
model makes accurate predictions regarding the time rate of change and spectral
energy distribution $P_M(k,t)$ of the magnetic field, where $k$ is the inverse
length scale. In particular, solutions for $P_M(k,t)$ that are exponentially
growing in time exist when the microphysical diffusion coefficient $\eta$ is
small relative to the turbulent diffusivity, a condition which is met when the
Reynolds number of the flow is sufficiently large.

The Kraichnan-Kazantsev dynamo predicts for the unstable growing solutions, (1)
that $P_M(k,t) \propto k^{3/2}$ and peaks at the resistive cutoff, and (2) that
the magnetic energy exponentiates at the fastest eddy turnover time of the
turbulent hydrodynamic cascade. The first numerical experiments to confirm the
basic features of the Kraichnan-Kazantsev dynamo were by
\citet{Meneguzzi1981}. At the time, direct numerical simulations were
exceedingly expensive, and those simulations were limited to grid sizes of
$32^3$ and $64^3$. Since then, many groups have continued this investigation
utilising high performance parallel computing architectures, and modern numerical
schemes for obtaining solutions to the MHD equation.

Although the original work by Kazantsev predicted that the magnetic Reynolds
number $\RM$ had to be ``sufficiently large'' for small scale dynamo action to
occur, the exact value of the critical $\RM$ needed to be found through
numerical experiments. For the case of $\PM = 1$, \citet{Haugen2004c} found that
the critical magnetic Reynolds number $\RM_{\rm crit}$ for dynamo action was
about $35$. For smaller $\PM$, it was found that $\RM_{\rm crit}$ increases as
$\PM$ decreases. This may be partially understood as having to do with the flow
properties in the resistive interval. When the $\PM$ is very large, the
resistive scale lies deep below the viscous cutoff, so that magnetic energy
peaks where the flow is very smooth. However, when $\PM$ is smaller
than one, the resistive scale moves into the turbulent inertial range of the
flow, and the velocity field is \emph{rough} at the resistive
scale. \citet{Haugen2004b} found that $\RM_{\rm crit}$ increased to
$70$ when the flow is supersonic, even when $\PM = 1$. \citet{Federrath2011}
found that for compressible flows the dynamo growth rate depends upon how the
turbulence is forced; vortically driven flows produced more efficient dynamo
than did dilatational forcing. It was argued analytically by \citet{Boldyrev2004}
that in the small $\PM$ regime, the critical magnetic Reynolds number should be
a strong function of the \emph{roughness} exponent $\alpha$ of the velocity
field, $v_\ell \propto \ell^\alpha$, such that smaller $\alpha$ (a rougher field)
requires a larger $\RM$ to produce small-scale kinematic dynamo.

A central question around small-scale dynamo is the nature of its
saturation. The exponentially growing solution of the Kazentsev model
holds only as long as the dynamo is in the kinematic regime, so that
magnetic back-reaction on the flow is negligible. Once the field
attains a strength high enough to compete with the fluid inertia, the
kinematic assumption breaks down, and some sort of saturation behaviour
is expected. This saturation is a highly nonlinear problem, and thus
very difficult to study analytically. \citet{Schekochihin2002a} argued
that in the non-linear stage, at least for very large $\PM$, magnetic
energy would remain concentrated in the sub-viscous range, moving
toward the inertial range on a resistive rather than dynamical
time. Were this to be the case, equipartition of the
  turbulent kinetic and magnetic energy densities at the largest
  scales would require an asymptotically long time to be established.
However, this scenario has not been observed in the numerical
literature. Instead, the emerging consensus (other differences aside)
\citep{Cho2000a, Brandenburg2003, Haugen2004, Schleicher2013,
  Schekochihin2004a, Maron2001a, Zrake2011, Zrake2013} is that
magnetic energy shifts into the inertial range regardless of the
magnetic Prandtl number. Once there, the field attains
  coherence at increasing scales until finally reaching the
  energy-containing scale of turbulence, thus establishing
  scale-by-scale equipartition. Both the kinetic and magnetic energy
follow a Kolmogorov spectrum, with the magnetic energy per unit
wavelength exceeding that of the kinetic energy by a factor of around
$2$.

The time-scale for the flow to fully establish scale-by-scale equipartition is
still not completely clear. It was argued by \citet{Beresnyak2012} that
non-linear small-scale dynamo exhibits universality in the sense that a constant
fraction $C_E$ of the work done by turbulent pumping accumulates in magnetic
energy. In this scenario, scale-by-scale equipartition should be attained after
a time $1 / C_E$. Careful modelling of the approach to saturation
by \citet{Zrake2013} finds that indeed scale-by-scale equipartition is reached
after a number of dynamical times that is independent of the Reynolds number.

\section{Conclusions}

In this chapter we have reviewed progress made on the origin of
magnetic fields in intermediate and massive main sequence stars and in
compact stars.

Observations of magnetic fields in white dwarfs may indicate that the
distribution is bi-modal, exhibiting a paucity of objects between
$10^5-10^6$\,G, although further sensitive
  spectropolarimetric surveys conducted on 8\,m class telescopes are
  needed in order to confirm this finding. Should this bimodality be
confirmed, then two different channels for field formation and
evolution could be at play. In particular, one could identify weakly
magnetic main sequence stars such as Vega and Sirius as the
progenitors of weakly magnetic white dwarfs belonging to the low field
component. This would lend some support to the fossil field
hypothesis for the origin of fields in compact stars.

However, the viability of the fossil field hypothesis
  has been recently questioned because of the absence of magnetic
  white dwarfs paired with non-degenerate stars.  These observations
  mirror the dearth of magnetic non-degenerate A, B and O stars in
  short-period binary systems. It is not clear whether there is any
  connection or analogy to be made between fields on the main sequence and
  compact star phases, but a merger scenario for field formation has been suggested to
  overcome some of the problems raised by recent observations.

Here we stress that the pairing properties of magnetic
  white dwarfs do not necessarily invalidate the viability of the
  fossil scenario since more than one formation channel could be at
  work. However, the complete lack, rather than just a dearth, of
  magnetic white dwarfs paired with non-degenerate companions in
  non-interacting systems, raises strong doubts on whether the fossil
  route can be the main channel for the formation of magnetic white
  dwarfs.  If magnetic white dwarfs were the descendants of magnetic
  main sequence stars, then magnetic white dwarfs should be commonly
  found in binary systems to reflect the incidence of binarity
  observed in (magnetic) stars of F to late B spectral types. However,
  there is \emph{not one single example} of a magnetic white dwarf
  paired with a non-degenerate companion in a non-interacting binary.

  In order to explain this curious finding, it has been proposed that
  fields in isolated highly magnetic white dwarfs could be generated
  during common envelope evolution when the two components of a binary
  system merge. 

  The observational situation surrounding birth magnetic fields of
  neutron stars remains frustratingly uncertain, because what
  measurements exist are indirect. Population synthesis models point,
  on balance, to moderate birth fields $\sim 10^{12.65\pm
    0.55}$\,G. Zeeman spectropolarimetry has revealed strongly
  magnetised (i.e., a few thousand gauss) progenitors across a range
  of spectral classes, but even so it seems that the numbers may be
  insufficient to account for the magnetar birth rate under the fossil
  field scenario. At least for magnetars, therefore, a proto-neutron
  star dynamo is favoured, driven by differential rotation and the
  Tayler or magneto-rotational instabilities rather than thermal
  convection \citep{s09}. Emergence (through buoyancy), stability
  (through helicity), and ongoing activity (through superfluid
  turbulence) play roles in determining the strength of the
  (observable) surface field. Future gravitational wave observations
  offer the best prospects for direct internal field
  measurements. Until these observations become
    available we can only say that, as for magnetic white dwarfs, it is
    possible that both fossil and dynamo processes are at work in
    different neutron stars.

We have also reviewed the theory of magnetohydrodynamic turbulence
because its presence underpins the operation of stellar dynamos and
controls diffusive magnetic flux transport in stars.  But MHD
turbulence poses a great theoretical challenge due to its inherent
complexities, and many features of its operation are yet to be
satisfactorily explained. Advances in numerical algorithms and
computing resources have yielded considerable gains in the empirical
grounding of various phenomenological pictures. 

\section*{Acknowledgements}

The authors wish to thank the International Space Science Institute (ISSI) for hospitality and support. This paper was supported by an Australian Research Council Discovery Project grant. Andrew Melatos thanks Alpha Mastrano for extensive assistance with typesetting Sect.\,\ref{ns-magnetism}. We would like to thank the Referee for carefully reading our manuscript and for making many valuable comments that have helped us improve this review paper.

\bibliographystyle{spr-mp-nameyear-cnd}  
\bibliography{ferrario,melatos,zrake}                

\begin{thebibliography}{259}
\ifx \bisbn   \undefined \def \bisbn  #1{ISBN #1}\fi
\ifx \binits  \undefined \def \binits#1{#1} \fi
\ifx \bauthor  \undefined \def \bauthor#1{#1} \fi
\ifx \batitle  \undefined \def \batitle#1{#1} \fi
\ifx \bjtitle  \undefined \def \bjtitle#1{#1}\fi
\ifx \bvolume  \undefined \def \bvolume#1{\textbf{#1}}\fi
\ifx \byear  \undefined \def \byear#1{#1} \fi
\ifx \bissue  \undefined \def \bissue#1{#1} \fi
\ifx \bfpage  \undefined \def \bfpage#1{#1} \fi
\ifx \blpage  \undefined \def \blpage #1{#1} \fi
\ifx \burl  \undefined \def \burl#1{\textsf{#1}} \fi
\ifx \doiurl  \undefined \def \doiurl#1{\textsf{#1}} \fi
\ifx \betal  \undefined \def \betal{\textit{et al.}} \fi
\ifx \binstitute  \undefined \def \binstitute#1{#1} \fi
\ifx \binstitutionaled  \undefined \def \binstitutionaled#1{#1} \fi
\ifx \bctitle  \undefined \def \bctitle#1{#1} \fi
\ifx \beditor  \undefined \def \beditor#1{#1} \fi
\ifx \bpublisher  \undefined \def \bpublisher#1{#1} \fi
\ifx \bbtitle  \undefined \def \bbtitle#1{#1} \fi
\ifx \bedition  \undefined \def \bedition#1{#1} \fi
\ifx \bseriesno  \undefined \def \bseriesno#1{#1} \fi
\ifx \blocation  \undefined \def \blocation#1{#1} \fi
\ifx \bsertitle  \undefined \def \bsertitle#1{#1} \fi
\ifx \bsnm \undefined \def \bsnm#1{#1} \fi
\ifx \bsuffix \undefined \def \bsuffix#1{#1} \fi
\ifx \bparticle \undefined \def \bparticle#1{#1} \fi
\ifx \barticle \undefined \def \barticle#1{#1} \fi
\ifx \bconfdate \undefined \def \bconfdate #1{#1} \fi
\ifx \botherref \undefined \def \botherref #1{#1} \fi
\ifx \url \undefined \def \url#1{\textsf{#1}} \fi
\ifx \bchapter \undefined \def \bchapter#1{#1} \fi
\ifx \bbook \undefined \def \bbook#1{#1} \fi
\ifx \bcomment \undefined \def \bcomment#1{#1} \fi
\ifx \oauthor \undefined \def \oauthor#1{#1} \fi
\ifx \citeauthoryear \undefined \def \citeauthoryear#1{#1} \fi
\ifx \endbibitem  \undefined \def \endbibitem {}\fi
\ifx \bconflocation  \undefined \def \bconflocation#1{#1} \fi
\ifx \arxivurl  \undefined \def \arxivurl#1{\textsf{#1}} \fi

\bibitem[\protect\citeauthoryear{{Aasi} et~al.}{2014}]{aetal14}
\begin{barticle}
\bauthor{\bsnm{{Aasi}}, \binits{J.}},
\bauthor{\bsnm{{Abadie}}, \binits{J.}},
\bauthor{\bsnm{{Abbott}}, \binits{B.P.}},
\bauthor{\bsnm{{Abbott}}, \binits{R.}},
\bauthor{\bsnm{{Abbott}}, \binits{T.}},
\bauthor{\bsnm{{Abernathy}}, \binits{M.R.}},
\bauthor{\bsnm{{Accadia}}, \binits{T.}},
\bauthor{\bsnm{{Acernese}}, \binits{F.}},
\bauthor{\bsnm{{Adams}}, \binits{C.}},
\bauthor{\bsnm{{Adams}}, \binits{T.}},
\bauthor{\bparticle{et} \bsnm{al.}}:
\bjtitle{\apj}
\bvolume{785},
\bfpage{119}
(\byear{2014}).
\arxivurl{1309.4027}.
doi:\doiurl{10.1088/0004-637X/785/2/119}
\end{barticle}
\endbibitem

\bibitem[\protect\citeauthoryear{{Abadie} et~al.}{2011}]{aetal11}
\begin{barticle}
\bauthor{\bsnm{{Abadie}}, \binits{J.}},
\bauthor{\bsnm{{Abbott}}, \binits{B.P.}},
\bauthor{\bsnm{{Abbott}}, \binits{R.}},
\bauthor{\bsnm{{Abernathy}}, \binits{M.}},
\bauthor{\bsnm{{Accadia}}, \binits{T.}},
\bauthor{\bsnm{{Acernese}}, \binits{F.}},
\bauthor{\bsnm{{Adams}}, \binits{C.}},
\bauthor{\bsnm{{Adhikari}}, \binits{R.}},
\bauthor{\bsnm{{Affeldt}}, \binits{C.}},
\bauthor{\bsnm{{Allen}}, \binits{B.}},
\bauthor{\bparticle{et} \bsnm{al.}}:
\bjtitle{\apj}
\bvolume{737},
\bfpage{93}
(\byear{2011}).
\arxivurl{1104.2712}.
doi:\doiurl{10.1088/0004-637X/737/2/93}
\end{barticle}
\endbibitem

\bibitem[\protect\citeauthoryear{{Abbott} et~al.}{2010}]{aetal10}
\begin{barticle}
\bauthor{\bsnm{{Abbott}}, \binits{B.P.}},
\bauthor{\bsnm{{Abbott}}, \binits{R.}},
\bauthor{\bsnm{{Acernese}}, \binits{F.}},
\bauthor{\bsnm{{Adhikari}}, \binits{R.}},
\bauthor{\bsnm{{Ajith}}, \binits{P.}},
\bauthor{\bsnm{{Allen}}, \binits{B.}},
\bauthor{\bsnm{{Allen}}, \binits{G.}},
\bauthor{\bsnm{{Alshourbagy}}, \binits{M.}},
\bauthor{\bsnm{{Amin}}, \binits{R.S.}},
\bauthor{\bsnm{{Anderson}}, \binits{S.B.}},
\bauthor{\bparticle{et} \bsnm{al.}}:
\bjtitle{\apj}
\bvolume{713},
\bfpage{671}
(\byear{2010}).
\arxivurl{0909.3583}.
doi:\doiurl{10.1088/0004-637X/713/1/671}
\end{barticle}
\endbibitem

\bibitem[\protect\citeauthoryear{{Abney} and {Epstein}}{1996}]{ae96}
\begin{barticle}
\bauthor{\bsnm{{Abney}}, \binits{M.}},
\bauthor{\bsnm{{Epstein}}, \binits{R.I.}}:
\bjtitle{Journal of Fluid Mechanics}
\bvolume{312},
\bfpage{327}
(\byear{1996}).
\arxivurl{astro-ph/9504051}.
doi:\doiurl{10.1017/S0022112096002030}
\end{barticle}
\endbibitem

\bibitem[\protect\citeauthoryear{{Akg{\"u}n} et~al.}{2013}]{aetal13}
\begin{barticle}
\bauthor{\bsnm{{Akg{\"u}n}}, \binits{T.}},
\bauthor{\bsnm{{Reisenegger}}, \binits{A.}},
\bauthor{\bsnm{{Mastrano}}, \binits{A.}},
\bauthor{\bsnm{{Marchant}}, \binits{P.}}:
\bjtitle{\mnras}
\bvolume{433},
\bfpage{2445}
(\byear{2013}).
\arxivurl{1302.0273}.
doi:\doiurl{10.1093/mnras/stt913}
\end{barticle}
\endbibitem

\bibitem[\protect\citeauthoryear{{Alecian} et~al.}{2008}]{Alecian2008}
\begin{barticle}
\bauthor{\bsnm{{Alecian}}, \binits{E.}},
\bauthor{\bsnm{{Catala}}, \binits{C.}},
\bauthor{\bsnm{{Wade}}, \binits{G.A.}},
\bauthor{\bsnm{{Donati}}, \binits{J.-F.}},
\bauthor{\bsnm{{Petit}}, \binits{P.}},
\bauthor{\bsnm{{Landstreet}}, \binits{J.D.}},
\bauthor{\bsnm{{B{\"o}hm}}, \binits{T.}},
\bauthor{\bsnm{{Bouret}}, \binits{J.-C.}},
\bauthor{\bsnm{{Bagnulo}}, \binits{S.}},
\bauthor{\bsnm{{Folsom}}, \binits{C.}},
\bauthor{\bsnm{{Grunhut}}, \binits{J.}},
\bauthor{\bsnm{{Silvester}}, \binits{J.}}:
\bjtitle{\mnras}
\bvolume{385},
\bfpage{391}
(\byear{2008}).
\arxivurl{0712.1746}.
doi:\doiurl{10.1111/j.1365-2966.2008.12842.x}
\end{barticle}
\endbibitem

\bibitem[\protect\citeauthoryear{{Alecian} et~al.}{2009}]{Alecian2009}
\begin{barticle}
\bauthor{\bsnm{{Alecian}}, \binits{E.}},
\bauthor{\bsnm{{Wade}}, \binits{G.A.}},
\bauthor{\bsnm{{Catala}}, \binits{C.}},
\bauthor{\bsnm{{Bagnulo}}, \binits{S.}},
\bauthor{\bsnm{{B{\"o}hm}}, \binits{T.}},
\bauthor{\bsnm{{Bouret}}, \binits{J.-C.}},
\bauthor{\bsnm{{Donati}}, \binits{J.-F.}},
\bauthor{\bsnm{{Folsom}}, \binits{C.P.}},
\bauthor{\bsnm{{Grunhut}}, \binits{J.}},
\bauthor{\bsnm{{Landstreet}}, \binits{J.D.}}:
\bjtitle{\mnras}
\bvolume{400},
\bfpage{354}
(\byear{2009}).
\arxivurl{0907.5113}.
doi:\doiurl{10.1111/j.1365-2966.2009.15460.x}
\end{barticle}
\endbibitem

\bibitem[\protect\citeauthoryear{{Alecian} et~al.}{2013a}]{Alecian2013a}
\begin{barticle}
\bauthor{\bsnm{{Alecian}}, \binits{E.}},
\bauthor{\bsnm{{Wade}}, \binits{G.A.}},
\bauthor{\bsnm{{Catala}}, \binits{C.}},
\bauthor{\bsnm{{Grunhut}}, \binits{J.H.}},
\bauthor{\bsnm{{Landstreet}}, \binits{J.D.}},
\bauthor{\bsnm{{Bagnulo}}, \binits{S.}},
\bauthor{\bsnm{{B{\"o}hm}}, \binits{T.}},
\bauthor{\bsnm{{Folsom}}, \binits{C.P.}},
\bauthor{\bsnm{{Marsden}}, \binits{S.}},
\bauthor{\bsnm{{Waite}}, \binits{I.}}:
\bjtitle{\mnras}
\bvolume{429},
\bfpage{1001}
(\byear{2013}a).
\arxivurl{1211.2907}.
doi:\doiurl{10.1093/mnras/sts383}
\end{barticle}
\endbibitem

\bibitem[\protect\citeauthoryear{{Alecian} et~al.}{2013b}]{Alecian2013b}
\begin{barticle}
\bauthor{\bsnm{{Alecian}}, \binits{E.}},
\bauthor{\bsnm{{Neiner}}, \binits{C.}},
\bauthor{\bsnm{{Mathis}}, \binits{S.}},
\bauthor{\bsnm{{Catala}}, \binits{C.}},
\bauthor{\bsnm{{Kochukhov}}, \binits{O.}},
\bauthor{\bsnm{{Landstreet}}, \binits{J.}}:
\bjtitle{\aap}
\bvolume{549},
\bfpage{8}
(\byear{2013}b).
\arxivurl{1301.1804}.
doi:\doiurl{10.1051/0004-6361/201220796}
\end{barticle}
\endbibitem

\bibitem[\protect\citeauthoryear{{Alecian} et~al.}{2014}]{Alecian2014}
\begin{botherref}
\oauthor{\bsnm{{Alecian}}, \binits{E.}},
\oauthor{\bsnm{{Neiner}}, \binits{C.}},
\oauthor{\bsnm{{Wade}}, \binits{G.A.}},
\oauthor{\bsnm{{Mathis}}, \binits{S.}},
\oauthor{\bsnm{{Bohlender}}, \binits{D.}},
\oauthor{\bsnm{{C{\'e}bron}}, \binits{D.}},
\oauthor{\bsnm{{Folsom}}, \binits{C.}},
\oauthor{\bsnm{{Grunhut}}, \binits{J.}},
\oauthor{\bsnm{{Le Bouquin}}, \binits{J.-B.}},
\oauthor{\bsnm{{Petit}}, \binits{V.}},
\oauthor{\bsnm{{Sana}}, \binits{H.}},
\oauthor{\bsnm{{Tkachenko}}, \binits{A.}},
\oauthor{\bsnm{{ud-Doula}}, \binits{A.}},
\oauthor{\bsnm{{the BinaMIcS collaboration}}}:
ArXiv e-prints
(2014).
\arxivurl{1409.1094}
\end{botherref}
\endbibitem

\bibitem[\protect\citeauthoryear{{Armaza} et~al.}{2013}]{aetal13b}
\begin{botherref}
\oauthor{\bsnm{{Armaza}}, \binits{C.}},
\oauthor{\bsnm{{Reisenegger}}, \binits{A.}},
\oauthor{\bsnm{{Alejandro Valdivia}}, \binits{J.}},
\oauthor{\bsnm{{Marchant}}, \binits{P.}}:
ArXiv e-prints
(2013).
\arxivurl{1305.0592}
\end{botherref}
\endbibitem

\bibitem[\protect\citeauthoryear{{Auri{\`e}re} et~al.}{2007}]{Auriere2007}
\begin{barticle}
\bauthor{\bsnm{{Auri{\`e}re}}, \binits{M.}},
\bauthor{\bsnm{{Wade}}, \binits{G.A.}},
\bauthor{\bsnm{{Silvester}}, \binits{J.}},
\bauthor{\bsnm{{Ligni{\`e}res}}, \binits{F.}},
\bauthor{\bsnm{{Bagnulo}}, \binits{S.}},
\bauthor{\bsnm{{Bale}}, \binits{K.}},
\bauthor{\bsnm{{Dintrans}}, \binits{B.}},
\bauthor{\bsnm{{Donati}}, \binits{J.F.}},
\bauthor{\bsnm{{Folsom}}, \binits{C.P.}},
\bauthor{\bsnm{{Gruberbauer}}, \binits{M.}},
\bauthor{\bsnm{{Hui Bon Hoa}}, \binits{A.}},
\bauthor{\bsnm{{Jeffers}}, \binits{S.}},
\bauthor{\bsnm{{Johnson}}, \binits{N.}},
\bauthor{\bsnm{{Landstreet}}, \binits{J.D.}},
\bauthor{\bsnm{{L{\`e}bre}}, \binits{A.}},
\bauthor{\bsnm{{Lueftinger}}, \binits{T.}},
\bauthor{\bsnm{{Marsden}}, \binits{S.}},
\bauthor{\bsnm{{Mouillet}}, \binits{D.}},
\bauthor{\bsnm{{Naseri}}, \binits{S.}},
\bauthor{\bsnm{{Paletou}}, \binits{F.}},
\bauthor{\bsnm{{Petit}}, \binits{P.}},
\bauthor{\bsnm{{Power}}, \binits{J.}},
\bauthor{\bsnm{{Rincon}}, \binits{F.}},
\bauthor{\bsnm{{Strasser}}, \binits{S.}},
\bauthor{\bsnm{{Toqu{\'e}}}, \binits{N.}}:
\bjtitle{\aap}
\bvolume{475},
\bfpage{1053}
(\byear{2007}).
\arxivurl{0710.1554}.
doi:\doiurl{10.1051/0004-6361:20078189}
\end{barticle}
\endbibitem

\bibitem[\protect\citeauthoryear{{Auri{\`e}re} et~al.}{2008}]{Auriere2008}
\begin{barticle}
\bauthor{\bsnm{{Auri{\`e}re}}, \binits{M.}},
\bauthor{\bsnm{{Konstantinova-Antova}}, \binits{R.}},
\bauthor{\bsnm{{Petit}}, \binits{P.}},
\bauthor{\bsnm{{Charbonnel}}, \binits{C.}},
\bauthor{\bsnm{{Dintrans}}, \binits{B.}},
\bauthor{\bsnm{{Ligni{\`e}res}}, \binits{F.}},
\bauthor{\bsnm{{Roudier}}, \binits{T.}},
\bauthor{\bsnm{{Alecian}}, \binits{E.}},
\bauthor{\bsnm{{Donati}}, \binits{J.F.}},
\bauthor{\bsnm{{Landstreet}}, \binits{J.D.}},
\bauthor{\bsnm{{Wade}}, \binits{G.A.}}:
\bjtitle{\aap}
\bvolume{491},
\bfpage{499}
(\byear{2008}).
\arxivurl{0810.2228}.
doi:\doiurl{10.1051/0004-6361:200810502}
\end{barticle}
\endbibitem

\bibitem[\protect\citeauthoryear{{Auri{\`e}re} et~al.}{2010}]{Auriere2010}
\begin{barticle}
\bauthor{\bsnm{{Auri{\`e}re}}, \binits{M.}},
\bauthor{\bsnm{{Wade}}, \binits{G.A.}},
\bauthor{\bsnm{{Ligni{\`e}res}}, \binits{F.}},
\bauthor{\bsnm{{Hui-Bon-Hoa}}, \binits{A.}},
\bauthor{\bsnm{{Landstreet}}, \binits{J.D.}},
\bauthor{\bsnm{{Iliev}}, \binits{I.K.}},
\bauthor{\bsnm{{Donati}}, \binits{J.-F.}},
\bauthor{\bsnm{{Petit}}, \binits{P.}},
\bauthor{\bsnm{{Roudier}}, \binits{T.}},
\bauthor{\bsnm{{Th{\'e}ado}}, \binits{S.}}:
\bjtitle{\aap}
\bvolume{523},
\bfpage{40}
(\byear{2010}).
\arxivurl{1008.3086}.
doi:\doiurl{10.1051/0004-6361/201014848}
\end{barticle}
\endbibitem

\bibitem[\protect\citeauthoryear{{Auriere} et~al.}{2013}]{Auriere2013}
\begin{botherref}
\oauthor{\bsnm{{Auriere}}, \binits{M.}},
\oauthor{\bsnm{{Lignieres}}, \binits{F.}},
\oauthor{\bsnm{{Konstantinova-Antova}}, \binits{R.}},
\oauthor{\bsnm{{Charbonnel}}, \binits{C.}},
\oauthor{\bsnm{{Petit}}, \binits{P.}},
\oauthor{\bsnm{{Tsvetkova}}, \binits{S.}},
\oauthor{\bsnm{{Wade}}, \binits{G.A.}}:
ArXiv e-prints
(2013).
\arxivurl{1310.6942}
\end{botherref}
\endbibitem

\bibitem[\protect\citeauthoryear{{Aznar Cuadrado}
  et~al.}{2004}]{AznarCuadrado2004}
\begin{barticle}
\bauthor{\bsnm{{Aznar Cuadrado}}, \binits{R.}},
\bauthor{\bsnm{{Jordan}}, \binits{S.}},
\bauthor{\bsnm{{Napiwotzki}}, \binits{R.}},
\bauthor{\bsnm{{Schmid}}, \binits{H.M.}},
\bauthor{\bsnm{{Solanki}}, \binits{S.K.}},
\bauthor{\bsnm{{Mathys}}, \binits{G.}}:
\bjtitle{\aap}
\bvolume{423},
\bfpage{1081}
(\byear{2004}).
\arxivurl{astro-ph/0405308}.
doi:\doiurl{10.1051/0004-6361:20040355}
\end{barticle}
\endbibitem

\bibitem[\protect\citeauthoryear{{Baade}}{1942}]{Baade1942}
\begin{barticle}
\bauthor{\bsnm{{Baade}}, \binits{W.}}:
\bjtitle{\apj}
\bvolume{96},
\bfpage{188}
(\byear{1942}).
doi:\doiurl{10.1086/144446}
\end{barticle}
\endbibitem

\bibitem[\protect\citeauthoryear{{Baade} and {Zwicky}}{1934}]{BaadeZwicky1934}
\begin{barticle}
\bauthor{\bsnm{{Baade}}, \binits{W.}},
\bauthor{\bsnm{{Zwicky}}, \binits{F.}}:
\bjtitle{Proceedings of the National Academy of Science}
\bvolume{20},
\bfpage{254}
(\byear{1934}).
doi:\doiurl{10.1073/pnas.20.5.254}
\end{barticle}
\endbibitem

\bibitem[\protect\citeauthoryear{{Babcock}}{1947}]{Babcock1947}
\begin{barticle}
\bauthor{\bsnm{{Babcock}}, \binits{H.W.}}:
\bjtitle{Contributions from the Mount Wilson Observatory / Carnegie Institution
  of Washington}
\bvolume{727},
\bfpage{1}
(\byear{1947})
\end{barticle}
\endbibitem

\bibitem[\protect\citeauthoryear{{Babcock}}{1958}]{Babcock1958}
\begin{barticle}
\bauthor{\bsnm{{Babcock}}, \binits{H.W.}}:
\bjtitle{\apjs}
\bvolume{3},
\bfpage{141}
(\byear{1958}).
doi:\doiurl{10.1086/190035}
\end{barticle}
\endbibitem

\bibitem[\protect\citeauthoryear{{Bagnulo} et~al.}{2006a}]{betal06}
\begin{barticle}
\bauthor{\bsnm{{Bagnulo}}, \binits{S.}},
\bauthor{\bsnm{{Landstreet}}, \binits{J.D.}},
\bauthor{\bsnm{{Mason}}, \binits{E.}},
\bauthor{\bsnm{{Andretta}}, \binits{V.}},
\bauthor{\bsnm{{Silaj}}, \binits{J.}},
\bauthor{\bsnm{{Wade}}, \binits{G.A.}}:
\bjtitle{\aap}
\bvolume{450},
\bfpage{777}
(\byear{2006}a).
\arxivurl{astro-ph/0601516}.
doi:\doiurl{10.1051/0004-6361:20054223}
\end{barticle}
\endbibitem

\bibitem[\protect\citeauthoryear{{Bagnulo} et~al.}{2006b}]{Bagnulo2006}
\begin{barticle}
\bauthor{\bsnm{{Bagnulo}}, \binits{S.}},
\bauthor{\bsnm{{Landstreet}}, \binits{J.D.}},
\bauthor{\bsnm{{Mason}}, \binits{E.}},
\bauthor{\bsnm{{Andretta}}, \binits{V.}},
\bauthor{\bsnm{{Silaj}}, \binits{J.}},
\bauthor{\bsnm{{Wade}}, \binits{G.A.}}:
\bjtitle{\aap}
\bvolume{450},
\bfpage{777}
(\byear{2006}b).
\arxivurl{astro-ph/0601516}.
doi:\doiurl{10.1051/0004-6361:20054223}
\end{barticle}
\endbibitem

\bibitem[\protect\citeauthoryear{Banerjee and Jedamzik}{2004}]{Banerjee2004}
\begin{barticle}
\bauthor{\bsnm{Banerjee}, \binits{R.}},
\bauthor{\bsnm{Jedamzik}, \binits{K.}}:
\bjtitle{Physical Review D}
\bvolume{70}(\bissue{12}),
\bfpage{123003}
(\byear{2004}).
doi:\doiurl{10.1103/PhysRevD.70.123003}
\end{barticle}
\endbibitem

\bibitem[\protect\citeauthoryear{{Barstow} et~al.}{1995}]{Barstow1995}
\begin{barticle}
\bauthor{\bsnm{{Barstow}}, \binits{M.A.}},
\bauthor{\bsnm{{Jordan}}, \binits{S.}},
\bauthor{\bsnm{{O'Donoghue}}, \binits{D.}},
\bauthor{\bsnm{{Burleigh}}, \binits{M.R.}},
\bauthor{\bsnm{{Napiwotzki}}, \binits{R.}},
\bauthor{\bsnm{{Harrop-Allin}}, \binits{M.K.}}:
\bjtitle{\mnras}
\bvolume{277},
\bfpage{971}
(\byear{1995})
\end{barticle}
\endbibitem

\bibitem[\protect\citeauthoryear{{Barsukov} et~al.}{2013}]{bgt13}
\begin{barticle}
\bauthor{\bsnm{{Barsukov}}, \binits{D.P.}},
\bauthor{\bsnm{{Goglichidze}}, \binits{O.A.}},
\bauthor{\bsnm{{Tsygan}}, \binits{A.I.}}:
\bjtitle{\mnras}
\bvolume{432},
\bfpage{520}
(\byear{2013}).
\arxivurl{1303.5672}.
doi:\doiurl{10.1093/mnras/stt501}
\end{barticle}
\endbibitem

\bibitem[\protect\citeauthoryear{{Barsukov} et~al.}{2014}]{bgt14}
\begin{barticle}
\bauthor{\bsnm{{Barsukov}}, \binits{D.P.}},
\bauthor{\bsnm{{Goglichidze}}, \binits{O.A.}},
\bauthor{\bsnm{{Tsygan}}, \binits{A.I.}}:
\bjtitle{Journal of Physics Conference Series}
\bvolume{496}(\bissue{1}),
\bfpage{012013}
(\byear{2014}).
\arxivurl{1402.1444}.
doi:\doiurl{10.1088/1742-6596/496/1/012013}
\end{barticle}
\endbibitem

\bibitem[\protect\citeauthoryear{Bekenstein and Oron}{2000}]{Bekenstein2000}
\begin{barticle}
\bauthor{\bsnm{Bekenstein}, \binits{J.}},
\bauthor{\bsnm{Oron}, \binits{A.}}:
\bjtitle{Physical Review E}
\bvolume{62}(\bissue{4}),
\bfpage{5594}
(\byear{2000}).
doi:\doiurl{10.1103/PhysRevE.62.5594}
\end{barticle}
\endbibitem

\bibitem[\protect\citeauthoryear{Bekenstein}{1987}]{Bekenstein1987}
\begin{barticle}
\bauthor{\bsnm{Bekenstein}, \binits{J.D.}}:
\bjtitle{The Astrophysical Journal}
\bvolume{319},
\bfpage{207}
(\byear{1987}).
doi:\doiurl{10.1086/165447}
\end{barticle}
\endbibitem

\bibitem[\protect\citeauthoryear{Berera and Linkmann}{2014}]{Berera2014}
\begin{botherref}
\oauthor{\bsnm{Berera}, \binits{A.}},
\oauthor{\bsnm{Linkmann}, \binits{M.}}:
{Inverse cascades and the evolution of decaying magnetohydrodynamic turbulence}
(2014).
\arxivurl{1405.6756}
\end{botherref}
\endbibitem

\bibitem[\protect\citeauthoryear{Beresnyak}{2012}]{Beresnyak2012}
\begin{barticle}
\bauthor{\bsnm{Beresnyak}, \binits{A.}}:
\bjtitle{Phys Rev Lett}
\bvolume{108},
\bfpage{35002}
(\byear{2012})
\end{barticle}
\endbibitem

\bibitem[\protect\citeauthoryear{{Biskamp}}{1997}]{b97}
\begin{bbook}
\bauthor{\bsnm{{Biskamp}}, \binits{D.}}:
\bbtitle{{Nonlinear Magnetohydrodynamics}},
(\byear{1997})
\end{bbook}
\endbibitem

\bibitem[\protect\citeauthoryear{{Blackett}}{1947}]{Blackett1947}
\begin{barticle}
\bauthor{\bsnm{{Blackett}}, \binits{P.M.S.}}:
\bjtitle{\nat}
\bvolume{159},
\bfpage{658}
(\byear{1947}).
doi:\doiurl{10.1038/159658a0}
\end{barticle}
\endbibitem

\bibitem[\protect\citeauthoryear{{Blaz{\`e}re} et~al.}{2014}]{Blazere2014}
\begin{bchapter}
\bauthor{\bsnm{{Blaz{\`e}re}}, \binits{A.}},
\bauthor{\bsnm{{Petit}}, \binits{P.}},
\bauthor{\bsnm{{Ligni{\`e}res}}, \binits{F.}},
\bauthor{\bsnm{{Auri{\`e}re}}, \binits{M.}},
\bauthor{\bsnm{{B{\"o}hm}}, \binits{T.}},
\bauthor{\bsnm{{Wade}}, \binits{G.}}:
In: \beditor{\bsnm{{Ballet}}, \binits{J.}},
\beditor{\bsnm{{Martins}}, \binits{F.}},
\beditor{\bsnm{{Bournaud}}, \binits{F.}},
\beditor{\bsnm{{Monier}}, \binits{R.}},
\beditor{\bsnm{{Reyl{\'e}}}, \binits{C.}} (eds.)
\bbtitle{SF2A-2014: Proceedings of the Annual meeting of the French Society of
  Astronomy and Astrophysics},
p. \bfpage{463}
(\byear{2014}).
\arxivurl{1410.1412}
\end{bchapter}
\endbibitem

\bibitem[\protect\citeauthoryear{{Bogomazov} and
  {Tutukov}}{2009}]{Bogomazov2009}
\begin{barticle}
\bauthor{\bsnm{{Bogomazov}}, \binits{A.I.}},
\bauthor{\bsnm{{Tutukov}}, \binits{A.V.}}:
\bjtitle{Astronomy Reports}
\bvolume{53},
\bfpage{214}
(\byear{2009}).
\arxivurl{0901.4899}.
doi:\doiurl{10.1134/S1063772909030032}
\end{barticle}
\endbibitem

\bibitem[\protect\citeauthoryear{Boldyrev and Cattaneo}{2004}]{Boldyrev2004}
\begin{barticle}
\bauthor{\bsnm{Boldyrev}, \binits{S.}},
\bauthor{\bsnm{Cattaneo}, \binits{F.}}:
\bjtitle{Phys Rev Lett}
\bvolume{92},
\bfpage{144501}
(\byear{2004})
\end{barticle}
\endbibitem

\bibitem[\protect\citeauthoryear{{Bonanno} et~al.}{2003}]{bru03}
\begin{barticle}
\bauthor{\bsnm{{Bonanno}}, \binits{A.}},
\bauthor{\bsnm{{Rezzolla}}, \binits{L.}},
\bauthor{\bsnm{{Urpin}}, \binits{V.}}:
\bjtitle{\aap}
\bvolume{410},
\bfpage{33}
(\byear{2003}).
\arxivurl{astro-ph/0309783}.
doi:\doiurl{10.1051/0004-6361:20031459}
\end{barticle}
\endbibitem

\bibitem[\protect\citeauthoryear{{Bonanno} et~al.}{2005}]{bub05}
\begin{barticle}
\bauthor{\bsnm{{Bonanno}}, \binits{A.}},
\bauthor{\bsnm{{Urpin}}, \binits{V.}},
\bauthor{\bsnm{{Belvedere}}, \binits{G.}}:
\bjtitle{\aap}
\bvolume{440},
\bfpage{199}
(\byear{2005}).
\arxivurl{astro-ph/0504328}.
doi:\doiurl{10.1051/0004-6361:20042098}
\end{barticle}
\endbibitem

\bibitem[\protect\citeauthoryear{{Borra} et~al.}{1982}]{Borra1982}
\begin{barticle}
\bauthor{\bsnm{{Borra}}, \binits{E.F.}},
\bauthor{\bsnm{{Landstreet}}, \binits{J.D.}},
\bauthor{\bsnm{{Mestel}}, \binits{L.}}:
\bjtitle{\araa}
\bvolume{20},
\bfpage{191}
(\byear{1982}).
doi:\doiurl{10.1146/annurev.aa.20.090182.001203}
\end{barticle}
\endbibitem

\bibitem[\protect\citeauthoryear{{Braithwaite}}{2006}]{b06}
\begin{barticle}
\bauthor{\bsnm{{Braithwaite}}, \binits{J.}}:
\bjtitle{\aap}
\bvolume{449},
\bfpage{451}
(\byear{2006}).
\arxivurl{astro-ph/0509693}.
doi:\doiurl{10.1051/0004-6361:20054241}
\end{barticle}
\endbibitem

\bibitem[\protect\citeauthoryear{{Braithwaite}}{2009}]{b09}
\begin{barticle}
\bauthor{\bsnm{{Braithwaite}}, \binits{J.}}:
\bjtitle{\mnras}
\bvolume{397},
\bfpage{763}
(\byear{2009}).
\arxivurl{0810.1049}.
doi:\doiurl{10.1111/j.1365-2966.2008.14034.x}
\end{barticle}
\endbibitem

\bibitem[\protect\citeauthoryear{{Braithwaite} and
  {Cantiello}}{2013}]{BraithwaiteCantiello2013}
\begin{barticle}
\bauthor{\bsnm{{Braithwaite}}, \binits{J.}},
\bauthor{\bsnm{{Cantiello}}, \binits{M.}}:
\bjtitle{\mnras}
\bvolume{428},
\bfpage{2789}
(\byear{2013}).
\arxivurl{1201.5646}.
doi:\doiurl{10.1093/mnras/sts109}
\end{barticle}
\endbibitem

\bibitem[\protect\citeauthoryear{{Braithwaite} and
  {Nordlund}}{2006a}]{BraithwaiteNordlund2006}
\begin{barticle}
\bauthor{\bsnm{{Braithwaite}}, \binits{J.}},
\bauthor{\bsnm{{Nordlund}}, \binits{{\AA}.}}:
\bjtitle{\aap}
\bvolume{450},
\bfpage{1077}
(\byear{2006}a).
\arxivurl{astro-ph/0510316}.
doi:\doiurl{10.1051/0004-6361:20041980}
\end{barticle}
\endbibitem

\bibitem[\protect\citeauthoryear{{Braithwaite} and {Nordlund}}{2006b}]{bn06}
\begin{barticle}
\bauthor{\bsnm{{Braithwaite}}, \binits{J.}},
\bauthor{\bsnm{{Nordlund}}, \binits{{\AA}.}}:
\bjtitle{\aap}
\bvolume{450},
\bfpage{1077}
(\byear{2006}b).
\arxivurl{astro-ph/0510316}.
doi:\doiurl{10.1051/0004-6361:20041980}
\end{barticle}
\endbibitem

\bibitem[\protect\citeauthoryear{{Braithwaite} and {Spruit}}{2004a}]{bs04}
\begin{barticle}
\bauthor{\bsnm{{Braithwaite}}, \binits{J.}},
\bauthor{\bsnm{{Spruit}}, \binits{H.C.}}:
\bjtitle{\nat}
\bvolume{431},
\bfpage{819}
(\byear{2004}a).
\arxivurl{astro-ph/0502043}.
doi:\doiurl{10.1038/nature02934}
\end{barticle}
\endbibitem

\bibitem[\protect\citeauthoryear{{Braithwaite} and
  {Spruit}}{2004b}]{BraithwaiteSpruit2004}
\begin{barticle}
\bauthor{\bsnm{{Braithwaite}}, \binits{J.}},
\bauthor{\bsnm{{Spruit}}, \binits{H.C.}}:
\bjtitle{\nat}
\bvolume{431},
\bfpage{819}
(\byear{2004}b).
\arxivurl{astro-ph/0502043}.
doi:\doiurl{10.1038/nature02934}
\end{barticle}
\endbibitem

\bibitem[\protect\citeauthoryear{Brandenburg and
  Subramanian}{2005}]{Brandenburg2005}
\begin{barticle}
\bauthor{\bsnm{Brandenburg}, \binits{A.}},
\bauthor{\bsnm{Subramanian}, \binits{K.}}:
\bjtitle{Physics Reports}
\bvolume{417}(\bissue{1-4}),
\bfpage{1}
(\byear{2005})
\end{barticle}
\endbibitem

\bibitem[\protect\citeauthoryear{Brandenburg et~al.}{2003}]{Brandenburg2003}
\begin{botherref}
\oauthor{\bsnm{Brandenburg}, \binits{A.}},
\oauthor{\bsnm{Haugen}, \binits{N.E.L.}},
\oauthor{\bsnm{Dobler}, \binits{W.}}:
eprint arXiv,
3371
(2003)
\end{botherref}
\endbibitem

\bibitem[\protect\citeauthoryear{Brandenburg et~al.}{2014}]{Brandenburg2014}
\begin{botherref}
\oauthor{\bsnm{Brandenburg}, \binits{A.}},
\oauthor{\bsnm{Kahniashvili}, \binits{T.}},
\oauthor{\bsnm{Tevzadze}, \binits{A.G.}}:
eprint arXiv:1404.2238
(2014)
\end{botherref}
\endbibitem

\bibitem[\protect\citeauthoryear{{Brethouwer} et~al.}{2007}]{betal07b}
\begin{barticle}
\bauthor{\bsnm{{Brethouwer}}, \binits{G.}},
\bauthor{\bsnm{{Billant}}, \binits{P.}},
\bauthor{\bsnm{{Lindborg}}, \binits{E.}},
\bauthor{\bsnm{{Chomaz}}, \binits{J.-M.}}:
\bjtitle{Journal of Fluid Mechanics}
\bvolume{585},
\bfpage{343}
(\byear{2007}).
doi:\doiurl{10.1017/S0022112007006854}
\end{barticle}
\endbibitem

\bibitem[\protect\citeauthoryear{{Briggs} et~al.}{2015}]{Briggs2015}
\begin{barticle}
\bauthor{\bsnm{{Briggs}}, \binits{G.P.}},
\bauthor{\bsnm{{Ferrario}}, \binits{L.}},
\bauthor{\bsnm{{Tout}}, \binits{C.A.}},
\bauthor{\bsnm{{Wickramasinghe}}, \binits{D.T.}},
\bauthor{\bsnm{{Hurley}}, \binits{J.R.}}:
\bjtitle{\mnras}
\bvolume{447},
\bfpage{1713}
(\byear{2015}).
\arxivurl{1412.5662}.
doi:\doiurl{10.1093/mnras/stu2539}
\end{barticle}
\endbibitem

\bibitem[\protect\citeauthoryear{{Broderick} and {Narayan}}{2008}]{bn08}
\begin{barticle}
\bauthor{\bsnm{{Broderick}}, \binits{A.E.}},
\bauthor{\bsnm{{Narayan}}, \binits{R.}}:
\bjtitle{\mnras}
\bvolume{383},
\bfpage{943}
(\byear{2008}).
\arxivurl{astro-ph/0702128}.
doi:\doiurl{10.1111/j.1365-2966.2007.12634.x}
\end{barticle}
\endbibitem

\bibitem[\protect\citeauthoryear{{Burrows} et~al.}{2007}]{betal07}
\begin{barticle}
\bauthor{\bsnm{{Burrows}}, \binits{A.}},
\bauthor{\bsnm{{Dessart}}, \binits{L.}},
\bauthor{\bsnm{{Livne}}, \binits{E.}},
\bauthor{\bsnm{{Ott}}, \binits{C.D.}},
\bauthor{\bsnm{{Murphy}}, \binits{J.}}:
\bjtitle{\apj}
\bvolume{664},
\bfpage{416}
(\byear{2007}).
\arxivurl{astro-ph/0702539}.
doi:\doiurl{10.1086/519161}
\end{barticle}
\endbibitem

\bibitem[\protect\citeauthoryear{{Carrier} et~al.}{2002}]{Carrier2002}
\begin{barticle}
\bauthor{\bsnm{{Carrier}}, \binits{F.}},
\bauthor{\bsnm{{North}}, \binits{P.}},
\bauthor{\bsnm{{Udry}}, \binits{S.}},
\bauthor{\bsnm{{Babel}}, \binits{J.}}:
\bjtitle{\aap}
\bvolume{394},
\bfpage{151}
(\byear{2002}).
\arxivurl{astro-ph/0208082}.
doi:\doiurl{10.1051/0004-6361:20021122}
\end{barticle}
\endbibitem

\bibitem[\protect\citeauthoryear{{Chamel}}{2013}]{c13}
\begin{barticle}
\bauthor{\bsnm{{Chamel}}, \binits{N.}}:
\bjtitle{Physical Review Letters}
\bvolume{110}(\bissue{1}),
\bfpage{011101}
(\byear{2013}).
\arxivurl{1210.8177}.
doi:\doiurl{10.1103/PhysRevLett.110.011101}
\end{barticle}
\endbibitem

\bibitem[\protect\citeauthoryear{{Charbonneau}}{2014}]{Charbonneau2014}
\begin{barticle}
\bauthor{\bsnm{{Charbonneau}}, \binits{P.}}:
\bjtitle{\araa}
\bvolume{52},
\bfpage{251}
(\byear{2014}).
doi:\doiurl{10.1146/annurev-astro-081913-040012}
\end{barticle}
\endbibitem

\bibitem[\protect\citeauthoryear{Chen et~al.}{2006}]{Chen2006}
\begin{barticle}
\bauthor{\bsnm{Chen}, \binits{S.}},
\bauthor{\bsnm{Eyink}, \binits{G.L.}},
\bauthor{\bsnm{Wan}, \binits{M.}},
\bauthor{\bsnm{Xiao}, \binits{Z.}}:
\bjtitle{Phys Rev Lett}
\bvolume{97},
\bfpage{144505}
(\byear{2006})
\end{barticle}
\endbibitem

\bibitem[\protect\citeauthoryear{Cho and Vishniac}{2000}]{Cho2000a}
\begin{barticle}
\bauthor{\bsnm{Cho}, \binits{J.}},
\bauthor{\bsnm{Vishniac}, \binits{E.T.}}:
\bjtitle{The Astrophysical Journal}
\bvolume{538},
\bfpage{217}
(\byear{2000})
\end{barticle}
\endbibitem

\bibitem[\protect\citeauthoryear{Christensson et~al.}{2001}]{Christensson2001}
\begin{barticle}
\bauthor{\bsnm{Christensson}, \binits{M.}},
\bauthor{\bsnm{Hindmarsh}, \binits{M.}},
\bauthor{\bsnm{Brandenburg}, \binits{A.}}:
\bjtitle{Physical Review E}
\bvolume{64}(\bissue{5}),
\bfpage{056405}
(\byear{2001}).
doi:\doiurl{10.1103/PhysRevE.64.056405}
\end{barticle}
\endbibitem

\bibitem[\protect\citeauthoryear{{Chung} and {Matheou}}{2012}]{cm12}
\begin{barticle}
\bauthor{\bsnm{{Chung}}, \binits{D.}},
\bauthor{\bsnm{{Matheou}}, \binits{G.}}:
\bjtitle{Journal of Fluid Mechanics}
\bvolume{696},
\bfpage{434}
(\byear{2012}).
doi:\doiurl{10.1017/jfm.2012.59}
\end{barticle}
\endbibitem

\bibitem[\protect\citeauthoryear{{Ciolfi} and {Rezzolla}}{2013}]{cr13}
\begin{barticle}
\bauthor{\bsnm{{Ciolfi}}, \binits{R.}},
\bauthor{\bsnm{{Rezzolla}}, \binits{L.}}:
\bjtitle{\mnras}
\bvolume{435},
\bfpage{43}
(\byear{2013}).
\arxivurl{1306.2803}.
doi:\doiurl{10.1093/mnrasl/slt092}
\end{barticle}
\endbibitem

\bibitem[\protect\citeauthoryear{{Ciolfi} et~al.}{2011}]{cetal11}
\begin{barticle}
\bauthor{\bsnm{{Ciolfi}}, \binits{R.}},
\bauthor{\bsnm{{Lander}}, \binits{S.K.}},
\bauthor{\bsnm{{Manca}}, \binits{G.M.}},
\bauthor{\bsnm{{Rezzolla}}, \binits{L.}}:
\bjtitle{\apjl}
\bvolume{736},
\bfpage{6}
(\byear{2011}).
\arxivurl{1105.3971}.
doi:\doiurl{10.1088/2041-8205/736/1/L6}
\end{barticle}
\endbibitem

\bibitem[\protect\citeauthoryear{{Cook} et~al.}{2003}]{css03}
\begin{barticle}
\bauthor{\bsnm{{Cook}}, \binits{J.N.}},
\bauthor{\bsnm{{Shapiro}}, \binits{S.L.}},
\bauthor{\bsnm{{Stephens}}, \binits{B.C.}}:
\bjtitle{\apj}
\bvolume{599},
\bfpage{1272}
(\byear{2003}).
\arxivurl{astro-ph/0310304}.
doi:\doiurl{10.1086/379283}
\end{barticle}
\endbibitem

\bibitem[\protect\citeauthoryear{{Cordes} and {Lazio}}{2002}]{cl02}
\begin{botherref}
\oauthor{\bsnm{{Cordes}}, \binits{J.M.}},
\oauthor{\bsnm{{Lazio}}, \binits{T.J.W.}}:
ArXiv Astrophysics e-prints
(2002).
\arxivurl{astro-ph/0207156}
\end{botherref}
\endbibitem

\bibitem[\protect\citeauthoryear{{Dall'Osso} and {Stella}}{2007}]{ds07}
\begin{barticle}
\bauthor{\bsnm{{Dall'Osso}}, \binits{S.}},
\bauthor{\bsnm{{Stella}}, \binits{L.}}:
\bjtitle{\apss}
\bvolume{308},
\bfpage{119}
(\byear{2007}).
\arxivurl{astro-ph/0702075}.
doi:\doiurl{10.1007/s10509-007-9323-0}
\end{barticle}
\endbibitem

\bibitem[\protect\citeauthoryear{{Dall'Osso} et~al.}{2009}]{dss09}
\begin{barticle}
\bauthor{\bsnm{{Dall'Osso}}, \binits{S.}},
\bauthor{\bsnm{{Shore}}, \binits{S.N.}},
\bauthor{\bsnm{{Stella}}, \binits{L.}}:
\bjtitle{\mnras}
\bvolume{398},
\bfpage{1869}
(\byear{2009}).
\arxivurl{0811.4311}.
doi:\doiurl{10.1111/j.1365-2966.2008.14054.x}
\end{barticle}
\endbibitem

\bibitem[\protect\citeauthoryear{{Donati} and
  {Landstreet}}{2009}]{DonatiLandstreet2009}
\begin{barticle}
\bauthor{\bsnm{{Donati}}, \binits{J.-F.}},
\bauthor{\bsnm{{Landstreet}}, \binits{J.D.}}:
\bjtitle{\araa}
\bvolume{47},
\bfpage{333}
(\byear{2009}).
\arxivurl{0904.1938}.
doi:\doiurl{10.1146/annurev-astro-082708-101833}
\end{barticle}
\endbibitem

\bibitem[\protect\citeauthoryear{{Duez} et~al.}{2006}]{detal06}
\begin{barticle}
\bauthor{\bsnm{{Duez}}, \binits{M.D.}},
\bauthor{\bsnm{{Liu}}, \binits{Y.T.}},
\bauthor{\bsnm{{Shapiro}}, \binits{S.L.}},
\bauthor{\bsnm{{Shibata}}, \binits{M.}},
\bauthor{\bsnm{{Stephens}}, \binits{B.C.}}:
\bjtitle{\prd}
\bvolume{73}(\bissue{10}),
\bfpage{104015}
(\byear{2006}).
\arxivurl{astro-ph/0605331}.
doi:\doiurl{10.1103/PhysRevD.73.104015}
\end{barticle}
\endbibitem

\bibitem[\protect\citeauthoryear{{Duez} and {Mathis}}{2010}]{Duez2010b}
\begin{barticle}
\bauthor{\bsnm{{Duez}}, \binits{V.}},
\bauthor{\bsnm{{Mathis}}, \binits{S.}}:
\bjtitle{\aap}
\bvolume{517},
\bfpage{58}
(\byear{2010}).
doi:\doiurl{10.1051/0004-6361/200913496}
\end{barticle}
\endbibitem

\bibitem[\protect\citeauthoryear{{Duez} et~al.}{2010a}]{Duez2010c}
\begin{barticle}
\bauthor{\bsnm{{Duez}}, \binits{V.}},
\bauthor{\bsnm{{Braithwaite}}, \binits{J.}},
\bauthor{\bsnm{{Mathis}}, \binits{S.}}:
\bjtitle{\apjl}
\bvolume{724},
\bfpage{34}
(\byear{2010}a).
\arxivurl{1009.5384}.
doi:\doiurl{10.1088/2041-8205/724/1/L34}
\end{barticle}
\endbibitem

\bibitem[\protect\citeauthoryear{{Duez} et~al.}{2010b}]{Duez2010a}
\begin{barticle}
\bauthor{\bsnm{{Duez}}, \binits{V.}},
\bauthor{\bsnm{{Mathis}}, \binits{S.}},
\bauthor{\bsnm{{Turck-Chi{\`e}ze}}, \binits{S.}}:
\bjtitle{\mnras}
\bvolume{402},
\bfpage{271}
(\byear{2010}b).
\arxivurl{0911.0788}.
doi:\doiurl{10.1111/j.1365-2966.2009.15955.x}
\end{barticle}
\endbibitem

\bibitem[\protect\citeauthoryear{{Duncan} and {Thompson}}{1992}]{dt92}
\begin{barticle}
\bauthor{\bsnm{{Duncan}}, \binits{R.C.}},
\bauthor{\bsnm{{Thompson}}, \binits{C.}}:
\bjtitle{\apjl}
\bvolume{392},
\bfpage{9}
(\byear{1992}).
doi:\doiurl{10.1086/186413}
\end{barticle}
\endbibitem

\bibitem[\protect\citeauthoryear{{Easson}}{1979a}]{e79b}
\begin{barticle}
\bauthor{\bsnm{{Easson}}, \binits{I.}}:
\bjtitle{\apj}
\bvolume{233},
\bfpage{711}
(\byear{1979}a).
doi:\doiurl{10.1086/157432}
\end{barticle}
\endbibitem

\bibitem[\protect\citeauthoryear{{Easson}}{1979b}]{e79a}
\begin{barticle}
\bauthor{\bsnm{{Easson}}, \binits{I.}}:
\bjtitle{\apj}
\bvolume{228},
\bfpage{257}
(\byear{1979}b).
doi:\doiurl{10.1086/156842}
\end{barticle}
\endbibitem

\bibitem[\protect\citeauthoryear{{Elkin} et~al.}{2010}]{Elkin2010}
\begin{barticle}
\bauthor{\bsnm{{Elkin}}, \binits{V.G.}},
\bauthor{\bsnm{{Mathys}}, \binits{G.}},
\bauthor{\bsnm{{Kurtz}}, \binits{D.W.}},
\bauthor{\bsnm{{Hubrig}}, \binits{S.}},
\bauthor{\bsnm{{Freyhammer}}, \binits{L.M.}}:
\bjtitle{\mnras}
\bvolume{402},
\bfpage{1883}
(\byear{2010}).
\arxivurl{0908.0849}.
doi:\doiurl{10.1111/j.1365-2966.2009.16015.x}
\end{barticle}
\endbibitem

\bibitem[\protect\citeauthoryear{Eyink}{2011}]{Eyink2011}
\begin{barticle}
\bauthor{\bsnm{Eyink}, \binits{G.L.}}:
\bjtitle{Physical Review E}
\bvolume{83}(\bissue{5}),
\bfpage{056405}
(\byear{2011}).
doi:\doiurl{10.1103/PhysRevE.83.056405}
\end{barticle}
\endbibitem

\bibitem[\protect\citeauthoryear{{Faucher-Gigu{\`e}re} and
  {Kaspi}}{2006}]{fk06}
\begin{barticle}
\bauthor{\bsnm{{Faucher-Gigu{\`e}re}}, \binits{C.-A.}},
\bauthor{\bsnm{{Kaspi}}, \binits{V.M.}}:
\bjtitle{ApJ}
\bvolume{643},
\bfpage{332}
(\byear{2006}).
\arxivurl{astro-ph/0512585}.
doi:\doiurl{10.1086/501516}
\end{barticle}
\endbibitem

\bibitem[\protect\citeauthoryear{{Faulkner} et~al.}{2004}]{fetal04}
\begin{barticle}
\bauthor{\bsnm{{Faulkner}}, \binits{A.J.}},
\bauthor{\bsnm{{Stairs}}, \binits{I.H.}},
\bauthor{\bsnm{{Kramer}}, \binits{M.}},
\bauthor{\bsnm{{Lyne}}, \binits{A.G.}},
\bauthor{\bsnm{{Hobbs}}, \binits{G.}},
\bauthor{\bsnm{{Possenti}}, \binits{A.}},
\bauthor{\bsnm{{Lorimer}}, \binits{D.R.}},
\bauthor{\bsnm{{Manchester}}, \binits{R.N.}},
\bauthor{\bsnm{{McLaughlin}}, \binits{M.A.}},
\bauthor{\bsnm{{D'Amico}}, \binits{N.}},
\bauthor{\bsnm{{Camilo}}, \binits{F.}},
\bauthor{\bsnm{{Burgay}}, \binits{M.}}:
\bjtitle{\mnras}
\bvolume{355},
\bfpage{147}
(\byear{2004}).
\arxivurl{astro-ph/0408228}.
doi:\doiurl{10.1111/j.1365-2966.2004.08310.x}
\end{barticle}
\endbibitem

\bibitem[\protect\citeauthoryear{Federrath et~al.}{2011}]{Federrath2011}
\begin{barticle}
\bauthor{\bsnm{Federrath}, \binits{C.}},
\bauthor{\bsnm{Chabrier}, \binits{G.}},
\bauthor{\bsnm{Schober}, \binits{J.}},
\bauthor{\bsnm{Banerjee}, \binits{R.}},
\bauthor{\bsnm{Klessen}, \binits{R.S.}},
\bauthor{\bsnm{Schleicher}, \binits{D.R.G.}}:
\bjtitle{Phys Rev Lett}
\bvolume{107},
\bfpage{114504}
(\byear{2011})
\end{barticle}
\endbibitem

\bibitem[\protect\citeauthoryear{{Ferrario}}{2012}]{Ferrario2012}
\begin{barticle}
\bauthor{\bsnm{{Ferrario}}, \binits{L.}}:
\bjtitle{\mnras}
\bvolume{426},
\bfpage{2500}
(\byear{2012}).
\arxivurl{1209.1427}.
doi:\doiurl{10.1111/j.1365-2966.2012.21836.x}
\end{barticle}
\endbibitem

\bibitem[\protect\citeauthoryear{{Ferrario} and {Wickramasinghe}}{2006a}]{fw06}
\begin{barticle}
\bauthor{\bsnm{{Ferrario}}, \binits{L.}},
\bauthor{\bsnm{{Wickramasinghe}}, \binits{D.}}:
\bjtitle{\mnras}
\bvolume{367},
\bfpage{1323}
(\byear{2006}a).
\arxivurl{astro-ph/0601258}.
doi:\doiurl{10.1111/j.1365-2966.2006.10058.x}
\end{barticle}
\endbibitem

\bibitem[\protect\citeauthoryear{{Ferrario} and
  {Wickramasinghe}}{2006b}]{Ferrario2006}
\begin{barticle}
\bauthor{\bsnm{{Ferrario}}, \binits{L.}},
\bauthor{\bsnm{{Wickramasinghe}}, \binits{D.}}:
\bjtitle{\mnras}
\bvolume{367},
\bfpage{1323}
(\byear{2006}b).
\arxivurl{astro-ph/0601258}.
doi:\doiurl{10.1111/j.1365-2966.2006.10058.x}
\end{barticle}
\endbibitem

\bibitem[\protect\citeauthoryear{{Ferrario} and {Wickramasinghe}}{2007}]{fw07}
\begin{barticle}
\bauthor{\bsnm{{Ferrario}}, \binits{L.}},
\bauthor{\bsnm{{Wickramasinghe}}, \binits{D.}}:
\bjtitle{\mnras}
\bvolume{375},
\bfpage{1009}
(\byear{2007}).
\arxivurl{astro-ph/0701444}.
doi:\doiurl{10.1111/j.1365-2966.2006.11365.x}
\end{barticle}
\endbibitem

\bibitem[\protect\citeauthoryear{{Ferrario} and {Wickramasinghe}}{2008a}]{fw08}
\begin{barticle}
\bauthor{\bsnm{{Ferrario}}, \binits{L.}},
\bauthor{\bsnm{{Wickramasinghe}}, \binits{D.}}:
\bjtitle{\mnras}
\bvolume{389},
\bfpage{66}
(\byear{2008}a).
\arxivurl{0807.2106}.
doi:\doiurl{10.1111/j.1745-3933.2008.00527.x}
\end{barticle}
\endbibitem

\bibitem[\protect\citeauthoryear{{Ferrario} and
  {Wickramasinghe}}{2008b}]{Ferrario2008}
\begin{barticle}
\bauthor{\bsnm{{Ferrario}}, \binits{L.}},
\bauthor{\bsnm{{Wickramasinghe}}, \binits{D.}}:
\bjtitle{\mnras}
\bvolume{389},
\bfpage{66}
(\byear{2008}b).
\arxivurl{0807.2106}.
doi:\doiurl{10.1111/j.1745-3933.2008.00527.x}
\end{barticle}
\endbibitem

\bibitem[\protect\citeauthoryear{{Ferrario} et~al.}{1997}]{Ferrario1997}
\begin{barticle}
\bauthor{\bsnm{{Ferrario}}, \binits{L.}},
\bauthor{\bsnm{{Vennes}}, \binits{S.}},
\bauthor{\bsnm{{Wickramasinghe}}, \binits{D.T.}},
\bauthor{\bsnm{{Bailey}}, \binits{J.A.}},
\bauthor{\bsnm{{Christian}}, \binits{D.J.}}:
\bjtitle{\mnras}
\bvolume{292},
\bfpage{205}
(\byear{1997})
\end{barticle}
\endbibitem

\bibitem[\protect\citeauthoryear{{Ferrario} et~al.}{2009}]{Ferrario2009}
\begin{barticle}
\bauthor{\bsnm{{Ferrario}}, \binits{L.}},
\bauthor{\bsnm{{Pringle}}, \binits{J.E.}},
\bauthor{\bsnm{{Tout}}, \binits{C.A.}},
\bauthor{\bsnm{{Wickramasinghe}}, \binits{D.T.}}:
\bjtitle{\mnras}
\bvolume{400},
\bfpage{71}
(\byear{2009}).
doi:\doiurl{10.1111/j.1745-3933.2009.00765.x}
\end{barticle}
\endbibitem

\bibitem[\protect\citeauthoryear{{Flowers} and {Ruderman}}{1977}]{fr77}
\begin{barticle}
\bauthor{\bsnm{{Flowers}}, \binits{E.}},
\bauthor{\bsnm{{Ruderman}}, \binits{M.A.}}:
\bjtitle{\apj}
\bvolume{215},
\bfpage{302}
(\byear{1977}).
doi:\doiurl{10.1086/155359}
\end{barticle}
\endbibitem

\bibitem[\protect\citeauthoryear{{Fossati} et~al.}{2014}]{Fossati2014}
\begin{botherref}
\oauthor{\bsnm{{Fossati}}, \binits{L.}},
\oauthor{\bsnm{{Castro}}, \binits{N.}},
\oauthor{\bsnm{{Morel}}, \binits{T.}},
\oauthor{\bsnm{{Langer}}, \binits{N.}},
\oauthor{\bsnm{{Briquet}}, \binits{M.}},
\oauthor{\bsnm{{Carroll}}, \binits{T.A.}},
\oauthor{\bsnm{{Hubrig}}, \binits{S.}},
\oauthor{\bsnm{{Nieva}}, \binits{M.F.}},
\oauthor{\bsnm{{Oskinova}}, \binits{L.M.}},
\oauthor{\bsnm{{Przybilla}}, \binits{N.}},
\oauthor{\bsnm{{Schneider}}, \binits{F.R.N.}},
\oauthor{\bsnm{{Scholler}}, \binits{M.}},
\oauthor{\bsnm{{Simon-Diaz}}, \binits{S.}},
\oauthor{\bsnm{{Ilyin}}, \binits{I.}},
\oauthor{\bsnm{{de Koter}}, \binits{A.}},
\oauthor{\bsnm{{Reisenegger}}, \binits{A.}},
\oauthor{\bsnm{{Sana}}, \binits{H.}},
\oauthor{\bsnm{{the BOB collaboration}}}:
ArXiv e-prints
(2014).
\arxivurl{1411.6490}
\end{botherref}
\endbibitem

\bibitem[\protect\citeauthoryear{{Freyhammer} et~al.}{2008}]{Freyhammer2008}
\begin{barticle}
\bauthor{\bsnm{{Freyhammer}}, \binits{L.M.}},
\bauthor{\bsnm{{Elkin}}, \binits{V.G.}},
\bauthor{\bsnm{{Kurtz}}, \binits{D.W.}},
\bauthor{\bsnm{{Mathys}}, \binits{G.}},
\bauthor{\bsnm{{Martinez}}, \binits{P.}}:
\bjtitle{\mnras}
\bvolume{389},
\bfpage{441}
(\byear{2008}).
\arxivurl{0806.2773}.
doi:\doiurl{10.1111/j.1365-2966.2008.13595.x}
\end{barticle}
\endbibitem

\bibitem[\protect\citeauthoryear{Frisch et~al.}{1975}]{Frisch1975}
\begin{barticle}
\bauthor{\bsnm{Frisch}, \binits{U.}},
\bauthor{\bsnm{Pouquet}, \binits{A.}},
\bauthor{\bsnm{Leorat}, \binits{J.}},
\bauthor{\bsnm{Mazure}, \binits{A.}}:
\bjtitle{Journal of Fluid Mechanics}
\bvolume{68},
\bfpage{769}
(\byear{1975})
\end{barticle}
\endbibitem

\bibitem[\protect\citeauthoryear{{Gaensler} et~al.}{2005}]{Gaensler2005}
\begin{barticle}
\bauthor{\bsnm{{Gaensler}}, \binits{B.M.}},
\bauthor{\bsnm{{McClure-Griffiths}}, \binits{N.M.}},
\bauthor{\bsnm{{Oey}}, \binits{M.S.}},
\bauthor{\bsnm{{Haverkorn}}, \binits{M.}},
\bauthor{\bsnm{{Dickey}}, \binits{J.M.}},
\bauthor{\bsnm{{Green}}, \binits{A.J.}}:
\bjtitle{\apjl}
\bvolume{620},
\bfpage{95}
(\byear{2005}).
\arxivurl{astro-ph/0501563}.
doi:\doiurl{10.1086/428725}
\end{barticle}
\endbibitem

\bibitem[\protect\citeauthoryear{{Garc{\'{\i}}a-Berro}
  et~al.}{2012}]{garcia2012}
\begin{barticle}
\bauthor{\bsnm{{Garc{\'{\i}}a-Berro}}, \binits{E.}},
\bauthor{\bsnm{{Lor{\'e}n-Aguilar}}, \binits{P.}},
\bauthor{\bsnm{{Aznar-Sigu{\'a}n}}, \binits{G.}},
\bauthor{\bsnm{{Torres}}, \binits{S.}},
\bauthor{\bsnm{{Camacho}}, \binits{J.}},
\bauthor{\bsnm{{Althaus}}, \binits{L.G.}},
\bauthor{\bsnm{{C{\'o}rsico}}, \binits{A.H.}},
\bauthor{\bsnm{{K{\"u}lebi}}, \binits{B.}},
\bauthor{\bsnm{{Isern}}, \binits{J.}}:
\bjtitle{\apj}
\bvolume{749},
\bfpage{25}
(\byear{2012}).
\arxivurl{1202.0461}.
doi:\doiurl{10.1088/0004-637X/749/1/25}
\end{barticle}
\endbibitem

\bibitem[\protect\citeauthoryear{{Geppert} et~al.}{2012}]{getal12}
\begin{bchapter}
\bauthor{\bsnm{{Geppert}}, \binits{U.}},
\bauthor{\bsnm{{Gil}}, \binits{J.}},
\bauthor{\bsnm{{Melikidze}}, \binits{G.}},
\bauthor{\bsnm{{Pons}}, \binits{J.}},
\bauthor{\bsnm{{Vigan{\`o}}}, \binits{D.}}:
In: \beditor{\bsnm{{Lewandowski}}, \binits{W.}},
\beditor{\bsnm{{Maron}}, \binits{O.}},
\beditor{\bsnm{{Kijak}}, \binits{J.}} (eds.)
\bbtitle{Electromagnetic Radiation from Pulsars and Magnetars}.
\bsertitle{Astronomical Society of the Pacific Conference Series},
vol. \bseriesno{466},
p. \bfpage{187}
(\byear{2012}).
\arxivurl{1206.1790}
\end{bchapter}
\endbibitem

\bibitem[\protect\citeauthoryear{{Glaberson} et~al.}{1974}]{gjo74}
\begin{barticle}
\bauthor{\bsnm{{Glaberson}}, \binits{W.I.}},
\bauthor{\bsnm{{Johnson}}, \binits{W.W.}},
\bauthor{\bsnm{{Ostermeier}}, \binits{R.M.}}:
\bjtitle{Physical Review Letters}
\bvolume{33},
\bfpage{1197}
(\byear{1974}).
doi:\doiurl{10.1103/PhysRevLett.33.1197}
\end{barticle}
\endbibitem

\bibitem[\protect\citeauthoryear{{Glampedakis} et~al.}{2009}]{gaj09}
\begin{barticle}
\bauthor{\bsnm{{Glampedakis}}, \binits{K.}},
\bauthor{\bsnm{{Andersson}}, \binits{N.}},
\bauthor{\bsnm{{Jones}}, \binits{D.I.}}:
\bjtitle{\mnras}
\bvolume{394},
\bfpage{1908}
(\byear{2009}).
\arxivurl{0801.4638}.
doi:\doiurl{10.1111/j.1365-2966.2008.13995.x}
\end{barticle}
\endbibitem

\bibitem[\protect\citeauthoryear{{Glampedakis} et~al.}{2012}]{gal12}
\begin{barticle}
\bauthor{\bsnm{{Glampedakis}}, \binits{K.}},
\bauthor{\bsnm{{Andersson}}, \binits{N.}},
\bauthor{\bsnm{{Lander}}, \binits{S.K.}}:
\bjtitle{\mnras}
\bvolume{420},
\bfpage{1263}
(\byear{2012}).
\arxivurl{1106.6330}.
doi:\doiurl{10.1111/j.1365-2966.2011.20112.x}
\end{barticle}
\endbibitem

\bibitem[\protect\citeauthoryear{{Goedbloed} and {Poedts}}{2004}]{gp04}
\begin{bbook}
\bauthor{\bsnm{{Goedbloed}}, \binits{J.P.H.}},
\bauthor{\bsnm{{Poedts}}, \binits{S.}}:
\bbtitle{{Principles of Magnetohydrodynamics}},
(\byear{2004})
\end{bbook}
\endbibitem

\bibitem[\protect\citeauthoryear{{Gourgouliatos} et~al.}{2013}]{getal13}
\begin{barticle}
\bauthor{\bsnm{{Gourgouliatos}}, \binits{K.N.}},
\bauthor{\bsnm{{Cumming}}, \binits{A.}},
\bauthor{\bsnm{{Reisenegger}}, \binits{A.}},
\bauthor{\bsnm{{Armaza}}, \binits{C.}},
\bauthor{\bsnm{{Lyutikov}}, \binits{M.}},
\bauthor{\bsnm{{Valdivia}}, \binits{J.A.}}:
\bjtitle{\mnras}
\bvolume{434},
\bfpage{2480}
(\byear{2013}).
\arxivurl{1305.6269}.
doi:\doiurl{10.1093/mnras/stt1195}
\end{barticle}
\endbibitem

\bibitem[\protect\citeauthoryear{{Grunhut} and {Wade}}{2013}]{Grunhut2013}
\begin{bchapter}
\bauthor{\bsnm{{Grunhut}}, \binits{J.H.}},
\bauthor{\bsnm{{Wade}}, \binits{G.A.}}:
In: \bbtitle{EAS Publications Series}.
\bsertitle{EAS Publications Series},
vol. \bseriesno{64},
p. \bfpage{67}
(\byear{2013}).
doi:\doiurl{10.1051/eas/1364009}
\end{bchapter}
\endbibitem

\bibitem[\protect\citeauthoryear{{Gunn} and {Ostriker}}{1970}]{go70}
\begin{barticle}
\bauthor{\bsnm{{Gunn}}, \binits{J.E.}},
\bauthor{\bsnm{{Ostriker}}, \binits{J.P.}}:
\bjtitle{ApJ}
\bvolume{160},
\bfpage{979}
(\byear{1970}).
doi:\doiurl{10.1086/150487}
\end{barticle}
\endbibitem

\bibitem[\protect\citeauthoryear{{Hale}}{1908}]{Hale1908}
\begin{barticle}
\bauthor{\bsnm{{Hale}}, \binits{G.E.}}:
\bjtitle{\apj}
\bvolume{28},
\bfpage{315}
(\byear{1908}).
doi:\doiurl{10.1086/141602}
\end{barticle}
\endbibitem

\bibitem[\protect\citeauthoryear{{Hale} et~al.}{1919}]{Hale1919}
\begin{barticle}
\bauthor{\bsnm{{Hale}}, \binits{G.E.}},
\bauthor{\bsnm{{Ellerman}}, \binits{F.}},
\bauthor{\bsnm{{Nicholson}}, \binits{S.B.}},
\bauthor{\bsnm{{Joy}}, \binits{A.H.}}:
\bjtitle{\apj}
\bvolume{49},
\bfpage{153}
(\byear{1919}).
doi:\doiurl{10.1086/142452}
\end{barticle}
\endbibitem

\bibitem[\protect\citeauthoryear{{Hartman} et~al.}{1997}]{hetal97}
\begin{barticle}
\bauthor{\bsnm{{Hartman}}, \binits{J.W.}},
\bauthor{\bsnm{{Bhattacharya}}, \binits{D.}},
\bauthor{\bsnm{{Wijers}}, \binits{R.}},
\bauthor{\bsnm{{Verbunt}}, \binits{F.}}:
\bjtitle{A\&A}
\bvolume{322},
\bfpage{477}
(\byear{1997})
\end{barticle}
\endbibitem

\bibitem[\protect\citeauthoryear{{Haskell} et~al.}{2009}]{hap09}
\begin{barticle}
\bauthor{\bsnm{{Haskell}}, \binits{B.}},
\bauthor{\bsnm{{Andersson}}, \binits{N.}},
\bauthor{\bsnm{{Passamonti}}, \binits{A.}}:
\bjtitle{\mnras}
\bvolume{397},
\bfpage{1464}
(\byear{2009}).
\arxivurl{0902.1149}.
doi:\doiurl{10.1111/j.1365-2966.2009.14963.x}
\end{barticle}
\endbibitem

\bibitem[\protect\citeauthoryear{{Haskell} et~al.}{2012}]{hps12}
\begin{barticle}
\bauthor{\bsnm{{Haskell}}, \binits{B.}},
\bauthor{\bsnm{{Pizzochero}}, \binits{P.M.}},
\bauthor{\bsnm{{Sidery}}, \binits{T.}}:
\bjtitle{\mnras}
\bvolume{420},
\bfpage{658}
(\byear{2012}).
\arxivurl{1107.5295}.
doi:\doiurl{10.1111/j.1365-2966.2011.20080.x}
\end{barticle}
\endbibitem

\bibitem[\protect\citeauthoryear{Haugen et~al.}{2004a}]{Haugen2004}
\begin{barticle}
\bauthor{\bsnm{Haugen}, \binits{N.E.L.}},
\bauthor{\bsnm{Brandenburg}, \binits{A.}},
\bauthor{\bsnm{Dobler}, \binits{W.}}:
\bjtitle{Astrophysics and Space Science}
\bvolume{292},
\bfpage{53}
(\byear{2004}a)
\end{barticle}
\endbibitem

\bibitem[\protect\citeauthoryear{Haugen et~al.}{2004b}]{Haugen2004c}
\begin{barticle}
\bauthor{\bsnm{Haugen}, \binits{N.E.}},
\bauthor{\bsnm{Brandenburg}, \binits{A.}},
\bauthor{\bsnm{Dobler}, \binits{W.}}:
\bjtitle{Physical Review E}
\bvolume{70},
\bfpage{16308}
(\byear{2004}b)
\end{barticle}
\endbibitem

\bibitem[\protect\citeauthoryear{Haugen et~al.}{2004c}]{Haugen2004b}
\begin{barticle}
\bauthor{\bsnm{Haugen}, \binits{N.E.L.}},
\bauthor{\bsnm{Brandenburg}, \binits{A.}},
\bauthor{\bsnm{Mee}, \binits{A.J.}}:
\bjtitle{Monthly Notices of the Royal Astronomical Society}
\bvolume{353},
\bfpage{947}
(\byear{2004}c)
\end{barticle}
\endbibitem

\bibitem[\protect\citeauthoryear{{Heiles}}{1997}]{Heiles1997}
\begin{barticle}
\bauthor{\bsnm{{Heiles}}, \binits{C.}}:
\bjtitle{\apjs}
\bvolume{111},
\bfpage{245}
(\byear{1997}).
doi:\doiurl{10.1086/313010}
\end{barticle}
\endbibitem

\bibitem[\protect\citeauthoryear{{Herbig}}{1960}]{Herbig1960}
\begin{barticle}
\bauthor{\bsnm{{Herbig}}, \binits{G.H.}}:
\bjtitle{\apjs}
\bvolume{4},
\bfpage{337}
(\byear{1960}).
doi:\doiurl{10.1086/190050}
\end{barticle}
\endbibitem

\bibitem[\protect\citeauthoryear{{Hewish} et~al.}{1968}]{Hewish1968}
\begin{barticle}
\bauthor{\bsnm{{Hewish}}, \binits{A.}},
\bauthor{\bsnm{{Bell}}, \binits{S.J.}},
\bauthor{\bsnm{{Pilkington}}, \binits{J.D.H.}},
\bauthor{\bsnm{{Scott}}, \binits{P.F.}},
\bauthor{\bsnm{{Collins}}, \binits{R.A.}}:
\bjtitle{\nat}
\bvolume{217},
\bfpage{709}
(\byear{1968}).
doi:\doiurl{10.1038/217709a0}
\end{barticle}
\endbibitem

\bibitem[\protect\citeauthoryear{{Hobbs} et~al.}{2004}]{hetal04}
\begin{barticle}
\bauthor{\bsnm{{Hobbs}}, \binits{G.}},
\bauthor{\bsnm{{Faulkner}}, \binits{A.}},
\bauthor{\bsnm{{Stairs}}, \binits{I.H.}},
\bauthor{\bsnm{{Camilo}}, \binits{F.}},
\bauthor{\bsnm{{Manchester}}, \binits{R.N.}},
\bauthor{\bsnm{{Lyne}}, \binits{A.G.}},
\bauthor{\bsnm{{Kramer}}, \binits{M.}},
\bauthor{\bsnm{{D'Amico}}, \binits{N.}},
\bauthor{\bsnm{{Kaspi}}, \binits{V.M.}},
\bauthor{\bsnm{{Possenti}}, \binits{A.}},
\bauthor{\bsnm{{McLaughlin}}, \binits{M.A.}},
\bauthor{\bsnm{{Lorimer}}, \binits{D.R.}},
\bauthor{\bsnm{{Burgay}}, \binits{M.}},
\bauthor{\bsnm{{Joshi}}, \binits{B.C.}},
\bauthor{\bsnm{{Crawford}}, \binits{F.}}:
\bjtitle{\mnras}
\bvolume{352},
\bfpage{1439}
(\byear{2004}).
\arxivurl{astro-ph/0405364}.
doi:\doiurl{10.1111/j.1365-2966.2004.08042.x}
\end{barticle}
\endbibitem

\bibitem[\protect\citeauthoryear{{Hubrig} et~al.}{2005}]{Hubrig2005}
\begin{barticle}
\bauthor{\bsnm{{Hubrig}}, \binits{S.}},
\bauthor{\bsnm{{Nesvacil}}, \binits{N.}},
\bauthor{\bsnm{{Sch{\"o}ller}}, \binits{M.}},
\bauthor{\bsnm{{North}}, \binits{P.}},
\bauthor{\bsnm{{Mathys}}, \binits{G.}},
\bauthor{\bsnm{{Kurtz}}, \binits{D.W.}},
\bauthor{\bsnm{{Wolff}}, \binits{B.}},
\bauthor{\bsnm{{Szeifert}}, \binits{T.}},
\bauthor{\bsnm{{Cunha}}, \binits{M.S.}},
\bauthor{\bsnm{{Elkin}}, \binits{V.G.}}:
\bjtitle{\aap}
\bvolume{440},
\bfpage{37}
(\byear{2005}).
doi:\doiurl{10.1051/0004-6361:200500164}
\end{barticle}
\endbibitem

\bibitem[\protect\citeauthoryear{{Hubrig} et~al.}{2009}]{Hubrig2009}
\begin{barticle}
\bauthor{\bsnm{{Hubrig}}, \binits{S.}},
\bauthor{\bsnm{{Stelzer}}, \binits{B.}},
\bauthor{\bsnm{{Sch{\"o}ller}}, \binits{M.}},
\bauthor{\bsnm{{Grady}}, \binits{C.}},
\bauthor{\bsnm{{Sch{\"u}tz}}, \binits{O.}},
\bauthor{\bsnm{{Pogodin}}, \binits{M.A.}},
\bauthor{\bsnm{{Cur{\'e}}}, \binits{M.}},
\bauthor{\bsnm{{Hamaguchi}}, \binits{K.}},
\bauthor{\bsnm{{Yudin}}, \binits{R.V.}}:
\bjtitle{\aap}
\bvolume{502},
\bfpage{283}
(\byear{2009}).
\arxivurl{0905.0864}.
doi:\doiurl{10.1051/0004-6361/200811533}
\end{barticle}
\endbibitem

\bibitem[\protect\citeauthoryear{{Hubrig} et~al.}{2011}]{Hubrig2011}
\begin{barticle}
\bauthor{\bsnm{{Hubrig}}, \binits{S.}},
\bauthor{\bsnm{{Sch{\"o}ller}}, \binits{M.}},
\bauthor{\bsnm{{Kharchenko}}, \binits{N.V.}},
\bauthor{\bsnm{{Langer}}, \binits{N.}},
\bauthor{\bsnm{{de Wit}}, \binits{W.J.}},
\bauthor{\bsnm{{Ilyin}}, \binits{I.}},
\bauthor{\bsnm{{Kholtygin}}, \binits{A.F.}},
\bauthor{\bsnm{{Piskunov}}, \binits{A.E.}},
\bauthor{\bsnm{{Przybilla}}, \binits{N.}},
\bauthor{\bsnm{{Magori Collaboration}}}:
\bjtitle{\aap}
\bvolume{528},
\bfpage{151}
(\byear{2011}).
\arxivurl{1102.2503}.
doi:\doiurl{10.1051/0004-6361/201016345}
\end{barticle}
\endbibitem

\bibitem[\protect\citeauthoryear{{Hubrig} et~al.}{2013}]{HubrigIlyin2013}
\begin{barticle}
\bauthor{\bsnm{{Hubrig}}, \binits{S.}},
\bauthor{\bsnm{{Ilyin}}, \binits{I.}},
\bauthor{\bsnm{{Sch{\"o}ller}}, \binits{M.}},
\bauthor{\bsnm{{Lo Curto}}, \binits{G.}}:
\bjtitle{Astronomische Nachrichten}
\bvolume{334},
\bfpage{1093}
(\byear{2013}).
\arxivurl{1307.0133}.
doi:\doiurl{10.1002/asna.201311948}
\end{barticle}
\endbibitem

\bibitem[\protect\citeauthoryear{{Hubrig} et~al.}{2014}]{Hubrig2014}
\begin{barticle}
\bauthor{\bsnm{{Hubrig}}, \binits{S.}},
\bauthor{\bsnm{{Fossati}}, \binits{L.}},
\bauthor{\bsnm{{Carroll}}, \binits{T.A.}},
\bauthor{\bsnm{{Castro}}, \binits{N.}},
\bauthor{\bsnm{{Gonz{\'a}lez}}, \binits{J.F.}},
\bauthor{\bsnm{{Ilyin}}, \binits{I.}},
\bauthor{\bsnm{{Przybilla}}, \binits{N.}},
\bauthor{\bsnm{{Sch{\"o}ller}}, \binits{M.}},
\bauthor{\bsnm{{Oskinova}}, \binits{L.M.}},
\bauthor{\bsnm{{Morel}}, \binits{T.}},
\bauthor{\bsnm{{Langer}}, \binits{N.}},
\bauthor{\bsnm{{Scholz}}, \binits{R.D.}},
\bauthor{\bsnm{{Kharchenko}}, \binits{N.V.}},
\bauthor{\bsnm{{Nieva}}, \binits{M.-F.}}:
\bjtitle{\aap}
\bvolume{564},
\bfpage{10}
(\byear{2014}).
\arxivurl{1403.0491}.
doi:\doiurl{10.1051/0004-6361/201423490}
\end{barticle}
\endbibitem

\bibitem[\protect\citeauthoryear{{Iida} et~al.}{2009}]{itn09}
\begin{barticle}
\bauthor{\bsnm{{Iida}}, \binits{O.}},
\bauthor{\bsnm{{Tsuzuki}}, \binits{N.}},
\bauthor{\bsnm{{Nagano}}, \binits{Y.}}:
\bjtitle{Theoretical and Computational Fluid Dynamics}
\bvolume{23},
\bfpage{109}
(\byear{2009}).
doi:\doiurl{10.1007/s00162-009-0101-1}
\end{barticle}
\endbibitem

\bibitem[\protect\citeauthoryear{{Ilkov} and {Soker}}{2013}]{Ilkov2013}
\begin{barticle}
\bauthor{\bsnm{{Ilkov}}, \binits{M.}},
\bauthor{\bsnm{{Soker}}, \binits{N.}}:
\bjtitle{\mnras}
\bvolume{428},
\bfpage{579}
(\byear{2013}).
\arxivurl{1208.0953}.
doi:\doiurl{10.1093/mnras/sts053}
\end{barticle}
\endbibitem

\bibitem[\protect\citeauthoryear{{Jahan-Miri}}{2010}]{jm10}
\begin{barticle}
\bauthor{\bsnm{{Jahan-Miri}}, \binits{M.}}:
\bjtitle{\apj}
\bvolume{725},
\bfpage{29}
(\byear{2010}).
\arxivurl{1009.5070}.
doi:\doiurl{10.1088/0004-637X/725/1/29}
\end{barticle}
\endbibitem

\bibitem[\protect\citeauthoryear{{Jordan} et~al.}{2007}]{Jordan2007}
\begin{barticle}
\bauthor{\bsnm{{Jordan}}, \binits{S.}},
\bauthor{\bsnm{{Aznar Cuadrado}}, \binits{R.}},
\bauthor{\bsnm{{Napiwotzki}}, \binits{R.}},
\bauthor{\bsnm{{Schmid}}, \binits{H.M.}},
\bauthor{\bsnm{{Solanki}}, \binits{S.K.}}:
\bjtitle{\aap}
\bvolume{462},
\bfpage{1097}
(\byear{2007}).
\arxivurl{astro-ph/0610875}.
doi:\doiurl{10.1051/0004-6361:20066163}
\end{barticle}
\endbibitem

\bibitem[\protect\citeauthoryear{Kahniashvili et~al.}{2010}]{Kahniashvili2010}
\begin{barticle}
\bauthor{\bsnm{Kahniashvili}, \binits{T.}},
\bauthor{\bsnm{Brandenburg}, \binits{A.}},
\bauthor{\bsnm{Tevzadze}, \binits{A.G.}},
\bauthor{\bsnm{Ratra}, \binits{B.}}:
\bjtitle{eprint arXiv}
\bvolume{1004},
\bfpage{3084}
(\byear{2010})
\end{barticle}
\endbibitem

\bibitem[\protect\citeauthoryear{Kalelkar and Pandit}{2004}]{Kalelkar2004}
\begin{barticle}
\bauthor{\bsnm{Kalelkar}, \binits{C.}},
\bauthor{\bsnm{Pandit}, \binits{R.}}:
\bjtitle{Physical Review E}
\bvolume{69}(\bissue{4}),
\bfpage{046304}
(\byear{2004}).
doi:\doiurl{10.1103/PhysRevE.69.046304}
\end{barticle}
\endbibitem

\bibitem[\protect\citeauthoryear{{Kawka} and {Vennes}}{2012}]{Kawka2012}
\begin{barticle}
\bauthor{\bsnm{{Kawka}}, \binits{A.}},
\bauthor{\bsnm{{Vennes}}, \binits{S.}}:
\bjtitle{\mnras}
\bvolume{425},
\bfpage{1394}
(\byear{2012}).
\arxivurl{1206.5113}.
doi:\doiurl{10.1111/j.1365-2966.2012.21574.x}
\end{barticle}
\endbibitem

\bibitem[\protect\citeauthoryear{{Kawka} et~al.}{2007}]{Kawka2007}
\begin{barticle}
\bauthor{\bsnm{{Kawka}}, \binits{A.}},
\bauthor{\bsnm{{Vennes}}, \binits{S.}},
\bauthor{\bsnm{{Schmidt}}, \binits{G.D.}},
\bauthor{\bsnm{{Wickramasinghe}}, \binits{D.T.}},
\bauthor{\bsnm{{Koch}}, \binits{R.}}:
\bjtitle{\apj}
\bvolume{654},
\bfpage{499}
(\byear{2007}).
\arxivurl{astro-ph/0609273}.
doi:\doiurl{10.1086/509072}
\end{barticle}
\endbibitem

\bibitem[\protect\citeauthoryear{Kazantsev}{1968}]{Kazantsev1968}
\begin{barticle}
\bauthor{\bsnm{Kazantsev}, \binits{A.P.}}:
\bjtitle{Soviet Physics JETP}
\bvolume{26},
\bfpage{1031}
(\byear{1968})
\end{barticle}
\endbibitem

\bibitem[\protect\citeauthoryear{{Keane} and {Kramer}}{2008}]{kk08}
\begin{barticle}
\bauthor{\bsnm{{Keane}}, \binits{E.F.}},
\bauthor{\bsnm{{Kramer}}, \binits{M.}}:
\bjtitle{\mnras}
\bvolume{391},
\bfpage{2009}
(\byear{2008}).
\arxivurl{0810.1512}.
doi:\doiurl{10.1111/j.1365-2966.2008.14045.x}
\end{barticle}
\endbibitem

\bibitem[\protect\citeauthoryear{{Kemp}}{1970}]{Kemp1970}
\begin{barticle}
\bauthor{\bsnm{{Kemp}}, \binits{J.C.}}:
\bjtitle{\apj}
\bvolume{162},
\bfpage{169}
(\byear{1970}).
doi:\doiurl{10.1086/150643}
\end{barticle}
\endbibitem

\bibitem[\protect\citeauthoryear{{Kemp} et~al.}{1970}]{Kempetal1970}
\begin{barticle}
\bauthor{\bsnm{{Kemp}}, \binits{J.C.}},
\bauthor{\bsnm{{Swedlund}}, \binits{J.B.}},
\bauthor{\bsnm{{Landstreet}}, \binits{J.D.}},
\bauthor{\bsnm{{Angel}}, \binits{J.R.P.}}:
\bjtitle{\apjl}
\bvolume{161},
\bfpage{77}
(\byear{1970}).
doi:\doiurl{10.1086/180574}
\end{barticle}
\endbibitem

\bibitem[\protect\citeauthoryear{{Kiel} et~al.}{2008}]{kbm08}
\begin{barticle}
\bauthor{\bsnm{{Kiel}}, \binits{P.D.}},
\bauthor{\bsnm{{Hurley}}, \binits{J.R.}},
\bauthor{\bsnm{{Bailes}}, \binits{M.}},
\bauthor{\bsnm{{Murray}}, \binits{J.R.}}:
\bjtitle{MNRAS}
\bvolume{388},
\bfpage{393}
(\byear{2008}).
\arxivurl{0805.0059}.
doi:\doiurl{10.1111/j.1365-2966.2008.13402.x}
\end{barticle}
\endbibitem

\bibitem[\protect\citeauthoryear{{Kochukhov}}{2006}]{Kochukhov2006}
\begin{barticle}
\bauthor{\bsnm{{Kochukhov}}, \binits{O.}}:
\bjtitle{\aap}
\bvolume{454},
\bfpage{321}
(\byear{2006}).
\arxivurl{astro-ph/0603831}.
doi:\doiurl{10.1051/0004-6361:20064932}
\end{barticle}
\endbibitem

\bibitem[\protect\citeauthoryear{{Kouveliotou} et~al.}{1998}]{ketal98}
\begin{barticle}
\bauthor{\bsnm{{Kouveliotou}}, \binits{C.}},
\bauthor{\bsnm{{Dieters}}, \binits{S.}},
\bauthor{\bsnm{{Strohmayer}}, \binits{T.}},
\bauthor{\bsnm{{van Paradijs}}, \binits{J.}},
\bauthor{\bsnm{{Fishman}}, \binits{G.J.}},
\bauthor{\bsnm{{Meegan}}, \binits{C.A.}},
\bauthor{\bsnm{{Hurley}}, \binits{K.}},
\bauthor{\bsnm{{Kommers}}, \binits{J.}},
\bauthor{\bsnm{{Smith}}, \binits{I.}},
\bauthor{\bsnm{{Frail}}, \binits{D.}},
\bauthor{\bsnm{{Murakami}}, \binits{T.}}:
\bjtitle{\nat}
\bvolume{393},
\bfpage{235}
(\byear{1998}).
doi:\doiurl{10.1038/30410}
\end{barticle}
\endbibitem

\bibitem[\protect\citeauthoryear{{Kramer} et~al.}{2003}]{ketal03}
\begin{barticle}
\bauthor{\bsnm{{Kramer}}, \binits{M.}},
\bauthor{\bsnm{{Bell}}, \binits{J.F.}},
\bauthor{\bsnm{{Manchester}}, \binits{R.N.}},
\bauthor{\bsnm{{Lyne}}, \binits{A.G.}},
\bauthor{\bsnm{{Camilo}}, \binits{F.}},
\bauthor{\bsnm{{Stairs}}, \binits{I.H.}},
\bauthor{\bsnm{{D'Amico}}, \binits{N.}},
\bauthor{\bsnm{{Kaspi}}, \binits{V.M.}},
\bauthor{\bsnm{{Hobbs}}, \binits{G.}},
\bauthor{\bsnm{{Morris}}, \binits{D.J.}},
\bauthor{\bsnm{{Crawford}}, \binits{F.}},
\bauthor{\bsnm{{Possenti}}, \binits{A.}},
\bauthor{\bsnm{{Joshi}}, \binits{B.C.}},
\bauthor{\bsnm{{McLaughlin}}, \binits{M.A.}},
\bauthor{\bsnm{{Lorimer}}, \binits{D.R.}},
\bauthor{\bsnm{{Faulkner}}, \binits{A.J.}}:
\bjtitle{\mnras}
\bvolume{342},
\bfpage{1299}
(\byear{2003}).
\arxivurl{astro-ph/0303473}.
doi:\doiurl{10.1046/j.1365-8711.2003.06637.x}
\end{barticle}
\endbibitem

\bibitem[\protect\citeauthoryear{Kulsrud and Zweibel}{2008}]{Kulsrud2008}
\begin{barticle}
\bauthor{\bsnm{Kulsrud}, \binits{R.M.}},
\bauthor{\bsnm{Zweibel}, \binits{E.G.}}:
\bjtitle{Reports on Progress in Physics}
\bvolume{71},
\bfpage{6901}
(\byear{2008})
\end{barticle}
\endbibitem

\bibitem[\protect\citeauthoryear{{Lander}}{2013}]{l13}
\begin{barticle}
\bauthor{\bsnm{{Lander}}, \binits{S.K.}}:
\bjtitle{Physical Review Letters}
\bvolume{110}(\bissue{7}),
\bfpage{071101}
(\byear{2013}).
\arxivurl{1211.3912}.
doi:\doiurl{10.1103/PhysRevLett.110.071101}
\end{barticle}
\endbibitem

\bibitem[\protect\citeauthoryear{{Lander}}{2014}]{l14}
\begin{barticle}
\bauthor{\bsnm{{Lander}}, \binits{S.K.}}:
\bjtitle{\mnras}
\bvolume{437},
\bfpage{424}
(\byear{2014}).
\arxivurl{1307.7020}.
doi:\doiurl{10.1093/mnras/stt1894}
\end{barticle}
\endbibitem

\bibitem[\protect\citeauthoryear{{Lander} and {Jones}}{2012}]{lj12}
\begin{barticle}
\bauthor{\bsnm{{Lander}}, \binits{S.K.}},
\bauthor{\bsnm{{Jones}}, \binits{D.I.}}:
\bjtitle{\mnras}
\bvolume{424},
\bfpage{482}
(\byear{2012}).
\arxivurl{1202.2339}.
doi:\doiurl{10.1111/j.1365-2966.2012.21213.x}
\end{barticle}
\endbibitem

\bibitem[\protect\citeauthoryear{{Lander} et~al.}{2012}]{lag12}
\begin{barticle}
\bauthor{\bsnm{{Lander}}, \binits{S.K.}},
\bauthor{\bsnm{{Andersson}}, \binits{N.}},
\bauthor{\bsnm{{Glampedakis}}, \binits{K.}}:
\bjtitle{\mnras}
\bvolume{419},
\bfpage{732}
(\byear{2012}).
\arxivurl{1106.6322}.
doi:\doiurl{10.1111/j.1365-2966.2011.19720.x}
\end{barticle}
\endbibitem

\bibitem[\protect\citeauthoryear{{Landstreet} et~al.}{2007a}]{letal07}
\begin{barticle}
\bauthor{\bsnm{{Landstreet}}, \binits{J.D.}},
\bauthor{\bsnm{{Bagnulo}}, \binits{S.}},
\bauthor{\bsnm{{Andretta}}, \binits{V.}},
\bauthor{\bsnm{{Fossati}}, \binits{L.}},
\bauthor{\bsnm{{Mason}}, \binits{E.}},
\bauthor{\bsnm{{Silaj}}, \binits{J.}},
\bauthor{\bsnm{{Wade}}, \binits{G.A.}}:
\bjtitle{\aap}
\bvolume{470},
\bfpage{685}
(\byear{2007}a).
\arxivurl{0706.0330}.
doi:\doiurl{10.1051/0004-6361:20077343}
\end{barticle}
\endbibitem

\bibitem[\protect\citeauthoryear{{Landstreet} et~al.}{2007b}]{Landstreet2007}
\begin{barticle}
\bauthor{\bsnm{{Landstreet}}, \binits{J.D.}},
\bauthor{\bsnm{{Bagnulo}}, \binits{S.}},
\bauthor{\bsnm{{Andretta}}, \binits{V.}},
\bauthor{\bsnm{{Fossati}}, \binits{L.}},
\bauthor{\bsnm{{Mason}}, \binits{E.}},
\bauthor{\bsnm{{Silaj}}, \binits{J.}},
\bauthor{\bsnm{{Wade}}, \binits{G.A.}}:
\bjtitle{\aap}
\bvolume{470},
\bfpage{685}
(\byear{2007}b).
\arxivurl{0706.0330}.
doi:\doiurl{10.1051/0004-6361:20077343}
\end{barticle}
\endbibitem

\bibitem[\protect\citeauthoryear{{Landstreet} et~al.}{2008}]{Landstreet2008}
\begin{barticle}
\bauthor{\bsnm{{Landstreet}}, \binits{J.D.}},
\bauthor{\bsnm{{Bagnulo}}, \binits{S.}},
\bauthor{\bsnm{{Andretta}}, \binits{V.}},
\bauthor{\bsnm{{Fossati}}, \binits{L.}},
\bauthor{\bsnm{{Mason}}, \binits{E.}},
\bauthor{\bsnm{{Silaj}}, \binits{J.}},
\bauthor{\bsnm{{Wade}}, \binits{G.A.}}:
\bjtitle{Contributions of the Astronomical Observatory Skalnate Pleso}
\bvolume{38},
\bfpage{391}
(\byear{2008})
\end{barticle}
\endbibitem

\bibitem[\protect\citeauthoryear{{Landstreet} et~al.}{2012}]{Landstreet2012}
\begin{barticle}
\bauthor{\bsnm{{Landstreet}}, \binits{J.D.}},
\bauthor{\bsnm{{Bagnulo}}, \binits{S.}},
\bauthor{\bsnm{{Valyavin}}, \binits{G.G.}},
\bauthor{\bsnm{{Fossati}}, \binits{L.}},
\bauthor{\bsnm{{Jordan}}, \binits{S.}},
\bauthor{\bsnm{{Monin}}, \binits{D.}},
\bauthor{\bsnm{{Wade}}, \binits{G.A.}}:
\bjtitle{\aap}
\bvolume{545},
\bfpage{30}
(\byear{2012}).
\arxivurl{1208.3650}.
doi:\doiurl{10.1051/0004-6361/201219829}
\end{barticle}
\endbibitem

\bibitem[\protect\citeauthoryear{{Lasky} et~al.}{2013}]{lbm13}
\begin{barticle}
\bauthor{\bsnm{{Lasky}}, \binits{P.D.}},
\bauthor{\bsnm{{Bennett}}, \binits{M.F.}},
\bauthor{\bsnm{{Melatos}}, \binits{A.}}:
\bjtitle{\prd}
\bvolume{87}(\bissue{6}),
\bfpage{063004}
(\byear{2013}).
\arxivurl{1302.6033}.
doi:\doiurl{10.1103/PhysRevD.87.063004}
\end{barticle}
\endbibitem

\bibitem[\protect\citeauthoryear{Lazarian and Cho}{2005}]{Lazarian2005}
\begin{barticle}
\bauthor{\bsnm{Lazarian}, \binits{A.}},
\bauthor{\bsnm{Cho}, \binits{J.}}:
\bjtitle{Physica Scripta}
\bvolume{2005},
\bfpage{32}
(\byear{2005}).
doi:\doiurl{10.1238/Physica.Topical.116a00032}
\end{barticle}
\endbibitem

\bibitem[\protect\citeauthoryear{{Liebert} et~al.}{2003}]{Liebert2003}
\begin{barticle}
\bauthor{\bsnm{{Liebert}}, \binits{J.}},
\bauthor{\bsnm{{Bergeron}}, \binits{P.}},
\bauthor{\bsnm{{Holberg}}, \binits{J.B.}}:
\bjtitle{\aj}
\bvolume{125},
\bfpage{348}
(\byear{2003}).
\arxivurl{astro-ph/0210319}.
doi:\doiurl{10.1086/345573}
\end{barticle}
\endbibitem

\bibitem[\protect\citeauthoryear{{Liebert} et~al.}{2005}]{Liebert2005}
\begin{barticle}
\bauthor{\bsnm{{Liebert}}, \binits{J.}},
\bauthor{\bsnm{{Wickramasinghe}}, \binits{D.T.}},
\bauthor{\bsnm{{Schmidt}}, \binits{G.D.}},
\bauthor{\bsnm{{Silvestri}}, \binits{N.M.}},
\bauthor{\bsnm{{Hawley}}, \binits{S.L.}},
\bauthor{\bsnm{{Szkody}}, \binits{P.}},
\bauthor{\bsnm{{Ferrario}}, \binits{L.}},
\bauthor{\bsnm{{Webbink}}, \binits{R.F.}},
\bauthor{\bsnm{{Oswalt}}, \binits{T.D.}},
\bauthor{\bsnm{{Smith}}, \binits{J.A.}},
\bauthor{\bsnm{{Lemagie}}, \binits{M.P.}}:
\bjtitle{\aj}
\bvolume{129},
\bfpage{2376}
(\byear{2005}).
doi:\doiurl{10.1086/429639}
\end{barticle}
\endbibitem

\bibitem[\protect\citeauthoryear{{Liebert} et~al.}{2015}]{Liebert2015}
\begin{barticle}
\bauthor{\bsnm{{Liebert}}, \binits{J.}},
\bauthor{\bsnm{{Ferrario}}, \binits{L.}},
\bauthor{\bsnm{{Wickramasinghe}}, \binits{D.T.}},
\bauthor{\bsnm{{Smith}}, \binits{P.}}:
\bjtitle{\apj}
\bvolume{1},
\bfpage{1}
(\byear{2015}).
\arxivurl{preprint}
\end{barticle}
\endbibitem

\bibitem[\protect\citeauthoryear{{Ligni{\`e}res} et~al.}{2009}]{Lignieres2009}
\begin{barticle}
\bauthor{\bsnm{{Ligni{\`e}res}}, \binits{F.}},
\bauthor{\bsnm{{Petit}}, \binits{P.}},
\bauthor{\bsnm{{B{\"o}hm}}, \binits{T.}},
\bauthor{\bsnm{{Auri{\`e}re}}, \binits{M.}}:
\bjtitle{\aap}
\bvolume{500},
\bfpage{41}
(\byear{2009}).
\arxivurl{0903.1247}.
doi:\doiurl{10.1051/0004-6361/200911996}
\end{barticle}
\endbibitem

\bibitem[\protect\citeauthoryear{{Ligni{\`e}res} et~al.}{2014}]{Lignieres2014}
\begin{bchapter}
\bauthor{\bsnm{{Ligni{\`e}res}}, \binits{F.}},
\bauthor{\bsnm{{Petit}}, \binits{P.}},
\bauthor{\bsnm{{Auri{\`e}re}}, \binits{M.}},
\bauthor{\bsnm{{Wade}}, \binits{G.A.}},
\bauthor{\bsnm{{B{\"o}hm}}, \binits{T.}}:
In: \bbtitle{IAU Symposium}.
\bsertitle{IAU Symposium},
vol. \bseriesno{302},
p. \bfpage{338}
(\byear{2014}).
\arxivurl{1402.5362}.
doi:\doiurl{10.1017/S1743921314002440}
\end{bchapter}
\endbibitem

\bibitem[\protect\citeauthoryear{{Link}}{2012a}]{l12}
\begin{barticle}
\bauthor{\bsnm{{Link}}, \binits{B.}}:
\bjtitle{\mnras}
\bvolume{421},
\bfpage{2682}
(\byear{2012}a).
\arxivurl{1111.0696}.
doi:\doiurl{10.1111/j.1365-2966.2012.20498.x}
\end{barticle}
\endbibitem

\bibitem[\protect\citeauthoryear{{Link}}{2012b}]{l12b}
\begin{barticle}
\bauthor{\bsnm{{Link}}, \binits{B.}}:
\bjtitle{\mnras}
\bvolume{422},
\bfpage{1640}
(\byear{2012}b).
\arxivurl{1105.4654}.
doi:\doiurl{10.1111/j.1365-2966.2012.20740.x}
\end{barticle}
\endbibitem

\bibitem[\protect\citeauthoryear{{Link}}{2013}]{l13b}
\begin{botherref}
\oauthor{\bsnm{{Link}}, \binits{B.}}:
ArXiv e-prints
(2013).
\arxivurl{1311.2499}
\end{botherref}
\endbibitem

\bibitem[\protect\citeauthoryear{{Lyne} et~al.}{2013}]{letal13}
\begin{barticle}
\bauthor{\bsnm{{Lyne}}, \binits{A.}},
\bauthor{\bsnm{{Graham-Smith}}, \binits{F.}},
\bauthor{\bsnm{{Weltevrede}}, \binits{P.}},
\bauthor{\bsnm{{Jordan}}, \binits{C.}},
\bauthor{\bsnm{{Stappers}}, \binits{B.}},
\bauthor{\bsnm{{Bassa}}, \binits{C.}},
\bauthor{\bsnm{{Kramer}}, \binits{M.}}:
\bjtitle{Science}
\bvolume{342},
\bfpage{598}
(\byear{2013}).
\arxivurl{1311.0408}.
doi:\doiurl{10.1126/science.1243254}
\end{barticle}
\endbibitem

\bibitem[\protect\citeauthoryear{{Macy}}{1974}]{m74}
\begin{barticle}
\bauthor{\bsnm{{Macy}}, \binits{W.W.} \bsuffix{Jr.}}:
\bjtitle{\apj}
\bvolume{190},
\bfpage{153}
(\byear{1974}).
doi:\doiurl{10.1086/152859}
\end{barticle}
\endbibitem

\bibitem[\protect\citeauthoryear{{Maeder} and {Meynet}}{2014}]{Maeder2014}
\begin{barticle}
\bauthor{\bsnm{{Maeder}}, \binits{A.}},
\bauthor{\bsnm{{Meynet}}, \binits{G.}}:
\bjtitle{\apj}
\bvolume{793},
\bfpage{123}
(\byear{2014}).
\arxivurl{1408.1192}.
doi:\doiurl{10.1088/0004-637X/793/2/123}
\end{barticle}
\endbibitem

\bibitem[\protect\citeauthoryear{{Manchester} et~al.}{2001}]{metal01}
\begin{barticle}
\bauthor{\bsnm{{Manchester}}, \binits{R.N.}},
\bauthor{\bsnm{{Lyne}}, \binits{A.G.}},
\bauthor{\bsnm{{Camilo}}, \binits{F.}},
\bauthor{\bsnm{{Bell}}, \binits{J.F.}},
\bauthor{\bsnm{{Kaspi}}, \binits{V.M.}},
\bauthor{\bsnm{{D'Amico}}, \binits{N.}},
\bauthor{\bsnm{{McKay}}, \binits{N.P.F.}},
\bauthor{\bsnm{{Crawford}}, \binits{F.}},
\bauthor{\bsnm{{Stairs}}, \binits{I.H.}},
\bauthor{\bsnm{{Possenti}}, \binits{A.}},
\bauthor{\bsnm{{Kramer}}, \binits{M.}},
\bauthor{\bsnm{{Sheppard}}, \binits{D.C.}}:
\bjtitle{\mnras}
\bvolume{328},
\bfpage{17}
(\byear{2001}).
\arxivurl{astro-ph/0106522}.
doi:\doiurl{10.1046/j.1365-8711.2001.04751.x}
\end{barticle}
\endbibitem

\bibitem[\protect\citeauthoryear{{Marchant} et~al.}{2011}]{mra11}
\begin{barticle}
\bauthor{\bsnm{{Marchant}}, \binits{P.}},
\bauthor{\bsnm{{Reisenegger}}, \binits{A.}},
\bauthor{\bsnm{{Akg{\"u}n}}, \binits{T.}}:
\bjtitle{\mnras}
\bvolume{415},
\bfpage{2426}
(\byear{2011}).
\arxivurl{1011.0661}.
doi:\doiurl{10.1111/j.1365-2966.2011.18874.x}
\end{barticle}
\endbibitem

\bibitem[\protect\citeauthoryear{{Markey} and {Tayler}}{1973}]{mt73}
\begin{barticle}
\bauthor{\bsnm{{Markey}}, \binits{P.}},
\bauthor{\bsnm{{Tayler}}, \binits{R.J.}}:
\bjtitle{\mnras}
\bvolume{163},
\bfpage{77}
(\byear{1973})
\end{barticle}
\endbibitem

\bibitem[\protect\citeauthoryear{Maron and Cowley}{2001}]{Maron2001a}
\begin{botherref}
\oauthor{\bsnm{Maron}, \binits{J.}},
\oauthor{\bsnm{Cowley}, \binits{S.}}:
eprint arXiv,
11008
(2001)
\end{botherref}
\endbibitem

\bibitem[\protect\citeauthoryear{{Mastrano} and {Melatos}}{2008}]{mm08}
\begin{barticle}
\bauthor{\bsnm{{Mastrano}}, \binits{A.}},
\bauthor{\bsnm{{Melatos}}, \binits{A.}}:
\bjtitle{\mnras}
\bvolume{387},
\bfpage{1735}
(\byear{2008}).
\arxivurl{0804.4056}.
doi:\doiurl{10.1111/j.1365-2966.2008.13365.x}
\end{barticle}
\endbibitem

\bibitem[\protect\citeauthoryear{{Mastrano} and {Melatos}}{2011}]{mm11}
\begin{barticle}
\bauthor{\bsnm{{Mastrano}}, \binits{A.}},
\bauthor{\bsnm{{Melatos}}, \binits{A.}}:
\bjtitle{\mnras}
\bvolume{417},
\bfpage{508}
(\byear{2011}).
\arxivurl{1108.0221}.
doi:\doiurl{10.1111/j.1365-2966.2011.19290.x}
\end{barticle}
\endbibitem

\bibitem[\protect\citeauthoryear{{Mastrano} and {Melatos}}{2012}]{mm12}
\begin{barticle}
\bauthor{\bsnm{{Mastrano}}, \binits{A.}},
\bauthor{\bsnm{{Melatos}}, \binits{A.}}:
\bjtitle{\mnras}
\bvolume{421},
\bfpage{760}
(\byear{2012}).
\arxivurl{1112.1542}.
doi:\doiurl{10.1111/j.1365-2966.2011.20350.x}
\end{barticle}
\endbibitem

\bibitem[\protect\citeauthoryear{{Mastrano} et~al.}{2013}]{mlm13}
\begin{barticle}
\bauthor{\bsnm{{Mastrano}}, \binits{A.}},
\bauthor{\bsnm{{Lasky}}, \binits{P.D.}},
\bauthor{\bsnm{{Melatos}}, \binits{A.}}:
\bjtitle{\mnras}
\bvolume{434},
\bfpage{1658}
(\byear{2013}).
\arxivurl{1306.4503}.
doi:\doiurl{10.1093/mnras/stt1131}
\end{barticle}
\endbibitem

\bibitem[\protect\citeauthoryear{{Mastrano} et~al.}{2011}]{metal11}
\begin{barticle}
\bauthor{\bsnm{{Mastrano}}, \binits{A.}},
\bauthor{\bsnm{{Melatos}}, \binits{A.}},
\bauthor{\bsnm{{Reisenegger}}, \binits{A.}},
\bauthor{\bsnm{{Akg{\"u}n}}, \binits{T.}}:
\bjtitle{\mnras}
\bvolume{417},
\bfpage{2288}
(\byear{2011}).
\arxivurl{1108.0219}.
doi:\doiurl{10.1111/j.1365-2966.2011.19410.x}
\end{barticle}
\endbibitem

\bibitem[\protect\citeauthoryear{{Mathys}}{1995}]{Mathys1995}
\begin{barticle}
\bauthor{\bsnm{{Mathys}}, \binits{G.}}:
\bjtitle{\aap}
\bvolume{293},
\bfpage{746}
(\byear{1995})
\end{barticle}
\endbibitem

\bibitem[\protect\citeauthoryear{{Melatos}}{2000}]{m00}
\begin{barticle}
\bauthor{\bsnm{{Melatos}}, \binits{A.}}:
\bjtitle{\mnras}
\bvolume{313},
\bfpage{217}
(\byear{2000}).
\arxivurl{astro-ph/0004035}.
doi:\doiurl{10.1046/j.1365-8711.2000.03031.x}
\end{barticle}
\endbibitem

\bibitem[\protect\citeauthoryear{{Melatos}}{2012}]{m12}
\begin{barticle}
\bauthor{\bsnm{{Melatos}}, \binits{A.}}:
\bjtitle{\apj}
\bvolume{761},
\bfpage{32}
(\byear{2012}).
\arxivurl{1210.5872}.
doi:\doiurl{10.1088/0004-637X/761/1/32}
\end{barticle}
\endbibitem

\bibitem[\protect\citeauthoryear{{Melatos} and {Peralta}}{2007}]{mp07}
\begin{barticle}
\bauthor{\bsnm{{Melatos}}, \binits{A.}},
\bauthor{\bsnm{{Peralta}}, \binits{C.}}:
\bjtitle{\apjl}
\bvolume{662},
\bfpage{99}
(\byear{2007}).
\arxivurl{0710.2455}.
doi:\doiurl{10.1086/518598}
\end{barticle}
\endbibitem

\bibitem[\protect\citeauthoryear{{Melatos} and {Priymak}}{2014}]{melpri14}
\begin{barticle}
\bauthor{\bsnm{{Melatos}}, \binits{A.}},
\bauthor{\bsnm{{Priymak}}, \binits{M.}}:
\bjtitle{\apj}
\bvolume{794},
\bfpage{170}
(\byear{2014}).
\arxivurl{1409.1375}.
doi:\doiurl{10.1088/0004-637X/794/2/170}
\end{barticle}
\endbibitem

\bibitem[\protect\citeauthoryear{{Melatos} et~al.}{2008}]{metal08}
\begin{barticle}
\bauthor{\bsnm{{Melatos}}, \binits{A.}},
\bauthor{\bsnm{{Peralta}}, \binits{C.}},
\bauthor{\bsnm{{Wyithe}}, \binits{J.S.B.}}:
\bjtitle{\apj}
\bvolume{672},
\bfpage{1103}
(\byear{2008}).
\arxivurl{0710.1021}.
doi:\doiurl{10.1086/523349}
\end{barticle}
\endbibitem

\bibitem[\protect\citeauthoryear{{Mendell}}{1998}]{m98}
\begin{barticle}
\bauthor{\bsnm{{Mendell}}, \binits{G.}}:
\bjtitle{\mnras}
\bvolume{296},
\bfpage{903}
(\byear{1998}).
\arxivurl{astro-ph/9702032}.
doi:\doiurl{10.1046/j.1365-8711.1998.01451.x}
\end{barticle}
\endbibitem

\bibitem[\protect\citeauthoryear{Meneguzzi et~al.}{1981}]{Meneguzzi1981}
\begin{barticle}
\bauthor{\bsnm{Meneguzzi}, \binits{M.}},
\bauthor{\bsnm{Frisch}, \binits{U.}},
\bauthor{\bsnm{Pouquet}, \binits{A.}}:
\bjtitle{Phys Rev Lett}
\bvolume{47},
\bfpage{1060}
(\byear{1981})
\end{barticle}
\endbibitem

\bibitem[\protect\citeauthoryear{{Mestel}}{1966}]{Mestel1966}
\begin{barticle}
\bauthor{\bsnm{{Mestel}}, \binits{L.}}:
\bjtitle{\mnras}
\bvolume{133},
\bfpage{265}
(\byear{1966})
\end{barticle}
\endbibitem

\bibitem[\protect\citeauthoryear{{Mestel} and {Moss}}{2010}]{MestelMoss2010}
\begin{barticle}
\bauthor{\bsnm{{Mestel}}, \binits{L.}},
\bauthor{\bsnm{{Moss}}, \binits{D.}}:
\bjtitle{\mnras}
\bvolume{405},
\bfpage{1845}
(\byear{2010}).
doi:\doiurl{10.1111/j.1365-2966.2010.16558.x}
\end{barticle}
\endbibitem

\bibitem[\protect\citeauthoryear{{Minkowski}}{1942}]{Minkowski1942}
\begin{barticle}
\bauthor{\bsnm{{Minkowski}}, \binits{R.}}:
\bjtitle{\apj}
\bvolume{96},
\bfpage{199}
(\byear{1942}).
doi:\doiurl{10.1086/144447}
\end{barticle}
\endbibitem

\bibitem[\protect\citeauthoryear{{Mitchell} et~al.}{2013}]{metal13}
\begin{botherref}
\oauthor{\bsnm{{Mitchell}}, \binits{J.P.}},
\oauthor{\bsnm{{Braithwaite}}, \binits{J.}},
\oauthor{\bsnm{{Langer}}, \binits{N.}},
\oauthor{\bsnm{{Reisenegger}}, \binits{A.}},
\oauthor{\bsnm{{Spruit}}, \binits{H.}}:
ArXiv e-prints
(2013).
\arxivurl{1310.2595}
\end{botherref}
\endbibitem

\bibitem[\protect\citeauthoryear{{Moffatt}}{1978}]{m78}
\begin{bbook}
\bauthor{\bsnm{{Moffatt}}, \binits{H.K.}}:
\bbtitle{{Magnetic Field Generation in Electrically Conducting Fluids}},
(\byear{1978})
\end{bbook}
\endbibitem

\bibitem[\protect\citeauthoryear{{Moiseenko} et~al.}{2006}]{mbka06}
\begin{barticle}
\bauthor{\bsnm{{Moiseenko}}, \binits{S.G.}},
\bauthor{\bsnm{{Bisnovatyi-Kogan}}, \binits{G.S.}},
\bauthor{\bsnm{{Ardeljan}}, \binits{N.V.}}:
\bjtitle{\mnras}
\bvolume{370},
\bfpage{501}
(\byear{2006}).
doi:\doiurl{10.1111/j.1365-2966.2006.10517.x}
\end{barticle}
\endbibitem

\bibitem[\protect\citeauthoryear{{Morris} et~al.}{2002}]{metal02}
\begin{barticle}
\bauthor{\bsnm{{Morris}}, \binits{D.J.}},
\bauthor{\bsnm{{Hobbs}}, \binits{G.}},
\bauthor{\bsnm{{Lyne}}, \binits{A.G.}},
\bauthor{\bsnm{{Stairs}}, \binits{I.H.}},
\bauthor{\bsnm{{Camilo}}, \binits{F.}},
\bauthor{\bsnm{{Manchester}}, \binits{R.N.}},
\bauthor{\bsnm{{Possenti}}, \binits{A.}},
\bauthor{\bsnm{{Bell}}, \binits{J.F.}},
\bauthor{\bsnm{{Kaspi}}, \binits{V.M.}},
\bauthor{\bsnm{{Amico}}, \binits{N.D.}},
\bauthor{\bsnm{{McKay}}, \binits{N.P.F.}},
\bauthor{\bsnm{{Crawford}}, \binits{F.}},
\bauthor{\bsnm{{Kramer}}, \binits{M.}}:
\bjtitle{MNRAS}
\bvolume{335},
\bfpage{275}
(\byear{2002}).
\arxivurl{astro-ph/0204238}.
doi:\doiurl{10.1046/j.1365-8711.2002.05551.x}
\end{barticle}
\endbibitem

\bibitem[\protect\citeauthoryear{{Moss}}{2001}]{Moss2001}
\begin{bchapter}
\bauthor{\bsnm{{Moss}}, \binits{D.}}:
In: \beditor{\bsnm{{Mathys}}, \binits{G.}},
\beditor{\bsnm{{Solanki}}, \binits{S.K.}},
\beditor{\bsnm{{Wickramasinghe}}, \binits{D.T.}} (eds.)
\bbtitle{Magnetic Fields Across the Hertzsprung-Russell Diagram}.
\bsertitle{Astronomical Society of the Pacific Conference Series},
vol. \bseriesno{248},
p. \bfpage{305}
(\byear{2001})
\end{bchapter}
\endbibitem

\bibitem[\protect\citeauthoryear{{Moss}}{2003}]{Moss2003}
\begin{barticle}
\bauthor{\bsnm{{Moss}}, \binits{D.}}:
\bjtitle{\aap}
\bvolume{403},
\bfpage{693}
(\byear{2003}).
doi:\doiurl{10.1051/0004-6361:20030431}
\end{barticle}
\endbibitem

\bibitem[\protect\citeauthoryear{{Muno} et~al.}{2006}]{Muno2006}
\begin{barticle}
\bauthor{\bsnm{{Muno}}, \binits{M.P.}},
\bauthor{\bsnm{{Clark}}, \binits{J.S.}},
\bauthor{\bsnm{{Crowther}}, \binits{P.A.}},
\bauthor{\bsnm{{Dougherty}}, \binits{S.M.}},
\bauthor{\bsnm{{de Grijs}}, \binits{R.}},
\bauthor{\bsnm{{Law}}, \binits{C.}},
\bauthor{\bsnm{{McMillan}}, \binits{S.L.W.}},
\bauthor{\bsnm{{Morris}}, \binits{M.R.}},
\bauthor{\bsnm{{Negueruela}}, \binits{I.}},
\bauthor{\bsnm{{Pooley}}, \binits{D.}},
\bauthor{\bsnm{{Portegies Zwart}}, \binits{S.}},
\bauthor{\bsnm{{Yusef-Zadeh}}, \binits{F.}}:
\bjtitle{\apjl}
\bvolume{636},
\bfpage{41}
(\byear{2006}).
\arxivurl{astro-ph/0509408}.
doi:\doiurl{10.1086/499776}
\end{barticle}
\endbibitem

\bibitem[\protect\citeauthoryear{{Nakabayashi}}{1983}]{n83}
\begin{barticle}
\bauthor{\bsnm{{Nakabayashi}}, \binits{K.}}:
\bjtitle{Journal of Fluid Mechanics}
\bvolume{132},
\bfpage{209}
(\byear{1983}).
doi:\doiurl{10.1017/S0022112083001561}
\end{barticle}
\endbibitem

\bibitem[\protect\citeauthoryear{{Neiner} and {Alecian}}{2013}]{Neiner2013}
\begin{bchapter}
\bauthor{\bsnm{{Neiner}}, \binits{C.}},
\bauthor{\bsnm{{Alecian}}, \binits{E.}}:
In: \beditor{\bsnm{{Pavlovski}}, \binits{K.}},
\beditor{\bsnm{{Tkachenko}}, \binits{A.}},
\beditor{\bsnm{{Torres}}, \binits{G.}} (eds.)
\bbtitle{EAS Publications Series}.
\bsertitle{EAS Publications Series},
vol. \bseriesno{64},
p. \bfpage{75}
(\byear{2013}).
\arxivurl{1311.2389}.
doi:\doiurl{10.1051/eas/1364010}
\end{bchapter}
\endbibitem

\bibitem[\protect\citeauthoryear{{Neiner} et~al.}{2014a}]{NeinerFolsom2014}
\begin{bchapter}
\bauthor{\bsnm{{Neiner}}, \binits{C.}},
\bauthor{\bsnm{{Folsom}}, \binits{C.P.}},
\bauthor{\bsnm{{Blazere}}, \binits{A.}}:
In: \beditor{\bsnm{{Ballet}}, \binits{J.}},
\beditor{\bsnm{{Martins}}, \binits{F.}},
\beditor{\bsnm{{Bournaud}}, \binits{F.}},
\beditor{\bsnm{{Monier}}, \binits{R.}},
\beditor{\bsnm{{Reyl{\'e}}}, \binits{C.}} (eds.)
\bbtitle{SF2A-2014: Proceedings of the Annual meeting of the French Society of
  Astronomy and Astrophysics},
p. \bfpage{163}
(\byear{2014}a).
\arxivurl{1410.2755}
\end{bchapter}
\endbibitem

\bibitem[\protect\citeauthoryear{{Neiner} et~al.}{2014b}]{Neiner2014}
\begin{barticle}
\bauthor{\bsnm{{Neiner}}, \binits{C.}},
\bauthor{\bsnm{{Monin}}, \binits{D.}},
\bauthor{\bsnm{{Leroy}}, \binits{B.}},
\bauthor{\bsnm{{Mathis}}, \binits{S.}},
\bauthor{\bsnm{{Bohlender}}, \binits{D.}}:
\bjtitle{\aap}
\bvolume{562},
\bfpage{59}
(\byear{2014}b).
\arxivurl{1312.3521}.
doi:\doiurl{10.1051/0004-6361/201323093}
\end{barticle}
\endbibitem

\bibitem[\protect\citeauthoryear{{Nordhaus} et~al.}{2007}]{Nordhaus2007}
\begin{barticle}
\bauthor{\bsnm{{Nordhaus}}, \binits{J.}},
\bauthor{\bsnm{{Blackman}}, \binits{E.G.}},
\bauthor{\bsnm{{Frank}}, \binits{A.}}:
\bjtitle{\mnras}
\bvolume{376},
\bfpage{599}
(\byear{2007}).
\arxivurl{astro-ph/0609726}.
doi:\doiurl{10.1111/j.1365-2966.2007.11417.x}
\end{barticle}
\endbibitem

\bibitem[\protect\citeauthoryear{{Nordhaus} et~al.}{2011}]{Nordhaus2011}
\begin{barticle}
\bauthor{\bsnm{{Nordhaus}}, \binits{J.}},
\bauthor{\bsnm{{Wellons}}, \binits{S.}},
\bauthor{\bsnm{{Spiegel}}, \binits{D.S.}},
\bauthor{\bsnm{{Metzger}}, \binits{B.D.}},
\bauthor{\bsnm{{Blackman}}, \binits{E.G.}}:
\bjtitle{Proceedings of the National Academy of Science}
\bvolume{108},
\bfpage{3135}
(\byear{2011}).
\arxivurl{1010.1529}.
doi:\doiurl{10.1073/pnas.1015005108}
\end{barticle}
\endbibitem

\bibitem[\protect\citeauthoryear{Olesen}{1997}]{Olesen1997}
\begin{barticle}
\bauthor{\bsnm{Olesen}, \binits{P.}}:
\bjtitle{Physics Letters B}
\bvolume{398}(\bissue{3-4}),
\bfpage{321}
(\byear{1997}).
doi:\doiurl{10.1016/S0370-2693(97)00235-9}
\end{barticle}
\endbibitem

\bibitem[\protect\citeauthoryear{{Ott} et~al.}{2006}]{oetal06}
\begin{barticle}
\bauthor{\bsnm{{Ott}}, \binits{C.D.}},
\bauthor{\bsnm{{Burrows}}, \binits{A.}},
\bauthor{\bsnm{{Thompson}}, \binits{T.A.}},
\bauthor{\bsnm{{Livne}}, \binits{E.}},
\bauthor{\bsnm{{Walder}}, \binits{R.}}:
\bjtitle{\apjs}
\bvolume{164},
\bfpage{130}
(\byear{2006}).
\arxivurl{astro-ph/0508462}.
doi:\doiurl{10.1086/500832}
\end{barticle}
\endbibitem

\bibitem[\protect\citeauthoryear{{Pacini}}{1968}]{Pacini1968}
\begin{barticle}
\bauthor{\bsnm{{Pacini}}, \binits{F.}}:
\bjtitle{\nat}
\bvolume{219},
\bfpage{145}
(\byear{1968}).
doi:\doiurl{10.1038/219145a0}
\end{barticle}
\endbibitem

\bibitem[\protect\citeauthoryear{{Passamonti} and {Lander}}{2013}]{pl13}
\begin{barticle}
\bauthor{\bsnm{{Passamonti}}, \binits{A.}},
\bauthor{\bsnm{{Lander}}, \binits{S.K.}}:
\bjtitle{\mnras}
\bvolume{429},
\bfpage{767}
(\byear{2013}).
\arxivurl{1210.2969}.
doi:\doiurl{10.1093/mnras/sts372}
\end{barticle}
\endbibitem

\bibitem[\protect\citeauthoryear{{Payne} and {Melatos}}{2004}]{pm04}
\begin{barticle}
\bauthor{\bsnm{{Payne}}, \binits{D.J.B.}},
\bauthor{\bsnm{{Melatos}}, \binits{A.}}:
\bjtitle{\mnras}
\bvolume{351},
\bfpage{569}
(\byear{2004}).
\arxivurl{astro-ph/0403173}.
doi:\doiurl{10.1111/j.1365-2966.2004.07798.x}
\end{barticle}
\endbibitem

\bibitem[\protect\citeauthoryear{{Peralta} et~al.}{2005}]{petal05}
\begin{barticle}
\bauthor{\bsnm{{Peralta}}, \binits{C.}},
\bauthor{\bsnm{{Melatos}}, \binits{A.}},
\bauthor{\bsnm{{Giacobello}}, \binits{M.}},
\bauthor{\bsnm{{Ooi}}, \binits{A.}}:
\bjtitle{\apj}
\bvolume{635},
\bfpage{1224}
(\byear{2005}).
\arxivurl{astro-ph/0509416}.
doi:\doiurl{10.1086/497899}
\end{barticle}
\endbibitem

\bibitem[\protect\citeauthoryear{{Peralta} et~al.}{2006}]{petal06}
\begin{barticle}
\bauthor{\bsnm{{Peralta}}, \binits{C.}},
\bauthor{\bsnm{{Melatos}}, \binits{A.}},
\bauthor{\bsnm{{Giacobello}}, \binits{M.}},
\bauthor{\bsnm{{Ooi}}, \binits{A.}}:
\bjtitle{\apj}
\bvolume{651},
\bfpage{1079}
(\byear{2006}).
\arxivurl{astro-ph/0607161}.
doi:\doiurl{10.1086/507576}
\end{barticle}
\endbibitem

\bibitem[\protect\citeauthoryear{{Petit} et~al.}{2011}]{Petit2011}
\begin{barticle}
\bauthor{\bsnm{{Petit}}, \binits{P.}},
\bauthor{\bsnm{{Ligni{\`e}res}}, \binits{F.}},
\bauthor{\bsnm{{Auri{\`e}re}}, \binits{M.}},
\bauthor{\bsnm{{Wade}}, \binits{G.A.}},
\bauthor{\bsnm{{Alina}}, \binits{D.}},
\bauthor{\bsnm{{Ballot}}, \binits{J.}},
\bauthor{\bsnm{{B{\"o}hm}}, \binits{T.}},
\bauthor{\bsnm{{Jouve}}, \binits{L.}},
\bauthor{\bsnm{{Oza}}, \binits{A.}},
\bauthor{\bsnm{{Paletou}}, \binits{F.}},
\bauthor{\bsnm{{Th{\'e}ado}}, \binits{S.}}:
\bjtitle{\aap}
\bvolume{532},
\bfpage{13}
(\byear{2011}).
\arxivurl{1106.5363}.
doi:\doiurl{10.1051/0004-6361/201117573}
\end{barticle}
\endbibitem

\bibitem[\protect\citeauthoryear{{Petit} et~al.}{2013}]{Petit2013}
\begin{barticle}
\bauthor{\bsnm{{Petit}}, \binits{V.}},
\bauthor{\bsnm{{Owocki}}, \binits{S.P.}},
\bauthor{\bsnm{{Wade}}, \binits{G.A.}},
\bauthor{\bsnm{{Cohen}}, \binits{D.H.}},
\bauthor{\bsnm{{Sundqvist}}, \binits{J.O.}},
\bauthor{\bsnm{{Gagn{\'e}}}, \binits{M.}},
\bauthor{\bsnm{{Ma{\'{\i}}z Apell{\'a}niz}}, \binits{J.}},
\bauthor{\bsnm{{Oksala}}, \binits{M.E.}},
\bauthor{\bsnm{{Bohlender}}, \binits{D.A.}},
\bauthor{\bsnm{{Rivinius}}, \binits{T.}},
\bauthor{\bsnm{{Henrichs}}, \binits{H.F.}},
\bauthor{\bsnm{{Alecian}}, \binits{E.}},
\bauthor{\bsnm{{Townsend}}, \binits{R.H.D.}},
\bauthor{\bsnm{{ud-Doula}}, \binits{A.}},
\bauthor{\bsnm{{MiMeS Collaboration}}}:
\bjtitle{\mnras}
\bvolume{429},
\bfpage{398}
(\byear{2013}).
\arxivurl{1211.0282}.
doi:\doiurl{10.1093/mnras/sts344}
\end{barticle}
\endbibitem

\bibitem[\protect\citeauthoryear{{Pons} et~al.}{2012}]{pvg12}
\begin{barticle}
\bauthor{\bsnm{{Pons}}, \binits{J.A.}},
\bauthor{\bsnm{{Vigan{\`o}}}, \binits{D.}},
\bauthor{\bsnm{{Geppert}}, \binits{U.}}:
\bjtitle{\aap}
\bvolume{547},
\bfpage{9}
(\byear{2012}).
\arxivurl{1209.2273}.
doi:\doiurl{10.1051/0004-6361/201220091}
\end{barticle}
\endbibitem

\bibitem[\protect\citeauthoryear{{Power} et~al.}{2007}]{Power2007}
\begin{bchapter}
\bauthor{\bsnm{{Power}}, \binits{J.}},
\bauthor{\bsnm{{Wade}}, \binits{G.A.}},
\bauthor{\bsnm{{Hanes}}, \binits{D.A.}},
\bauthor{\bsnm{{Aurier}}, \binits{M.}},
\bauthor{\bsnm{{Silvester}}, \binits{J.}}:
In: \beditor{\bsnm{{Romanyuk}}, \binits{I.I.}},
\beditor{\bsnm{{Kudryavtsev}}, \binits{D.O.}},
\beditor{\bsnm{{Neizvestnaya}}, \binits{O.M.}},
\beditor{\bsnm{{Shapoval}}, \binits{V.M.}} (eds.)
\bbtitle{Physics of Magnetic Stars},
p. \bfpage{89}
(\byear{2007}).
\arxivurl{astro-ph/0612557}
\end{bchapter}
\endbibitem

\bibitem[\protect\citeauthoryear{{Prendergast}}{1956}]{p56}
\begin{barticle}
\bauthor{\bsnm{{Prendergast}}, \binits{K.H.}}:
\bjtitle{\apj}
\bvolume{123},
\bfpage{498}
(\byear{1956}).
doi:\doiurl{10.1086/146186}
\end{barticle}
\endbibitem

\bibitem[\protect\citeauthoryear{{Preston}}{1970}]{Preston1970}
\begin{barticle}
\bauthor{\bsnm{{Preston}}, \binits{G.W.}}:
\bjtitle{\apjl}
\bvolume{160},
\bfpage{143}
(\byear{1970}).
doi:\doiurl{10.1086/180547}
\end{barticle}
\endbibitem

\bibitem[\protect\citeauthoryear{Priest}{2014}]{Priest2014}
\begin{bbook}
\bauthor{\bsnm{Priest}, \binits{E.}}:
\bbtitle{{Magnetohydrodynamics of the Sun}}.
\bpublisher{Cambridge University Press}, \blocation{???}
(\byear{2014}).
\burl{http://books.google.com/books?id=RhL7AgAAQBAJ\&pgis=1}
\end{bbook}
\endbibitem

\bibitem[\protect\citeauthoryear{{Prix} et~al.}{2002}]{pca02}
\begin{barticle}
\bauthor{\bsnm{{Prix}}, \binits{R.}},
\bauthor{\bsnm{{Comer}}, \binits{G.L.}},
\bauthor{\bsnm{{Andersson}}, \binits{N.}}:
\bjtitle{\aap}
\bvolume{381},
\bfpage{178}
(\byear{2002}).
\arxivurl{astro-ph/0107176}.
doi:\doiurl{10.1051/0004-6361:20011499}
\end{barticle}
\endbibitem

\bibitem[\protect\citeauthoryear{{Railton} et~al.}{2014}]{Railton2014}
\begin{barticle}
\bauthor{\bsnm{{Railton}}, \binits{A.D.}},
\bauthor{\bsnm{{Tout}}, \binits{C.A.}},
\bauthor{\bsnm{{Aarseth}}, \binits{S.J.}}:
\bjtitle{\pasa}
\bvolume{31},
\bfpage{17}
(\byear{2014}).
\arxivurl{1402.3323}.
doi:\doiurl{10.1017/pasa.2014.10}
\end{barticle}
\endbibitem

\bibitem[\protect\citeauthoryear{{Reisenegger} and {Goldreich}}{1992}]{rg92}
\begin{barticle}
\bauthor{\bsnm{{Reisenegger}}, \binits{A.}},
\bauthor{\bsnm{{Goldreich}}, \binits{P.}}:
\bjtitle{\apj}
\bvolume{395},
\bfpage{240}
(\byear{1992}).
doi:\doiurl{10.1086/171645}
\end{barticle}
\endbibitem

\bibitem[\protect\citeauthoryear{{Reynolds}}{2008}]{r08}
\begin{barticle}
\bauthor{\bsnm{{Reynolds}}, \binits{S.P.}}:
\bjtitle{\araa}
\bvolume{46},
\bfpage{89}
(\byear{2008}).
doi:\doiurl{10.1146/annurev.astro.46.060407.145237}
\end{barticle}
\endbibitem

\bibitem[\protect\citeauthoryear{{Reynolds} and {Chevalier}}{1981}]{rc81}
\begin{barticle}
\bauthor{\bsnm{{Reynolds}}, \binits{S.P.}},
\bauthor{\bsnm{{Chevalier}}, \binits{R.A.}}:
\bjtitle{\apj}
\bvolume{245},
\bfpage{912}
(\byear{1981}).
doi:\doiurl{10.1086/158868}
\end{barticle}
\endbibitem

\bibitem[\protect\citeauthoryear{{Ruderman}}{1972}]{r72}
\begin{barticle}
\bauthor{\bsnm{{Ruderman}}, \binits{M.}}:
\bjtitle{\araa}
\bvolume{10},
\bfpage{427}
(\byear{1972}).
doi:\doiurl{10.1146/annurev.aa.10.090172.002235}
\end{barticle}
\endbibitem

\bibitem[\protect\citeauthoryear{{Ruderman}}{1991}]{r91}
\begin{barticle}
\bauthor{\bsnm{{Ruderman}}, \binits{M.}}:
\bjtitle{\apj}
\bvolume{366},
\bfpage{261}
(\byear{1991}).
doi:\doiurl{10.1086/169558}
\end{barticle}
\endbibitem

\bibitem[\protect\citeauthoryear{{Ruderman} and {Sutherland}}{1973}]{rs73}
\begin{barticle}
\bauthor{\bsnm{{Ruderman}}, \binits{M.A.}},
\bauthor{\bsnm{{Sutherland}}, \binits{P.G.}}:
\bjtitle{Nature Physical Science}
\bvolume{246},
\bfpage{93}
(\byear{1973}).
doi:\doiurl{10.1038/physci246093a0}
\end{barticle}
\endbibitem

\bibitem[\protect\citeauthoryear{{R{\"u}diger} et~al.}{2009}]{retal09}
\begin{barticle}
\bauthor{\bsnm{{R{\"u}diger}}, \binits{G.}},
\bauthor{\bsnm{{Shalybkov}}, \binits{D.A.}},
\bauthor{\bsnm{{Schultz}}, \binits{M.}},
\bauthor{\bsnm{{Mond}}, \binits{M.}}:
\bjtitle{Astronomische Nachrichten}
\bvolume{330},
\bfpage{12}
(\byear{2009}).
\arxivurl{0803.1354}.
doi:\doiurl{10.1002/asna.200811042}
\end{barticle}
\endbibitem

\bibitem[\protect\citeauthoryear{Schekochihin
  et~al.}{2002a}]{Schekochihin2002a}
\begin{barticle}
\bauthor{\bsnm{Schekochihin}, \binits{A.A.}},
\bauthor{\bsnm{Cowley}, \binits{S.C.}},
\bauthor{\bsnm{Hammett}, \binits{G.W.}},
\bauthor{\bsnm{Maron}, \binits{J.L.}},
\bauthor{\bsnm{McWilliams}, \binits{J.C.}}:
\bjtitle{New Journal of Physics}
\bvolume{4}(\bissue{1}),
\bfpage{84}
(\byear{2002}a)
\end{barticle}
\endbibitem

\bibitem[\protect\citeauthoryear{Schekochihin et~al.}{2002b}]{Schekochihin2002}
\begin{barticle}
\bauthor{\bsnm{Schekochihin}, \binits{A.A.}},
\bauthor{\bsnm{Maron}, \binits{J.L.}},
\bauthor{\bsnm{Cowley}, \binits{S.C.}},
\bauthor{\bsnm{McWilliams}, \binits{J.C.}}:
\bjtitle{The Astrophysical Journal}
\bvolume{576}(\bissue{2}),
\bfpage{806}
(\byear{2002}b).
doi:\doiurl{10.1086/341814}
\end{barticle}
\endbibitem

\bibitem[\protect\citeauthoryear{Schekochihin et~al.}{2004}]{Schekochihin2004a}
\begin{barticle}
\bauthor{\bsnm{Schekochihin}, \binits{A.A.}},
\bauthor{\bsnm{Cowley}, \binits{S.C.}},
\bauthor{\bsnm{Maron}, \binits{J.L.}},
\bauthor{\bsnm{McWilliams}, \binits{J.C.}}:
\bjtitle{Phys Rev Lett}
\bvolume{92},
\bfpage{64501}
(\byear{2004})
\end{barticle}
\endbibitem

\bibitem[\protect\citeauthoryear{Schleicher et~al.}{2013}]{Schleicher2013}
\begin{barticle}
\bauthor{\bsnm{Schleicher}, \binits{D.R.G.}},
\bauthor{\bsnm{Schober}, \binits{J.}},
\bauthor{\bsnm{Federrath}, \binits{C.}},
\bauthor{\bsnm{Bovino}, \binits{S.}},
\bauthor{\bsnm{Schmidt}, \binits{W.}}:
\bjtitle{New Journal of Physics}
\bvolume{15}(\bissue{2}),
\bfpage{23017}
(\byear{2013})
\end{barticle}
\endbibitem

\bibitem[\protect\citeauthoryear{{Sch{\"o}ller} et~al.}{2011}]{Schoeller2011}
\begin{barticle}
\bauthor{\bsnm{{Sch{\"o}ller}}, \binits{M.}},
\bauthor{\bsnm{{Hubrig}}, \binits{S.}},
\bauthor{\bsnm{{Ilyin}}, \binits{I.}},
\bauthor{\bsnm{{Kharchenko}}, \binits{N.V.}},
\bauthor{\bsnm{{Briquet}}, \binits{M.}},
\bauthor{\bsnm{{Gonz{\'a}lez}}, \binits{J.F.}},
\bauthor{\bsnm{{Langer}}, \binits{N.}},
\bauthor{\bsnm{{Oskinova}}, \binits{L.M.}},
\bauthor{\bsnm{{MAGORI Collaboration}}}:
\bjtitle{Astronomische Nachrichten}
\bvolume{332},
\bfpage{994}
(\byear{2011}).
\arxivurl{1110.5303}.
doi:\doiurl{10.1002/asna.201111606}
\end{barticle}
\endbibitem

\bibitem[\protect\citeauthoryear{Shiromizu}{1998}]{Shiromizu1998}
\begin{barticle}
\bauthor{\bsnm{Shiromizu}, \binits{T.}}:
\bjtitle{Physics Letters B}
\bvolume{443}(\bissue{1-4}),
\bfpage{127}
(\byear{1998}).
doi:\doiurl{10.1016/S0370-2693(98)01348-3}
\end{barticle}
\endbibitem

\bibitem[\protect\citeauthoryear{{Shklovsky}}{1953}]{Shklovsky1953}
\begin{barticle}
\bauthor{\bsnm{{Shklovsky}}, \binits{I.S.}}:
\bjtitle{Doklady Akad. Nauk U.S.S.R.}
\bvolume{90},
\bfpage{983}
(\byear{1953})
\end{barticle}
\endbibitem

\bibitem[\protect\citeauthoryear{{Shternin} and {Yakovlev}}{2008}]{sy08}
\begin{barticle}
\bauthor{\bsnm{{Shternin}}, \binits{P.S.}},
\bauthor{\bsnm{{Yakovlev}}, \binits{D.G.}}:
\bjtitle{\prd}
\bvolume{78}(\bissue{6}),
\bfpage{063006}
(\byear{2008}).
\arxivurl{0808.2018}.
doi:\doiurl{10.1103/PhysRevD.78.063006}
\end{barticle}
\endbibitem

\bibitem[\protect\citeauthoryear{Simon and Weiss}{1997}]{Simon1997}
\begin{barticle}
\bauthor{\bsnm{Simon}, \binits{G.W.}},
\bauthor{\bsnm{Weiss}, \binits{N.O.}}:
\bjtitle{The Astrophysical Journal}
\bvolume{489}(\bissue{2}),
\bfpage{960}
(\byear{1997}).
doi:\doiurl{10.1086/304800}
\end{barticle}
\endbibitem

\bibitem[\protect\citeauthoryear{Son}{1999}]{Son1999}
\begin{barticle}
\bauthor{\bsnm{Son}, \binits{D.}}:
\bjtitle{Physical Review D}
\bvolume{59}(\bissue{6}),
\bfpage{063008}
(\byear{1999}).
doi:\doiurl{10.1103/PhysRevD.59.063008}
\end{barticle}
\endbibitem

\bibitem[\protect\citeauthoryear{{Spruit} and {Phinney}}{1998}]{sp98}
\begin{barticle}
\bauthor{\bsnm{{Spruit}}, \binits{H.}},
\bauthor{\bsnm{{Phinney}}, \binits{E.S.}}:
\bjtitle{\nat}
\bvolume{393},
\bfpage{139}
(\byear{1998}).
\arxivurl{astro-ph/9803201}.
doi:\doiurl{10.1038/30168}
\end{barticle}
\endbibitem

\bibitem[\protect\citeauthoryear{{Spruit}}{2009}]{s09}
\begin{bchapter}
\bauthor{\bsnm{{Spruit}}, \binits{H.C.}}:
In: \beditor{\bsnm{{Strassmeier}}, \binits{K.G.}},
\beditor{\bsnm{{Kosovichev}}, \binits{A.G.}},
\beditor{\bsnm{{Beckman}}, \binits{J.E.}} (eds.)
\bbtitle{IAU Symposium}.
\bsertitle{IAU Symposium},
vol. \bseriesno{259},
p. \bfpage{61}
(\byear{2009}).
doi:\doiurl{10.1017/S1743921309030075}
\end{bchapter}
\endbibitem

\bibitem[\protect\citeauthoryear{{Srinivasan} et~al.}{1990}]{setal90}
\begin{barticle}
\bauthor{\bsnm{{Srinivasan}}, \binits{G.}},
\bauthor{\bsnm{{Bhattacharya}}, \binits{D.}},
\bauthor{\bsnm{{Muslimov}}, \binits{A.G.}},
\bauthor{\bsnm{{Tsygan}}, \binits{A.J.}}:
\bjtitle{Current Science}
\bvolume{59},
\bfpage{31}
(\byear{1990})
\end{barticle}
\endbibitem

\bibitem[\protect\citeauthoryear{{Staelin} and
  {Reifenstein}}{1968}]{Staelin1968}
\begin{barticle}
\bauthor{\bsnm{{Staelin}}, \binits{D.H.}},
\bauthor{\bsnm{{Reifenstein}}, \binits{E.C.} \bsuffix{III}}:
\bjtitle{Science}
\bvolume{162},
\bfpage{1481}
(\byear{1968}).
doi:\doiurl{10.1126/science.162.3861.1481}
\end{barticle}
\endbibitem

\bibitem[\protect\citeauthoryear{{Stairs} et~al.}{2000}]{sls00}
\begin{barticle}
\bauthor{\bsnm{{Stairs}}, \binits{I.H.}},
\bauthor{\bsnm{{Lyne}}, \binits{A.G.}},
\bauthor{\bsnm{{Shemar}}, \binits{S.L.}}:
\bjtitle{\nat}
\bvolume{406},
\bfpage{484}
(\byear{2000}).
doi:\doiurl{10.1038/35020010}
\end{barticle}
\endbibitem

\bibitem[\protect\citeauthoryear{{Stella} et~al.}{2005}]{setal05}
\begin{barticle}
\bauthor{\bsnm{{Stella}}, \binits{L.}},
\bauthor{\bsnm{{Dall'Osso}}, \binits{S.}},
\bauthor{\bsnm{{Israel}}, \binits{G.L.}},
\bauthor{\bsnm{{Vecchio}}, \binits{A.}}:
\bjtitle{\apjl}
\bvolume{634},
\bfpage{165}
(\byear{2005}).
\arxivurl{astro-ph/0511068}.
doi:\doiurl{10.1086/498685}
\end{barticle}
\endbibitem

\bibitem[\protect\citeauthoryear{Taub}{1959}]{Taub1959}
\begin{barticle}
\bauthor{\bsnm{Taub}, \binits{A.H.}}:
\bjtitle{Archive for Rational Mechanics and Analysis}
\bvolume{3},
\bfpage{312}
(\byear{1959})
\end{barticle}
\endbibitem

\bibitem[\protect\citeauthoryear{{Tayler}}{1973}]{t73}
\begin{barticle}
\bauthor{\bsnm{{Tayler}}, \binits{R.J.}}:
\bjtitle{\mnras}
\bvolume{161},
\bfpage{365}
(\byear{1973})
\end{barticle}
\endbibitem

\bibitem[\protect\citeauthoryear{Tevzadze et~al.}{2012}]{Tevzadze2012}
\begin{barticle}
\bauthor{\bsnm{Tevzadze}, \binits{A.G.}},
\bauthor{\bsnm{Kisslinger}, \binits{L.}},
\bauthor{\bsnm{Brandenburg}, \binits{A.}},
\bauthor{\bsnm{Kahniashvili}, \binits{T.}}:
\bjtitle{The Astrophysical Journal}
\bvolume{759}(\bissue{1}),
\bfpage{54}
(\byear{2012}).
doi:\doiurl{10.1088/0004-637X/759/1/54}
\end{barticle}
\endbibitem

\bibitem[\protect\citeauthoryear{{Thompson} and {Duncan}}{1993}]{td93}
\begin{barticle}
\bauthor{\bsnm{{Thompson}}, \binits{C.}},
\bauthor{\bsnm{{Duncan}}, \binits{R.C.}}:
\bjtitle{\apj}
\bvolume{408},
\bfpage{194}
(\byear{1993}).
doi:\doiurl{10.1086/172580}
\end{barticle}
\endbibitem

\bibitem[\protect\citeauthoryear{Tobias et~al.}{2011}]{Tobias2011}
\begin{barticle}
\bauthor{\bsnm{Tobias}, \binits{S.M.}},
\bauthor{\bsnm{Cattaneo}, \binits{F.}},
\bauthor{\bsnm{Boldyrev}, \binits{S.}}:
\bjtitle{eprint arXiv}
\bvolume{1103},
\bfpage{3138}
(\byear{2011})
\end{barticle}
\endbibitem

\bibitem[\protect\citeauthoryear{{Tout} et~al.}{2004}]{twf04}
\begin{barticle}
\bauthor{\bsnm{{Tout}}, \binits{C.A.}},
\bauthor{\bsnm{{Wickramasinghe}}, \binits{D.T.}},
\bauthor{\bsnm{{Ferrario}}, \binits{L.}}:
\bjtitle{\mnras}
\bvolume{355},
\bfpage{13}
(\byear{2004}).
doi:\doiurl{10.1111/j.1365-2966.2004.08482.x}
\end{barticle}
\endbibitem

\bibitem[\protect\citeauthoryear{{Tout} et~al.}{2008}]{Tout2008}
\begin{barticle}
\bauthor{\bsnm{{Tout}}, \binits{C.A.}},
\bauthor{\bsnm{{Wickramasinghe}}, \binits{D.T.}},
\bauthor{\bsnm{{Liebert}}, \binits{J.}},
\bauthor{\bsnm{{Ferrario}}, \binits{L.}},
\bauthor{\bsnm{{Pringle}}, \binits{J.E.}}:
\bjtitle{\mnras}
\bvolume{387},
\bfpage{897}
(\byear{2008}).
\arxivurl{0805.0115}.
doi:\doiurl{10.1111/j.1365-2966.2008.13291.x}
\end{barticle}
\endbibitem

\bibitem[\protect\citeauthoryear{{Tsubota} et~al.}{2003}]{tka03}
\begin{botherref}
\oauthor{\bsnm{{Tsubota}}, \binits{M.}},
\oauthor{\bsnm{{Kasamatsu}}, \binits{K.}},
\oauthor{\bsnm{{Araki}}, \binits{T.}}:
eprint arXiv:cond-mat/0309364
(2003).
\arxivurl{cond-mat/0309364}
\end{botherref}
\endbibitem

\bibitem[\protect\citeauthoryear{{Tsubota} et~al.}{2013}]{tkt12}
\begin{barticle}
\bauthor{\bsnm{{Tsubota}}, \binits{M.}},
\bauthor{\bsnm{{Kobayashi}}, \binits{M.}},
\bauthor{\bsnm{{Takeuchi}}, \binits{H.}}:
\bjtitle{\physrep}
\bvolume{522},
\bfpage{191}
(\byear{2013}).
\arxivurl{1208.0422}.
doi:\doiurl{10.1016/j.physrep.2012.09.007}
\end{barticle}
\endbibitem

\bibitem[\protect\citeauthoryear{{Tutukov} and {Fedorova}}{2010}]{Tutukov2010}
\begin{barticle}
\bauthor{\bsnm{{Tutukov}}, \binits{A.V.}},
\bauthor{\bsnm{{Fedorova}}, \binits{A.V.}}:
\bjtitle{Astronomy Reports}
\bvolume{54},
\bfpage{156}
(\byear{2010}).
doi:\doiurl{10.1134/S1063772910020083}
\end{barticle}
\endbibitem

\bibitem[\protect\citeauthoryear{{van Eysden} and {Melatos}}{2010}]{vem10}
\begin{barticle}
\bauthor{\bsnm{{van Eysden}}, \binits{C.A.}},
\bauthor{\bsnm{{Melatos}}, \binits{A.}}:
\bjtitle{\mnras}
\bvolume{409},
\bfpage{1253}
(\byear{2010}).
\arxivurl{1007.4360}.
doi:\doiurl{10.1111/j.1365-2966.2010.17387.x}
\end{barticle}
\endbibitem

\bibitem[\protect\citeauthoryear{{Vennes}}{1999}]{Vennes1999}
\begin{barticle}
\bauthor{\bsnm{{Vennes}}, \binits{S.}}:
\bjtitle{\apj}
\bvolume{525},
\bfpage{995}
(\byear{1999}).
doi:\doiurl{10.1086/307949}
\end{barticle}
\endbibitem

\bibitem[\protect\citeauthoryear{{Vigan{\`o}} et~al.}{2012}]{vpm12}
\begin{barticle}
\bauthor{\bsnm{{Vigan{\`o}}}, \binits{D.}},
\bauthor{\bsnm{{Pons}}, \binits{J.A.}},
\bauthor{\bsnm{{Miralles}}, \binits{J.A.}}:
\bjtitle{Computer Physics Communications}
\bvolume{183},
\bfpage{2042}
(\byear{2012}).
\arxivurl{1204.4707}.
doi:\doiurl{10.1016/j.cpc.2012.04.029}
\end{barticle}
\endbibitem

\bibitem[\protect\citeauthoryear{{Vigelius} et~al.}{2007}]{vetal07}
\begin{barticle}
\bauthor{\bsnm{{Vigelius}}, \binits{M.}},
\bauthor{\bsnm{{Melatos}}, \binits{A.}},
\bauthor{\bsnm{{Chatterjee}}, \binits{S.}},
\bauthor{\bsnm{{Gaensler}}, \binits{B.M.}},
\bauthor{\bsnm{{Ghavamian}}, \binits{P.}}:
\bjtitle{\mnras}
\bvolume{374},
\bfpage{793}
(\byear{2007}).
\arxivurl{astro-ph/0610454}.
doi:\doiurl{10.1111/j.1365-2966.2006.11193.x}
\end{barticle}
\endbibitem

\bibitem[\protect\citeauthoryear{Vincenzi}{2001}]{Vincenzi2001}
\begin{botherref}
\oauthor{\bsnm{Vincenzi}, \binits{D.}}:
eprint arXiv,
6090
(2001)
\end{botherref}
\endbibitem

\bibitem[\protect\citeauthoryear{{Vink} and {Kuiper}}{2006}]{vk06}
\begin{barticle}
\bauthor{\bsnm{{Vink}}, \binits{J.}},
\bauthor{\bsnm{{Kuiper}}, \binits{L.}}:
\bjtitle{\mnras}
\bvolume{370},
\bfpage{14}
(\byear{2006}).
\arxivurl{astro-ph/0604187}.
doi:\doiurl{10.1111/j.1745-3933.2006.00178.x}
\end{barticle}
\endbibitem

\bibitem[\protect\citeauthoryear{{Wade} and {the MiMeS
  Collaboration}}{2014}]{Wade2014}
\begin{botherref}
\oauthor{\bsnm{{Wade}}, \binits{G.A.}},
\oauthor{\bsnm{{the MiMeS Collaboration}}}:
ArXiv e-prints
(2014).
\arxivurl{1411.3604}
\end{botherref}
\endbibitem

\bibitem[\protect\citeauthoryear{{Wade} et~al.}{2011}]{Wade2011}
\begin{bchapter}
\bauthor{\bsnm{{Wade}}, \binits{G.A.}},
\bauthor{\bsnm{{Alecian}}, \binits{E.}},
\bauthor{\bsnm{{Bohlender}}, \binits{D.A.}},
\bauthor{\bsnm{{Bouret}}, \binits{J.-C.}},
\bauthor{\bsnm{{Cohen}}, \binits{D.H.}},
\bauthor{\bsnm{{Duez}}, \binits{V.}},
\bauthor{\bsnm{{Gagn{\'e}}}, \binits{M.}},
\bauthor{\bsnm{{Grunhut}}, \binits{J.H.}},
\bauthor{\bsnm{{Henrichs}}, \binits{H.F.}},
\bauthor{\bsnm{{Hill}}, \binits{N.R.}},
\bauthor{\bsnm{{Kochukhov}}, \binits{O.}},
\bauthor{\bsnm{{Mathis}}, \binits{S.}},
\bauthor{\bsnm{{Neiner}}, \binits{C.}},
\bauthor{\bsnm{{Oksala}}, \binits{M.E.}},
\bauthor{\bsnm{{Owocki}}, \binits{S.}},
\bauthor{\bsnm{{Petit}}, \binits{V.}},
\bauthor{\bsnm{{Shultz}}, \binits{M.}},
\bauthor{\bsnm{{Rivinius}}, \binits{T.}},
\bauthor{\bsnm{{Townsend}}, \binits{R.H.D.}},
\bauthor{\bsnm{{Vink}}, \binits{J.S.}},
\bauthor{\bsnm{{Vink}}}:
In: \beditor{\bsnm{{Neiner}}, \binits{C.}},
\beditor{\bsnm{{Wade}}, \binits{G.}},
\beditor{\bsnm{{Meynet}}, \binits{G.}},
\beditor{\bsnm{{Peters}}, \binits{G.}} (eds.)
\bbtitle{IAU Symposium}.
\bsertitle{IAU Symposium},
vol. \bseriesno{272},
p. \bfpage{118}
(\byear{2011}).
\arxivurl{1009.3563}.
doi:\doiurl{10.1017/S1743921311010131}
\end{bchapter}
\endbibitem

\bibitem[\protect\citeauthoryear{{Wade} et~al.}{2014a}]{WadeFolsom2014}
\begin{barticle}
\bauthor{\bsnm{{Wade}}, \binits{G.A.}},
\bauthor{\bsnm{{Folsom}}, \binits{C.P.}},
\bauthor{\bsnm{{Petit}}, \binits{P.}},
\bauthor{\bsnm{{Petit}}, \binits{V.}},
\bauthor{\bsnm{{Ligni{\`e}res}}, \binits{F.}},
\bauthor{\bsnm{{Auri{\`e}re}}, \binits{M.}},
\bauthor{\bsnm{{B{\"o}hm}}, \binits{T.}}:
\bjtitle{\mnras}
\bvolume{444},
\bfpage{1993}
(\byear{2014}a).
\arxivurl{1407.3991}.
doi:\doiurl{10.1093/mnras/stu1541}
\end{barticle}
\endbibitem

\bibitem[\protect\citeauthoryear{{Wade} et~al.}{2014b}]{Wade2013}
\begin{bchapter}
\bauthor{\bsnm{{Wade}}, \binits{G.A.}},
\bauthor{\bsnm{{Grunhut}}, \binits{J.}},
\bauthor{\bsnm{{Alecian}}, \binits{E.}},
\bauthor{\bsnm{{Neiner}}, \binits{C.}},
\bauthor{\bsnm{{Auri{\`e}re}}, \binits{M.}},
\bauthor{\bsnm{{Bohlender}}, \binits{D.A.}},
\bauthor{\bsnm{{David-Uraz}}, \binits{A.}},
\bauthor{\bsnm{{Folsom}}, \binits{C.}},
\bauthor{\bsnm{{Henrichs}}, \binits{H.F.}},
\bauthor{\bsnm{{Kochukhov}}, \binits{O.}},
\bauthor{\bsnm{{Mathis}}, \binits{S.}},
\bauthor{\bsnm{{Owocki}}, \binits{S.}},
\bauthor{\bsnm{{Petit}}, \binits{V.}},
\bauthor{\bsnm{{Petit}}}:
In: \bbtitle{IAU Symposium}.
\bsertitle{IAU Symposium},
vol. \bseriesno{302},
p. \bfpage{265}
(\byear{2014}b).
\arxivurl{1310.3965}.
doi:\doiurl{10.1017/S1743921314002233}
\end{bchapter}
\endbibitem

\bibitem[\protect\citeauthoryear{{Wang} et~al.}{2006}]{wlh06}
\begin{barticle}
\bauthor{\bsnm{{Wang}}, \binits{C.}},
\bauthor{\bsnm{{Lai}}, \binits{D.}},
\bauthor{\bsnm{{Han}}, \binits{J.L.}}:
\bjtitle{\apj}
\bvolume{639},
\bfpage{1007}
(\byear{2006}).
\arxivurl{astro-ph/0509484}.
doi:\doiurl{10.1086/499397}
\end{barticle}
\endbibitem

\bibitem[\protect\citeauthoryear{{Warszawski} and {Melatos}}{2013}]{wm13}
\begin{barticle}
\bauthor{\bsnm{{Warszawski}}, \binits{L.}},
\bauthor{\bsnm{{Melatos}}, \binits{A.}}:
\bjtitle{\mnras}
\bvolume{428},
\bfpage{1911}
(\byear{2013}).
\arxivurl{1210.2203}.
doi:\doiurl{10.1093/mnras/sts108}
\end{barticle}
\endbibitem

\bibitem[\protect\citeauthoryear{{Wickramasinghe} and
  {Ferrario}}{2005}]{Wick2005}
\begin{barticle}
\bauthor{\bsnm{{Wickramasinghe}}, \binits{D.T.}},
\bauthor{\bsnm{{Ferrario}}, \binits{L.}}:
\bjtitle{\mnras}
\bvolume{356},
\bfpage{1576}
(\byear{2005}).
doi:\doiurl{10.1111/j.1365-2966.2004.08603.x}
\end{barticle}
\endbibitem

\bibitem[\protect\citeauthoryear{{Wickramasinghe} et~al.}{2014}]{Wick2014}
\begin{barticle}
\bauthor{\bsnm{{Wickramasinghe}}, \binits{D.T.}},
\bauthor{\bsnm{{Tout}}, \binits{C.A.}},
\bauthor{\bsnm{{Ferrario}}, \binits{L.}}:
\bjtitle{\mnras}
\bvolume{437},
\bfpage{675}
(\byear{2014}).
\arxivurl{1310.2696}.
doi:\doiurl{10.1093/mnras/stt1910}
\end{barticle}
\endbibitem

\bibitem[\protect\citeauthoryear{{Woltjer}}{1964}]{w64}
\begin{barticle}
\bauthor{\bsnm{{Woltjer}}, \binits{L.}}:
\bjtitle{\apj}
\bvolume{140},
\bfpage{1309}
(\byear{1964}).
doi:\doiurl{10.1086/148028}
\end{barticle}
\endbibitem

\bibitem[\protect\citeauthoryear{{Woods}}{2008}]{w08}
\begin{bchapter}
\bauthor{\bsnm{{Woods}}, \binits{P.M.}}:
In: \beditor{\bsnm{{Bassa}}, \binits{C.}},
\beditor{\bsnm{{Wang}}, \binits{Z.}},
\beditor{\bsnm{{Cumming}}, \binits{A.}},
\beditor{\bsnm{{Kaspi}}, \binits{V.M.}} (eds.)
\bbtitle{40 Years of Pulsars: Millisecond Pulsars, Magnetars and More}.
\bsertitle{American Institute of Physics Conference Series},
vol. \bseriesno{983},
p. \bfpage{227}
(\byear{2008}).
doi:\doiurl{10.1063/1.2900149}
\end{bchapter}
\endbibitem

\bibitem[\protect\citeauthoryear{{Wright}}{1973}]{w73}
\begin{barticle}
\bauthor{\bsnm{{Wright}}, \binits{G.A.E.}}:
\bjtitle{\mnras}
\bvolume{162},
\bfpage{339}
(\byear{1973})
\end{barticle}
\endbibitem

\bibitem[\protect\citeauthoryear{{Zahn} et~al.}{2007}]{Zahn2007}
\begin{barticle}
\bauthor{\bsnm{{Zahn}}, \binits{J.-P.}},
\bauthor{\bsnm{{Brun}}, \binits{A.S.}},
\bauthor{\bsnm{{Mathis}}, \binits{S.}}:
\bjtitle{\aap}
\bvolume{474},
\bfpage{145}
(\byear{2007}).
\arxivurl{0707.3287}.
doi:\doiurl{10.1051/0004-6361:20077653}
\end{barticle}
\endbibitem

\bibitem[\protect\citeauthoryear{{Zhang}}{2003}]{Zhang2003}
\begin{bchapter}
\bauthor{\bsnm{{Zhang}}, \binits{B.}}:
In: \beditor{\bsnm{{Cheng}}, \binits{K.S.}},
\beditor{\bsnm{{Leung}}, \binits{K.C.}},
\beditor{\bsnm{{Li}}, \binits{T.P.}} (eds.)
\bbtitle{Astrophysics and Space Science Library}.
\bsertitle{Astrophysics and Space Science Library},
vol. \bseriesno{298},
p. \bfpage{27}
(\byear{2003}).
\arxivurl{astro-ph/0212016}
\end{bchapter}
\endbibitem

\bibitem[\protect\citeauthoryear{Zrake}{2014}]{Zrake2014}
\begin{barticle}
\bauthor{\bsnm{Zrake}, \binits{J.}}:
\bjtitle{The Astrophysical Journal}
\bvolume{794}(\bissue{2}),
\bfpage{26}
(\byear{2014}).
doi:\doiurl{10.1088/2041-8205/794/2/L26}
\end{barticle}
\endbibitem

\bibitem[\protect\citeauthoryear{Zrake and MacFadyen}{2011}]{Zrake2011}
\begin{barticle}
\bauthor{\bsnm{Zrake}, \binits{J.}},
\bauthor{\bsnm{MacFadyen}, \binits{A.I.}}:
\bjtitle{The Astrophysical Journal}
\bvolume{744}(\bissue{1}),
\bfpage{32}
(\byear{2011})
\end{barticle}
\endbibitem

\bibitem[\protect\citeauthoryear{Zrake and MacFadyen}{2013}]{Zrake2013}
\begin{barticle}
\bauthor{\bsnm{Zrake}, \binits{J.}},
\bauthor{\bsnm{MacFadyen}, \binits{A.I.}}:
\bjtitle{The Astrophysical Journal}
\bvolume{769}(\bissue{2}),
\bfpage{29}
(\byear{2013}).
doi:\doiurl{10.1088/2041-8205/769/2/L29}
\end{barticle}
\endbibitem

\end{thebibliography}

\end{document}